\documentstyle[preprint]{aastex}
\newcommand\wave[1]{\mbox{$\lambda$#1\,\AA}}
\newcommand\eps[1]{\mbox{log~$\epsilon$(#1)}} 
\def\kmsec{\mbox{km~s$^{\rm -1}$}}
\def\teff{\mbox{T$_{\rm eff}$}}
\def\logg{\mbox{log~{\it g}}}
\def\BmV0{\mbox{{\it (B-V)$^o$}}}
\def\VmK0{\mbox{{\it (V-K)$^o$}}}
\def\MV{\mbox{{\it M$_V$}}}
\def\MV0{\mbox{{\it M$_V^o$}}}
\def\mM{\mbox{{\it (m-M)$^o$}}}
\def\Msun{\mbox{$\mathcal{M}_{\odot}$}}

\def\etal{\mbox{{\it et al.}}}
\def\eg{\mbox{{\it e.g.}}}
\def\ie{\mbox{{\it i.e.}}}

\begin{document}

\received{}
\revised{}
\accepted{}

\shorttitle{Chemical Composition of Globular Cluster M5}
\shortauthors{Ivans \etal}

\title{New Analyses of Star-to-Star Abundance Variations Among Bright 
Giants in the Mildly Metal-Poor Globular Cluster M5\footnote{
Based in part on observations obtained with the W.~M.~Keck Observatory, 
which is operated by the California Association for Research in 
Astronomy, Inc., on behalf of the University of California, the
California Institute of Technology and the National Aeronautics and Space 
Administration.
}
}

\author{
Inese I.~Ivans\altaffilmark{2,3}, 
Robert P.~Kraft\altaffilmark{4}, 
Christopher Sneden\altaffilmark{2}, 
Graeme H.~Smith\altaffilmark{4},
R.~Michael Rich\altaffilmark{5},
Matthew Shetrone\altaffilmark{6}
}

\altaffiltext{2}{Department of Astronomy and McDonald Observatory,
University of Texas, RLM 15.308, Mail Code c1400, Austin, TX 78712; 
iivans@astro.as.utexas.edu; chris@verdi.as.utexas.edu}

\altaffiltext{3}{Research School of Astronomy \& Astrophysics, 
Australian National University, Mount Stromlo Observatory, Cotter 
Road, Weston ACT 2611, Australia}

\altaffiltext{4}{UCO/Lick Observatory, University of California, 
Santa Cruz, CA 95064; kraft@ucolick.org; graeme@ucolick.org}

\altaffiltext{5}{Dept.~of Physics \& Astronomy, UCLA, 
Math-Sciences 8979, Los Angeles, CA 90095-1562; rmr@astro.ucla.edu}

\altaffiltext{6}{McDonald Observatory, University of Texas, HC 75, 
Box 1337 L, Fort Davis, TX 79734; shetrone@astro.as.utexas.edu}

%%%%%%%%%%%%%%%%
\begin{abstract}
%%%%%%%%%%%%%%%%

We present a chemical composition analysis of 36 giant stars in the mildly 
metal-poor ($<$[Fe/H]$>$ = --1.21) globular cluster M5 (NGC~5904).  The 
analysis makes use of high resolution data acquired for 25 stars at the 
Keck~I telescope, as well as a re-analysis of the high resolution spectra 
for 13 stars acquired for an earlier study at Lick Observatory.  We 
employed two analysis techniques: one, adopting standard spectroscopic 
constraints, including setting the surface gravity from the ionization 
equilibrium of iron, and two, subsequent to investigating alternative
approaches, adopting an analysis consistent with the non-LTE precepts as 
recently described by Th\'evenin \& Idiart.  The abundance ratios we 
derive for magnesium, silicon, calcium, scandium, titanium, vanadium, nickel, 
barium and europium in M5 show no significant abundance variations and the 
ratios are comparable to those of halo field stars.  However, large 
variations are seen in the abundances of oxygen, sodium and aluminum, the 
elements that are sensitive to proton-capture nucleosynthesis.  These 
variations are  well-correlated with the CN bandstrength index S(3839).  
Surprisingly, in M5 the dependence of the abundance variations on \logg\ is 
in the opposite sense to that discovered in M13 by the Lick-Texas group 
where the relationship provided strong evidence in support of the 
evolutionary scenario.  The present analysis of M5 giants does not 
necessarily rule out an evolutionary scenario, but it provides no support 
for it either.  In comparing the abundances of M5 and M4 (NGC~6121), another 
mildly metal-poor ($<$[Fe/H]$>$ = --1.08) globular cluster, we find that 
silicon, aluminum, barium and lanthanum are overabundant in M4 with respect 
to what is seen in M5, confirming and expanding the results of previous 
studies.  In comparing the abundances between these two clusters and others 
having comparable metallicities, we find that the anti-correlations observed 
in M5 are similar to those found in more metal-poor clusters, M3, M10 and 
M13 ($<$[Fe/H]$>$ = --1.5 to --1.6), whereas the behavior in M4 is more like 
that of the more metal-rich globular cluster M71 ($<$[Fe/H]$> \sim$ --0.8).  
We conclude that among stars in Galactic globular clusters, there is no 
definitive ``single'' value of [el/Fe] at a given [Fe/H] for at least some 
alpha-capture, odd-Z and slow neutron-capture process elements, in this case, 
silicon, aluminum, barium and lanthanum.

\end{abstract}
\keywords{Galaxy: abundances --- globular clusters: general --- globular 
clusters individual (NGC~5904) --- stars: abundances --- stars: fundamental 
parameters}

%Section 1.0
%%%%%%%%%%%%%%%%%%%%%%
\section{Introduction}
%%%%%%%%%%%%%%%%%%%%%%

Large star-to-star abundance variations in the light elements C, N, O, Na, 
Mg, and Al are commonly found among the bright giant stars of metal-poor 
globular clusters.  Some star-to-star abundance variations exist in all 
metal-poor globular clusters in which the variations have been sought.  In 
clusters with sufficiently large sample sizes, N is typically 
anti-correlated with O and C, Na is anti-correlated with O, and Al is 
anti-correlated with Mg.  The reader is referred to reviews by Suntzeff 
(1993)\nocite{Su93}, Kraft (1994)\nocite{Kr94}, Briley \etal\ 
(1994)\nocite{BBHS94}, Da~Costa (1997)\nocite{DaC97}, Wallerstein \etal\ 
(1997)\nocite{Wetal97}, and Sneden (1999, 2000)\nocite{Sn99,Sn00} for 
detailed discussions of these abundance trends.  Except for 
anti-correlated behavior of N with respect to O and C, halo field giants 
do not exhibit the variations in Na, Mg, and Al that are seen among 
globular cluster giants (Pilachowski \etal\ 1996a\nocite{PSKL96}, Hanson 
\etal\ 1998\nocite{HSKF98}, Gratton \etal\ 2000\nocite{GSCB00}).

Most studies agree that the abundance anti-correlations found among cluster 
giants result from proton-capture nucleosynthesis that converts C and O into N,
Ne into Na, and Mg into Al in and above the hydrogen-burning shells of 
evolved stars (see \eg, Denissenkov \etal\ 1990\nocite{DDNW98}; Cavallo \& 
Nagar 2000\nocite{CN00} and references therein).  However it is less clear 
whether the synthesis takes place in the giants we presently observe (the 
``evolutionary'' scenario) or in a prior generation of more massive 
evolved stars (the ``primordial'' scenario) which selectively ``polluted'' 
the gas from which the present generation of stars was formed.  Evidence 
mounts that both scenarios are needed: a typical cluster contains main 
sequence stars already imprinted with variations in these elements, as 
studies of main sequence stars in 47~Tuc and NGC~6752 dramatically 
illustrate (Briley \etal\ 1995\nocite{Betal95}, Gratton \etal\ 
2001\nocite{Git01}).  These abundances may, however, be further modified 
when the stellar envelope is cycled through the H-burning shell as stars 
approach the red giant tip (see reviews by Briley \etal\ 
1994\nocite{BBHS94}, Kraft 2001\nocite{Kr01} and references therein).

In any given luminosity interval on the giant branch of a typical globular
cluster, there are stars with a range of Na, O, Mg, and Al abundances, 
usually exhibiting the anti-correlations noted above.  One possible 
expectation of the evolutionary scenario is that the distribution of these 
O and Na (or Mg and Al) abundances should change with advancing 
evolutionary state.  Thus as evolution proceeds, one might expect to find 
relatively more stars with low O and Mg and fewer with high O and Mg, and 
correspondingly more with high Na and Al and fewer with low Na and Al. 
This is indeed the case for M13 (Kraft \etal\ 1997\nocite{KSSSLP97},
Hanson \etal\ 1998\nocite{HSKF98}), in which there are different mean 
O, Na, Mg, and Al abundances for stars above and below \MV0~$\simeq$ 
--1.7 (or \logg~$\simeq$ 1.0), a point 0.8 mag below the red giant tip. 
For other clusters less is known because of flux limitations 
at faint magnitudes.  But even in M13, giants with \MV0~$>$ --1.7 exhibit 
the same spread and distribution of Na and Al abundances (Pilachowski 
\etal\ 1996b\nocite{PSKL96}, Cavallo \& Nagar 2000), independent of 
luminosity, to levels one magnitude below the horizontal branch (HB). 
In the more metal-poor clusters M92 and M15, there is no apparent 
change in the distribution of Na abundances with luminosity from the red 
giant tip to levels just above the HB (Sneden \etal\ 2000\nocite{SPK00}).
In the more metal-rich cluster M4, although the variations in C, N, O, 
Na, Mg, and Al are smaller than in M13, again the distributions show 
little dependence on evolutionary state (Ivans \etal\ 
1999\nocite{Ivetal99}, hereafter called I99-M4).

M5 is a mildly metal-poor cluster ($<$[Fe/H]$>$~= --1.4, Zinn \& West 
1984\nocite{ZW84}; $<$[Fe/H]$>$~= --1.17, Sneden \etal\ 
1992\nocite{SKPL92}; $<$[Fe/H]$>$~= --1.11, Carretta \& Gratton 
1997\nocite{CG97}) in which bright giants exhibit anti-correlated 
behavior of C and O with respect to N (Smith \etal\ 
1997\nocite{SSBCB97}), as well as an anti-correlation of O with Na 
(Sneden \etal\ 1992\nocite{SKPL92}, hereafter called S92-M5); Al and Mg 
abundance relationships have not been explored.  The cluster exhibits 
bimodal distributions of CN-strength on both the first ascent giant 
branch (``RGB''; Smith \& Norris 1983\nocite{SN83}) and asymptotic giant 
branch (``AGB''; Smith \& Norris 1993\nocite{SN93}).  At least one giant, 
IV-59, is known to have both high N and O, which again suggests the 
existence of primordial variations (Smith \etal\ 1997\nocite{SSBCB97}).  
However, is there evidence for an increase in the number of O-poor and 
Na-rich stars as evolutionary state advances?  Previous M5 sample sizes 
have been too small to explore whether in this cluster a shift with 
$M_V$ exists in the distribution of Na and O compatible with an 
evolutionary scenario.  We report here an exploration of this question, 
based on high resolution spectra of a sample of 36 giants ranging in 
luminosity from the RGB tip to \MV0~$\sim$ --0.5, \ie, about one 
magnitude above the HB. 

M4, a cluster with metallicity comparable to that of M5, has unusually 
high abundances of the $\alpha$-element Si, the light odd-Z element Al, 
and the $s$-process elements Ba and La (Brown \& Wallerstein 
1992\nocite{BW92}; I99-M4) in comparison to typical halo field giants of 
similar metallicity (see \eg, Gratton \& Sneden 1991\nocite{GS91}, 
1994\nocite{GS94}, Shetrone 1996\nocite{Sh96}), which follow an 
extrapolation of the abundance trends seen among halo stars of lower 
metallicities (see \eg, McWilliam \etal\ 1995\nocite{MPSS95}, Ryan 
\etal\ 1996\nocite{RNB96} and references therein).  We therefore compare 
[el/Fe]-ratios in M5 with those in M4 and the halo field, noting that M5 
pursues a galactic orbit with an apogalacticon in the outer reaches of 
the Galactic halo (Cudworth \& Hanson 1993\nocite{CH93}), where 
clusters having ``abnormal'' [el/Fe] ratios are sometimes found.

We introduce here for the first time an analysis of cluster [Fe/H] ratios 
based on an approach in which allowance is made for the over-ionization 
of Fe in the atmospheres of low-metallicity giants.  We estimate as well 
the effect of these non-local thermodynamic equilibrium (non-LTE) 
precepts on the derivation of [el/Fe] ratios.  Abundances based on more 
traditional methods of analysis are, however, retained so that the reader 
may judge the extent of the proposed modifications.

%Section 2.0
%%%%%%%%%%%%%%%%%%%%%%%%%%%%%%%%%%%%%%%%%%%%%%%%%%%%%%
\section{Observations, Reductions and EW Measurements} 
%%%%%%%%%%%%%%%%%%%%%%%%%%%%%%%%%%%%%%%%%%%%%%%%%%%%%%

Our prior study of 13 bright M5 giants (S92-M5) was based on high 
resolution (R~$\sim$ 30,000) spectra obtained with the Lick 3.0m
telescope and Hamilton coud\'e echelle spectrograph (Vogt 
1987\nocite{Vo87}).  The faintest stars observed in the Lick sample had 
V~$\sim$ 13.0, the practical limit for observations with 
signal-to-noise S/N~$\gtrsim$ 50 in reasonable integration times 
($\sim$120 minutes) using the 3.0m telescope.  However, stars near 
\MV0~$\sim$ --0.5 have $V$~$\sim$~14.0 in M5; obtaining spectra of 
high resolution and adequate S/N for such stars required use of the 
HIRES spectrograph of the Keck~I telescope (Vogt \etal\ 
1994\nocite{Voetal94}).

For the Keck observations, the entrance slit was set to a width of 
0.86$\arcsec$, which corresponds to a spectral resolving power of 
R~$\simeq$ 45,000 at the Tektronix 2048$\times$2048 pixel detector. 
In Table~\ref{m5.tab1}, we present an observing log of the 25 M5 
giants observed with HIRES, along with estimated S/N near \wave{6300},
values of $V^o$, \BmV0\ and \MV0\ for each star, assuming a reddening
$E(B-V)$~= 0.03 and a true distance modulus \mM~= 14.40 (Djorgovski 
1993\nocite{Dj93}).  We adopted the observed colors and magnitudes of 
Sandquist \etal\ (1996\nocite{SBSH96}; 2000, private communication)
for all but three stars which were unobserved in the Sandquist \etal\
study.  The photometry for star II-9 was taken from Cudworth 
(1979)\nocite{Cu79} and for star III-149, we used that of Rees 
(1993)\nocite{Re93}.  G2 is discussed below.

\begin{figure} %FIGURE 1
\epsscale{1.0}
\plotone{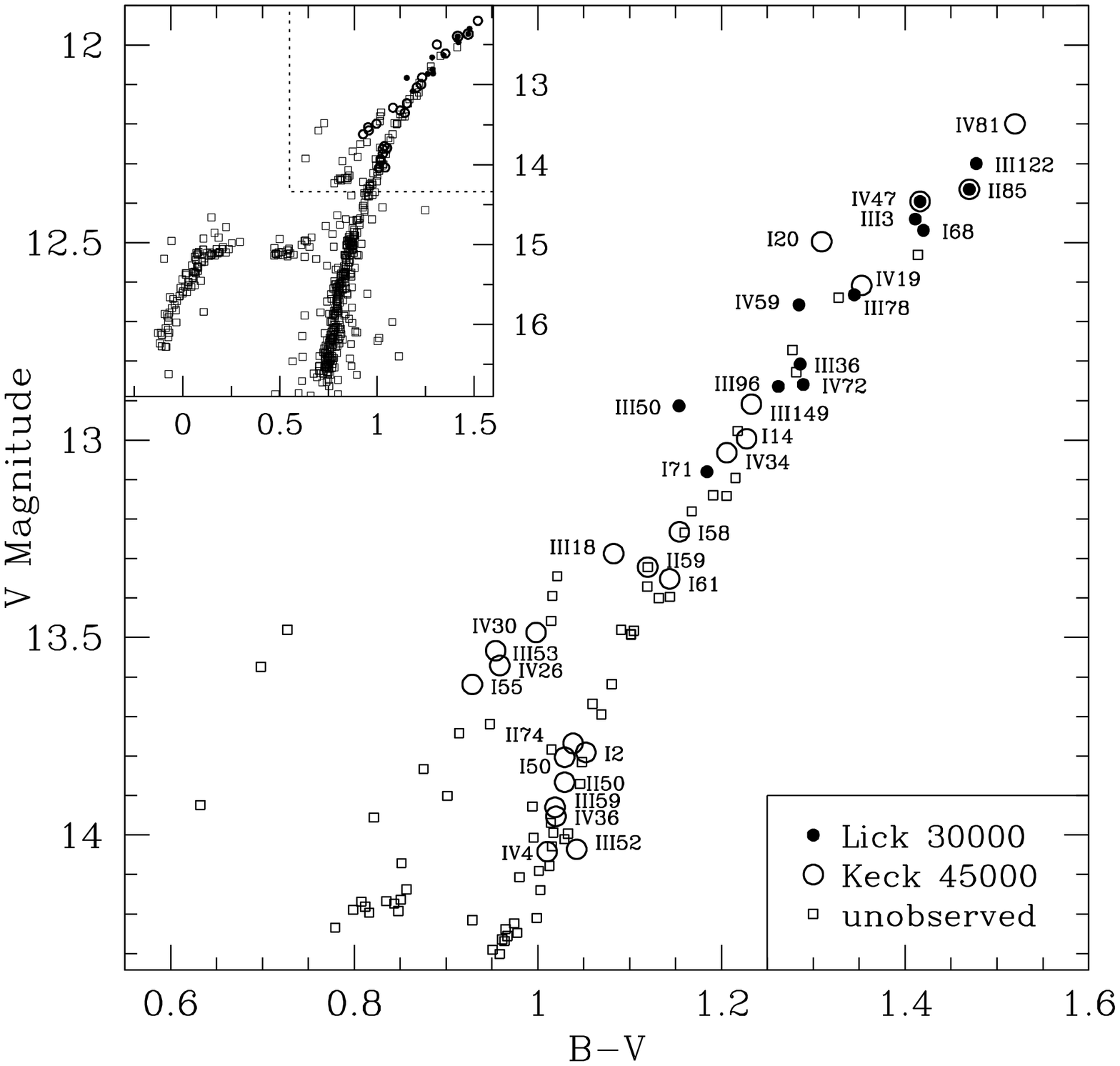}
\caption{CMD of M5 with photometry from Sandquist \etal\ (1996; 2000, 
private communication), showing the positions of our program stars on 
the AGB and RGB.  The symbols are given in the figure legend and 
correspond to the observatory and resolution used for each 
observation.  The inset diagram shows the program stars plotted in
relation to all Sandquist \etal\ stars of magnitude $\leq$ 16.9.
\label{m5.fig1}}
\end{figure}

Two of the stars observed at Lick (II-85 and IV-47) were re-observed
using HIRES at Keck~I, in order to study possible systematic offsets
in equivalent width (EW) and/or differences in analysis procedure 
between the Lick and Keck data.  Combining the two data sets, we are able 
to study abundances and abundance ratios in 36 M5 giants on both the RGB 
and AGB, ranging in luminosity from $M_{bol}$~$\simeq$ --1.0 to --3.4,
corresponding to an effective temperature range of \teff~$\simeq$ 4750~K 
to 3900~K.  Of these 36 stars, 34 are proper motion members of M5 
according to the catalog of Rees (1993)\nocite{Re93}; the remainder have 
colors, magnitudes, abundances, and radial velocities compatible with 
membership (discussed further in the next paragraph).  Eight of the stars 
are members of the AGB; a small fraction of stars brighter than $V$~= 
12.8 ($\sim$20\%, based on comparative lifetimes), in the region of the 
color-magnitude diagram where the RGB and AGB cannot be distinguished,
may also be AGB members.  Of 30 stars observed by us for which the CN 
strength index S(3839) has previously been measured by Smith \& Norris 
(1983\nocite{SN83}, 1993\nocite{SN93}), Briley \& Smith 
(1993\nocite{BS93}), and Smith \etal\ (1997\nocite{SSBCB97}), we observed 
14 CN-strong stars and 16 CN-weak stars.  Altogether, we observed 29\% of 
the 118 giants brighter than $V$~= 14.1 that were cataloged by Rees, plus 
two additional members not in the catalog.  Our sample is thus reasonably 
representative of the population of bright M5 giants. In 
Figure~\ref{m5.fig1} we exhibit the color-magnitude array of the 
brighter stars in M5, based on the CCD photometry of Sandquist \etal\ 
(1996\nocite{SBSH96}; 2000, private communication).  This figure 
illustrates the evolutionary domain of our program stars.  The labels in 
the figure identify the stars of this study plus those of S92-M5.

Two stars listed in Table~\ref{m5.tab1} require special comment. 
The one designated as G2, close to the central region of the cluster, is 
not to be found in published photometry but was revealed as a bright 
red star in a 2$\micron$ image of the cluster kindly obtained by 
Kirk Gilmore using the Lick 1.0m telescope.  G2 can be seen in the map 
(Figure~12) of Buonanno \etal\ (1981)\nocite{BCF81} and its estimated 
position is $\alpha$(1950)~= 15$^h$16$^m$04$^s$, $\delta$(1950)~= 
+02$\arcdeg$14$\arcmin$54$\arcsec$. It is a radial velocity member. 
A second star, listed here as ``III-149'' to prevent confusion, was
accidentally observed in the mistaken belief that it was III-147. 
It is actually the star, not numbered in Buonanno \etal, lying 
10$\arcsec$ west and 4$\arcsec$ south of III-147, essentially at the 
right-hand edge of Figure~12 of Buonanno {\it et al.}  It too is a 
radial velocity member of M5.

Processing of the raw spectra was carried out using the standard
IRAF software package\footnote{ 
IRAF is distributed by the National Optical Astronomy Observatories, 
which are operated by the Association of Universities for Research in 
Astronomy, Inc., under cooperative agreement with the National Science 
Foundation.}.  
The CCD frames were corrected for both bias and flat-field effects and 
the individual orders were extracted. Further analysis was performed 
using the SPECTRE code (Fitzpatrick \& Sneden 1987\nocite{FS87}); this 
involved continuum placement and normalization, cosmic ray removal, a 
wavelength calibration using stellar absorption lines within each order 
and removal of telluric absorption features using spectra of hot, 
rapidly-rotating, essentially featureless stars.  The interested reader 
will find additional details of our standard procedures in earlier 
papers by this group (see \eg, Sneden \etal\ 1991\nocite{SKPL91}, 
I99-M4).

Our nominal HIRES wavelength coverage is 5400~\AA~$\leq$ $\lambda$ 
$\leq$~6700~\AA, but the free spectral ranges of the echelle orders are 
larger than can be recorded by the 2048$\times$2048 pixel detector, so 
that features of some key elements are inevitably lost in the order 
interstices.  In the 1994/5 observations, the grating was set to permit 
both \ion{Al}{1} and \ion{Mg}{1} lines to be recorded in a study of M13 
giants.  Unfortunately, the radial velocities of M5 and M13 are 
sufficiently different that for the M13 grating setting, lines of 
\ion{Mg}{1} in M5 were shifted into the region between orders, and were 
therefore not recorded.\footnote{
During the observing runs discussed here, the Mauna Kea skies were
partially clouded, and observations therefore limited. It was deemed 
unwise in these conditions to shift back and forth between grating settings,
depending on temporal variations in transparency.}
On the other hand, the grating setting employed for the 1998 observations 
of M5 permitted observations of the \ion{Mg}{1} lines, but not the 
\ion{Al}{1} lines.

We measured EWs for all lines of interest by one of two techniques: 
direct integration of the flux across the observed line profile, or by 
adopting the EW of a Gaussian profile fitted to the line. 
The lines chosen for analysis in the \wave{5500} to \wave{6750} wavelength 
interval and their adopted $gf$-values are the same as those used in the 
previous paper of this series (I99-M4).  We list the atomic parameters and
corresponding reference for each line in Table~\ref{m5.tab10} in the 
Appendix, where further discussion of the linelist is to be found.  EWs of 
all measured lines can be obtained electronically by request to the 
authors.  They are also available at the Astronomical Data Center (ADC) at 
NASA Goddard Space Flight Center 
(http://adc.gsfc.nasa.gov/adc/archive\_search.html).  For the \ion{Na}{1} 
$\lambda\lambda$5682, 5688~\AA\ doublet we based the abundance of Na on a 
synthetic spectrum fit, rather than EW measurements, since the lines in 
question are blended with other metallic species.  We also checked by 
spectrum synthesis the O result obtained from the EW measurements of the 
[\ion{O}{1}] $\lambda\lambda$6300, 6364~\AA\ doublet, employing 
interpolated C and N abundances as a function of O from Smith \etal\ 
(1997).  Our vanadium abundances are derived from blended-line EW 
computations of $\lambda\lambda$6275, 6285~\AA\ for which we employed 
well-determined hyperfine structure components from McWilliam (2001, 
private communication), which are slightly revised from those of 
McWilliam \& Rich (1994)\nocite{MR94}, normalizing the $gf$-values to 
those adopted for these lines in previous studies by our group.  Finally, 
following the same Ba abundance analysis that was performed in I99-M4, 
the blended-line EW analysis of the lines at $\lambda\lambda$5854, 6142, 
6497~\AA\ includes both hyperfine and isotopic subcomponents adopted from 
McWilliam (1998)\nocite{Mc98}.  Similarly, in this study, we assume the 
solar abundance ratios among the $^{134 - 138}$Ba isotopes in the 
calculations.

%Section 3.0
%%%%%%%%%%%%%%%%%%%%%%%%%%%%%%%%%%%%%%%%%%%%%%%%%%%%%%%%%%%%%%%%%%%%%%
\section{Abundance Analysis: Critique of the Input Parameter Selection 
Process}
%%%%%%%%%%%%%%%%%%%%%%%%%%%%%%%%%%%%%%%%%%%%%%%%%%%%%%%%%%%%%%%%%%%%%%

%Section 3.1
%^^^^^^^^^^^^^^^^^^^^^^^^^^^^^^^^^^^^^^
\subsection{Standard Analysis Procedure}
%^^^^^^^^^^^^^^^^^^^^^^^^^^^^^^^^^^^^^^

The preliminary analysis of the observational data from the Keck~I HIRES
spectrograph followed the standard procedure of our earlier M5 paper 
(S92-M5).  In that study, the values of $V$ and $(B-V)$ given by Cudworth 
(1979)\nocite{Cu79}, and the relationship between \BmV0\ color and \teff\ 
that had been adopted by Cudworth were used to provide a first estimate 
of \teff\ for each program star.  The adopted color excess $E(B-V)$~= 
0.03 and true distance modulus \mM~= 14.03 were those recommended by 
Sandage \& Cacciari (1990)\nocite{SC90}.  In the present analysis, we 
used the values of $V$ and $(B-V)$ given by Sandquist \etal\ (1996; 2000, 
private communication).  We adopted the same color excess but the revised 
true distance modulus of \mM~= 14.40 (Djorgovski 1993\nocite{Dj93}) 
was employed in preliminary estimates of \logg\ obtained using 
the relationship obtained by combining the gravitation law with Stefan's 
law.  In S92-M5, the bolometric corrections of Bell \& Gustafsson 
(1978\nocite{BG78}) were employed but here we interpolated G.~Worthey's 
bolometric corrections (1994, private communication), in order to be
consistent with the previous paper of this series (see I99-M4 for 
details).

Armed with these preliminary estimates of \teff\ and \logg, we employed
the current version of the MOOG line analysis code (Sneden 
1973\nocite{Sn73}) to compute abundances from EWs on a line-by-line basis. 
For the various choices of \teff\ and \logg\ we calculated trial model 
atmospheres generated with the MARCS code (Gustafsson \etal\ 
1975)\nocite{GBEN75}.  Anticipating from S92-M5 that $<$[Fe/H]$>$ would be 
near --1.2, we took our input metallicity at [Fe/H]~= --1.0, in order to 
simulate an overall $\alpha$-element enhancement relative to Fe of 
$\sim$0.2 to 0.3 dex, since the models were originally calculated for 
[$\alpha$/Fe]~= 0.0.  A discussion of the validity of this approximation 
is found in Fulbright \& Kraft (1999\nocite{FK99}). 

Final model atmosphere parameters were determined, as before, by 
iteration, through satisfying the following requirements: (a) for \teff, 
that the abundances of individual \ion{Fe}{1} lines show no trend with 
excitation potential; (b) for microturbulent velocity $v_t$, that the 
\ion{Fe}{1} abundances show no trend with EW; and (c) for \logg, that the 
[Fe/H] abundance ratios derived from the \ion{Fe}{1} and \ion{Fe}{2} lines 
should not differ by more than 0.05~dex. The iterated model parameters are 
given in Table~\ref{m5.tab2}; the values listed for [Fe/H] in the 
traditional approach are a mean of determinations based on \ion{Fe}{1} and 
\ion{Fe}{2}.  It is important to note in the iterative process that the 
``final'' values of \teff\ and \logg\ may be fairly different from the
estimated ``input'' values.  Once \teff\ is set, \logg\ is constrained by 
the necessity to force close agreement in the [Fe/H] values determined from 
\ion{Fe}{1} and \ion{Fe}{2}.  Alternatively, once a \logg\ is found to 
satisfy the ionization equilibrium, the Teff is constrained to force 
agreement in the Fe~I abundances for lines of different excitation 
potentials.  Finally there is the additional constraint that [Fe/H] cannot 
be allowed to vary systematically over the range of \teff\ and \logg\ 
represented by the stars in the sample. 

Inspection of the preliminary atmospheric parameters in 
Table~\ref{m5.tab2} gives rise to concerns. First, giants lying in the 
same \teff\ range (3900~K to 4300~K) as those studied in S92-M5 have 
$<$\logg$>$ that is $\sim$0.5~dex lower than the values given in that paper. 
The increase in \mM\ from 14.03 to 14.40 should have instead lowered 
$<$\logg$>$ by only 0.15 dex.  Second, the mean [Fe/H] ratio is 0.15~dex 
lower among the six AGB stars than among the 13 RGB stars.  Thus for the 
AGB we derive $<$[Fe/H]$>$~= --1.45~$\pm$~0.01 ($\sigma$~= 0.03), whereas 
for the RGB we derive --1.30~$\pm$~0.01 ($\sigma$~= 0.04).  Since a real 
physical reduction in Fe abundance from the RGB to the AGB is surely not 
expected in stars of such low mass, the result clearly points to some 
inadequacy in our analysis procedure.  We list the [el/Fe] ratios derived 
in this traditional approach in the Appendix as Table~\ref{m5.tab11}
where the values of [el/Fe] are those based on the mean of \ion{Fe}{1} and 
\ion{Fe}{2} abundances.

We also investigated our abundance results in the context of possible
mass loss or chromospheric activity in the atmospheres of our giant stars.
In all but two of the Keck spectra, H-$\alpha$ was just barely recorded
on the blue edge of the chip.  Stars I-20, G2, IV-81, IV-19, and III-149 
all show some H-$\alpha$ emission in the blue wing.  Of these stars, I-20 
is apparently an AGB star; the rest are on the tip of the giant branch.  
Unfortunately, most of the Na~D doublet is unrecorded for these stars 
(the spectra are cut off redward of the blue wing of Na~D2).  The two
Keck spectra which are offset in wavelength from the rest do, however,
have the Na~D doublet recorded.  Stars II-85 and IV-47, both on the tip of 
the giant branch, show core shifts in both Na~D lines.  With a very 
conservative error of $\pm$2~\kmsec, we derive for the D2 lines a 
shift of --9.8~\kmsec\ and for D1, --7.6 and --9.7~\kmsec, for stars II-85 
and IV-47, respectively.  That the bluer D2 line may show a slightly higher 
blueshift is in accord with its formation higher in the atmosphere, and is 
thus more susceptible to any outward flows in the higher atmosphere regions 
(see \eg, Bates \etal\ 1993\nocite{BKM93} and references therein).

%Section 3.2
%^^^^^^^^^^^^^^^^^^^^^^^^^^
\subsection{A New Approach}
%^^^^^^^^^^^^^^^^^^^^^^^^^^

``For many years, the techniques used in stellar abundance determinations 
have remained essentially unchanged, despite a rather passionate controversy 
in the late fifties and early sixties [...] that departures from LTE could 
lead to abundances substantially different from those given by the
`classical' LTE approach'' (Dumont \etal\ 1975\nocite{DHJP75}).  
Detailed investigations since have confirmed not only the effect on derived 
abundances but also on the derived stellar parameters (see \eg, Hearnshaw 
1976\nocite{Hea76}, Luck \& Lambert 1985\nocite{LL85}, Fuhrmann \etal\ 
1997\nocite{FPFRG97}, Allende~Prieto \etal\ 1999\nocite{AGLG99}, Fulbright
2000\nocite{F2000}, the latter three studies based on {\it Hipparcos} [ESA 
1997]\nocite{ESA97} results).  The largest effect on stellar parameters is 
on the derived gravity: gravities derived by forcing ionization 
equilibrium (spectroscopic gravity) are lower than those derived by 
stellar parallaxes (trigonometric gravity) or by the evolutionary position 
in the HR-diagram (evolutionary gravity).  We confirm this gravity anomaly 
in our LTE analysis of the M5 giant stars and discuss the anomaly further 
in this section.

Recently, Th\'evenin \& Idiart (1999\nocite{TI99}, hereafter TI99) have 
explored in detail the problem of Fe over-ionization in the atmospheres of 
metal-poor stars.  For over-ionized atmospheres, application of standard LTE 
model atmospheres to abundance analysis of \ion{Fe}{1} will always lead to 
an underestimate of [Fe/H]. TI99 point out that at any given optical 
depth, the populations of the atomic levels of \ion{Fe}{1} are governed 
not by the local kinetic temperature but rather are modified by the outward 
leakage of UV photons into an atmosphere made progressively less opaque as 
metallicity is decreased.  The metallicity dependence is not surprising:  
it has been known for some time that for a given optical depth, lower 
metallicity stars have a larger physical depth (see \eg, Wallerstein 
1962\nocite{Wal62}) and the optical depth thus reaches hotter layers of 
the atmosphere. Fortunately, the abundance of Fe derived from \ion{Fe}{2} 
remains relatively unaffected, since in metal-poor stars, virtually all Fe 
is already in the form of \ion{Fe}{2}.  The TI99 calculations suggest that 
the reduction of [Fe/H] estimated from \ion{Fe}{1} relative to \ion{Fe}{2} 
amounts to about 0.1~dex at [Fe/H]~= --1.0 but rises to about 0.3~dex at 
[Fe/H]~= --2.5. For similar reasons, the leakage of UV photons should also 
become larger with lower atmospheric densities, \ie, surface gravities, 
at a given \teff, and with higher \teff\ values  at a given luminosity. 
Thus one might anticipate that should the TI99 effect be real, traditional 
analyses of AGB stars could well lead to lower overall estimates of [Fe/H] 
based on \ion{Fe}{1}, as compared with RGB stars. A smaller, but still 
noticeable, ``dragging down'' of [Fe/H] would occur when [Fe/H] is 
estimated from the mean of \ion{Fe}{1} and \ion{Fe}{2} determinations.  
Could this effect account for the anomalous apparent drop in [Fe/H] among 
M5 AGB stars?

We decided to test this possibility in three ways: first, by modifying our Fe
linelist to exclude all but the weakest lines; second, by modifying the 
linelist to exclude all but the highest excitation potential lines; and 
third, by modifying the procedure used to define the stellar parameters.  By
gradually culling the lines in descending order of equivalent width, we found
a small but steady increase in the microturbulent velocity required to 
satisfy the EW equilibrium constraint but, no significant 
change in either the \ion{Fe}{1} to \ion{Fe}{2} ratio or in the overall iron 
abundance.  Gradual deletions of lines in ascending order of excitation 
potential had no significant effect on the ratio or abundance and the changes 
in $v_t$ were up or down, depending on the subset of lines in use in a given 
trial.  Finally, we modified our procedure for estimating the input values of
\teff\ and \logg\ in the following way.  First, we set aside any reference to 
the \ion{Fe}{1} spectrum in estimating \teff\ and replaced it with values of 
\teff\ derived from \BmV0, using the calibration of Alonso \etal\ 
(1999\nocite{AAM99}; their Table~6, interpolating the computed table 
values).\footnote{
We have since verified that the {\em corrected} version of the formula in 
the caption of Table 2 of Alonso \etal\ yields similar results.} 
This scale, based on the Infra-Red Flux Method (``IRFM''; Blackwell \etal\ 
1990\nocite{BPAHS90} and references therein), applies to low-mass, 
metal-poor giants.  We then assigned to each star the value of \logg\ it 
should have, as predicted from stellar models coupled to stellar evolution. 
To each star on the RGB, we assigned a mass of 0.80~\Msun, and to each star 
on the AGB a mass of 0.70~\Msun, thus allowing for the mass loss expected 
in very late evolutionary stages.  We took $E(B-V)$~= 0.03, \mM~= 14.40, 
and calculated \logg\ from the relationship 
$g$~$\sim$~$\mathcal{M}\times$\teff$^4$/L, interpolating G.Worthey's 
bolometric corrections (1994, private communication).  The observed values 
of $V$ and $(B-V)$ used in S92-M5 were replaced by modern CCD-based values 
of Sandquist \etal\ (1996\nocite{SBSH96}; 2000, private communication). 

In Table~\ref{m5.tab2} we show a comparison of \teff\ and \logg\ values 
derived from the ``traditional'' approach based on the \ion{Fe}{1} and 
\ion{Fe}{2} line spectrum and the revised approach based on the Alonso 
\etal\ (1999)\nocite{AAM99} color versus \teff-scale and stellar 
evolutionary arguments.  For the 13 RGB stars and 6 AGB stars, we find 
$\delta$\teff~= +26~$\pm$~10~K and $\delta$\teff~= +35~$\pm$~18~K 
respectively, in the sense ``new'' $minus$ ``traditional''. 
The difference between \teff\ based on the \ion{Fe}{1} excitation plot 
and \teff\ based on the Alonso \etal\ color-\teff-scale is very small 
and the effect on the abundances of Fe derived from \ion{Fe}{1} and 
\ion{Fe}{2} is essentially negligible.  Thus an increase in \teff\ of 
30~K increases \eps{\ion{Fe}{1}} by 0.02~dex and decreases 
\eps{\ion{Fe}{2}} by 0.04~dex.  But the change in \logg\ has a more 
substantial effect;  we find $\delta$\logg~= +0.28~$\pm$~0.04 for the 
RGB sample and virtually the same result, $\delta$\logg~= 
+0.34~$\pm$~0.06 for the AGB sample, again in the sense ``new'' $minus$ 
``traditional''. Such a gravity change tends to drive the derived Fe 
abundances from \ion{Fe}{1} and \ion{Fe}{2} apart: for $\delta$\logg~= 
+0.30, we expect $\delta$\eps{\ion{Fe}{1}}~$\simeq$ --0.02 and 
$\delta$\eps{\ion{Fe}{2}}~$\simeq$ +0.15.  These changes are 
qualitatively what one would expect if the TI99 conjecture were in fact 
true.

In Table~\ref{m5.tab3} we tabulate the changes in \eps{\ion{Fe}{1}} 
and \eps{\ion{Fe}{2}} corresponding to small changes in the input 
parameters: \teff, \logg, $v_t$, [Fe/H], distance modulus, and stellar 
mass for a typical M5 RGB star: \teff~= 4325~K, \logg~= 1.08, $v_t$~= 
1.65.  From the arguments in the preceding paragraph plus inspection of 
this table, one can easily see the scope of the dilemma.  The 
\ion{Fe}{1} excitation plot yields essentially the same \teff\ values 
as the Alonso \etal\ (1999) \teff-scale, in turn based on the IRFM.  
It therefore seems unlikely this temperature scale is seriously in 
error.  If the disagreement between the spectroscopic gravities, based 
on forced agreement between \ion{Fe}{1} and \ion{Fe}{2} abundances, and 
the evolutionary gravities is due to a defect in estimating the latter, 
then the evolutionary \logg\ would need to be decreased by 
$\sim$0.3~dex.  This would in turn imply that the distance modulus of 
M5 is too small, and needs to be increased by 0.75~dex, \ie, to \mM~= 
15.15.  Such an increase in the distance modulus would be seriously at 
odds with recent estimates based on fitting of the M5 main sequence to 
the main sequence of mildly metal-poor subdwarfs having accurate {\it 
Hipparcos}-based parallaxes.  For example, Reid (1997)\nocite{Re97} 
finds \mM~= 14.45 from this approach, very close to the value adopted 
here.  M5 also contains many RR Lyrae variables for which $<V^o>$~= 
15.02 if $E(B-V)$~= 0.03 (Jones \etal\ 1988)\nocite{JCL88}.  Assuming 
that halo field and globular cluster RR Lyraes are analogs of each 
other, these are expected to have $<$\MV0$>$~= +0.7~$\pm$~0.1 
(Layden \etal\ 1996\nocite{LHHKH96}, based on their 
Figure~7) in which case \mM~$\sim$ 14.3, again close to our assumed 
value.  A modulus of 15.15 would cause the RR Lyraes of M5 to be 
unacceptably bright.

\begin{figure} %FIGURE 2
\epsscale{1.0}
\plotone{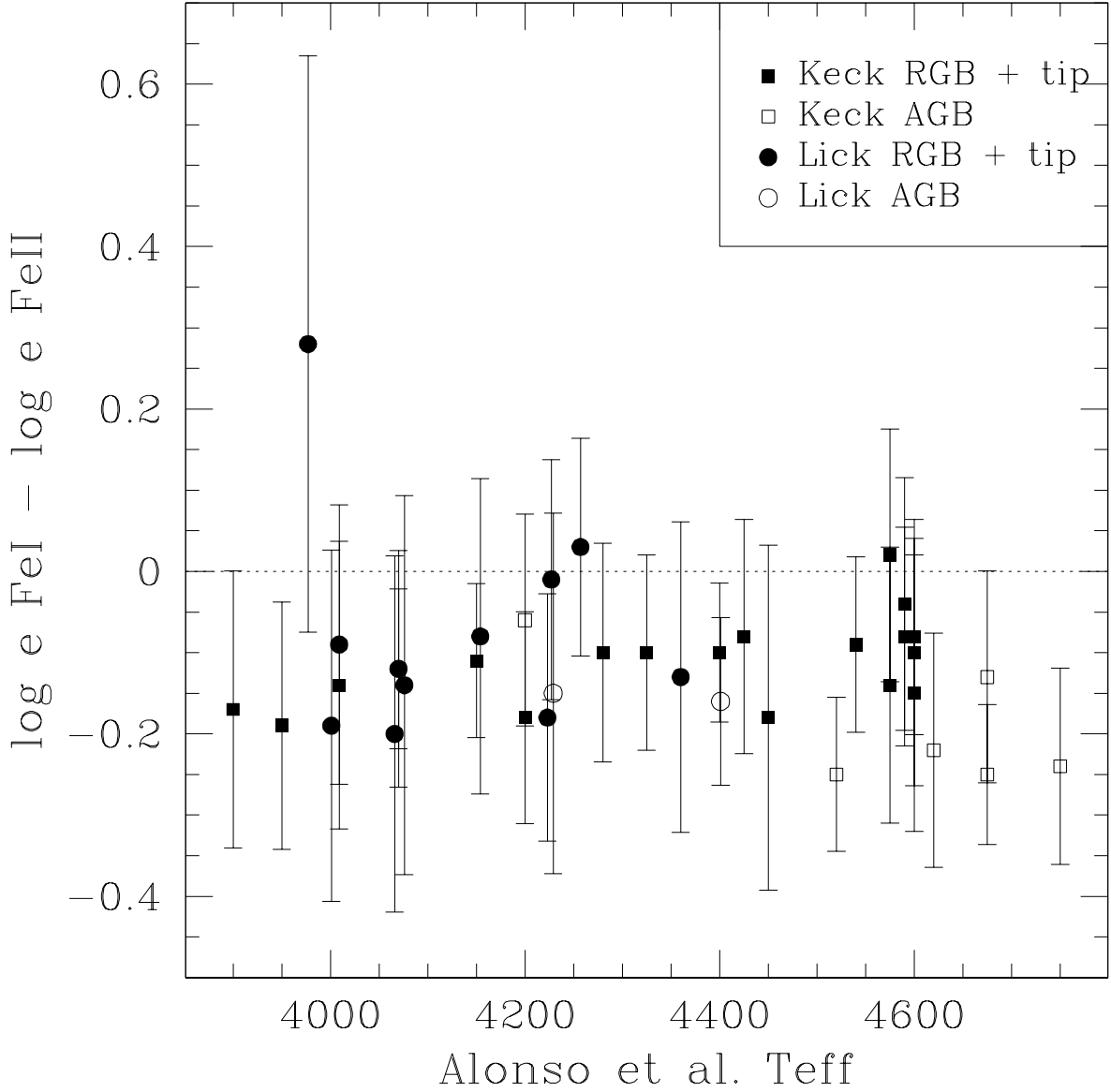}
\caption{Log $\epsilon$ (\ion{Fe}{1}) $minus$ log $\epsilon$ (\ion{Fe}{2}) 
as a function of \teff\ (as derived from the calibration of Alonso \etal\ 
1999) for our M5 program stars.  The symbols in the figure correspond to 
AGB or RGB and ``tip'' stars as well as the observatory used for each 
observation.  The error bars correspond to ($\sigma_{FeI}^{2}$ + 
$\sigma_{FeII}^{2}$)$^{1/2}$. 
\label{m5.fig2}}
\end{figure}

\begin{figure}  %FIGURE 3
\epsscale{1.0}
\plotone{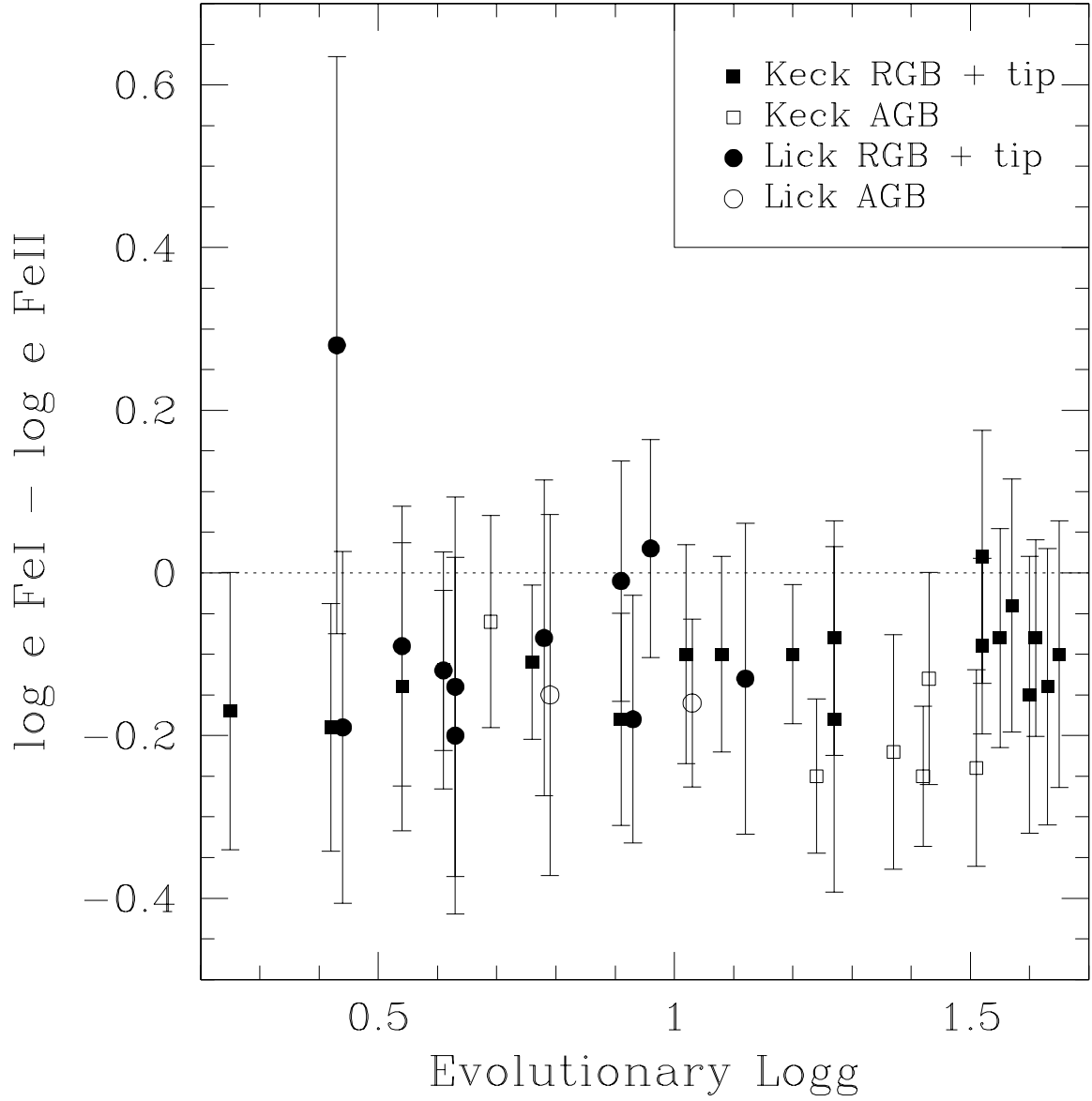}
\caption{Log $\epsilon$ (\ion{Fe}{1}) $minus$ log $\epsilon$ 
(\ion{Fe}{2}) as function of \logg~(evolutionary).  The symbols 
correspond to those of Figure~\ref{m5.fig2}.
\label{m5.fig3}}
\end{figure}

\begin{figure}  %FIGURE 4
\epsscale{1.0}
\plotone{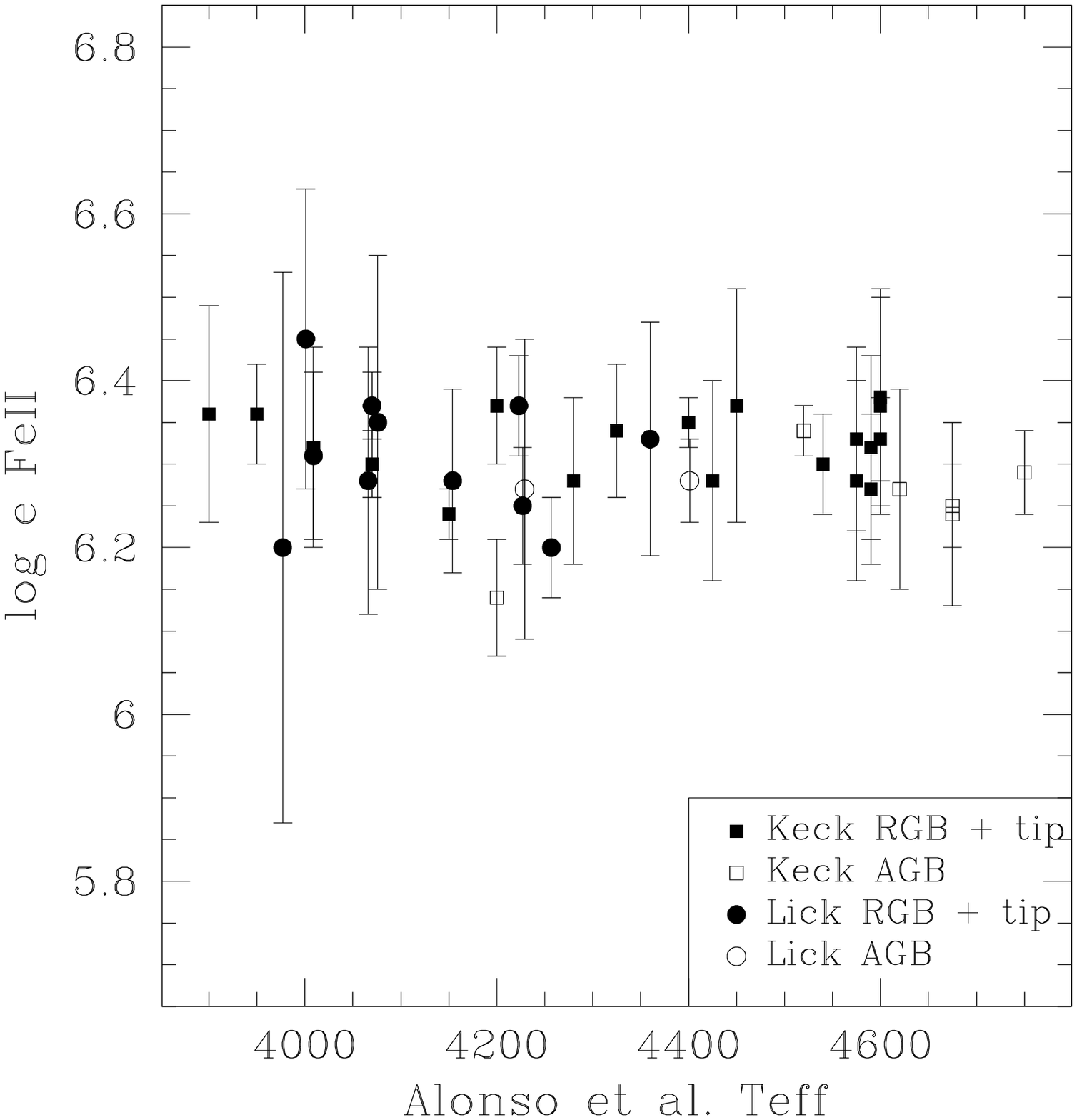}
\caption{Log $\epsilon$ (\ion{Fe}{2}) as function of \teff\ (Alonso \etal\ 
1999 scale).  The symbols correspond to those of Figure~\ref{m5.fig2}.
\label{m5.fig4}}
\end{figure}

Returning to Table~\ref{m5.tab2}, we tabulate values of [Fe/H] 
determined independently from \ion{Fe}{2} and \ion{Fe}{1}, based on our 
revised procedure.  We now $assume$ that [Fe/H] is correctly given by the 
\ion{Fe}{2} value.  In that case, for the 13 RGB stars, $<$[Fe/H]$>$~= 
--1.20~$\pm$~0.01, ($\sigma$~= 0.04) and for the 6 AGB stars, 
$<$[Fe/H]$>$~= --1.26~$\pm$~0.04 ($\sigma$~= 0.07).  In comparison to the 
results from the ``traditional'' analysis, the difference in Fe abundance 
between the AGB and RGB has been reduced from 0.15 to 0.06~dex; the 
latter difference is close to a 1-$\sigma$ combined error and thus is 
acceptable.  The over-ionization of Fe follows from a comparison of 
[Fe/H] determined from \ion{Fe}{1} and \ion{Fe}{2}.  Thus for the 13 RGB 
stars, $<\delta$[Fe/H]$>$~= --0.09~$\pm$~0.01 ($\sigma$~= 0.05) and for 
the 6 AGB stars, $<\delta$[Fe/H]$>$~= --0.18~$\pm$0.03 ($\sigma$~= 
0.08), in the sense \ion{Fe}{1} $minus$ \ion{Fe}{2}.\footnote{
Adoption of the larger distance modulus of 14.62, based on Hipparcos 
subdwarfs, favored by Gratton \etal\ (1997\nocite{Git97}), reduces 
slightly the differences between [Fe/H] based on \ion{Fe}{1} and [Fe/H] 
based on \ion{Fe}{2}. For the 13 RGB  and 6 AGB stars, the differences 
become --0.06 and -0.15 dex, respectively. The difference of the 
differences remains, of course, the same.}
As anticipated, the depression of \ion{Fe}{1} abundances relative to those
of \ion{Fe}{2} is more severe for AGB stars as compared to RGB stars. 
In Figures~\ref{m5.fig2}, \ref{m5.fig3} and \ref{m5.fig4}, we 
illustrate the difference, $\delta$\eps{Fe} = 
\eps{\ion{Fe}{1}}~--~\eps{\ion{Fe}{2}}, as functions of \teff\ (Alonso 
\etal\ 1999\nocite{AAM99}) and \logg\ (evolutionary), as well as 
\eps{\ion{Fe}{2}} as a function of \teff\ for all of our program 
stars.\footnote{
These figures include the entire sample of stars discussed in this paper.  
See also \S6.}
One AGB star (I-20) remains somewhat anomalous in having an unusually low 
\ion{Fe}{2} abundance; we return to this star later.

%Section 3.3
%^^^^^^^^^^^^^^^^^^^^^^^^^^^^^^^^^^
\subsection{Alternative Approaches}
%^^^^^^^^^^^^^^^^^^^^^^^^^^^^^^^^^^

In the preceding, we adopted the \teff-scale of Alonso \etal, which 
follows an empirical approach based on the IRFM.  However, examining the 
offsets in Fe abundance exhibited in Table~\ref{m5.tab3}, we see that 
agreement between \eps{\ion{Fe}{1}} and \eps{\ion{Fe}{2}} could also be 
achieved if we increased \teff\ by $\sim$60K and $\sim$120K for RGB and 
AGB stars (along with accompanying small increases in \logg), 
respectively.  However, from Table~\ref{m5.tab2}, we see that 
adoption of these increases would exascerbate the difference between 
the \teff\ obtained from the \ion{Fe}{1} excitation plot and the newly 
corrected \teff, by +72K and +160K for RGB and AGB stars.

\teff-scales other than that of Alonso \etal\ (1999\nocite{AAM99}) can be 
found in the literature.  A recent version is that of Gratton \etal\ 
(2000\nocite{GSCB00}).  We can compare values of \teff\ using metal-poor 
giants common to the two investigations; there are five such field giants 
shared between Gratton \etal\ (2000\nocite{GSCB00}) and Alonso \etal\ 
(1999\nocite{AAM99}), from which we find $\delta$\teff\ = +71K $\pm$ 68K, 
with the Gratton \etal\ scale the hotter of the two.  However, conflicting 
evidence is found from a study of near-UV fluxes and flux distributions of 
metal-poor stars by Allende~Prieto \& Lambert (2000\nocite{AL00}). Their 
investigation contains 15 stars in common with Alonso \etal\ 
(1996\nocite{AAM96}) and having [Fe/H] $<$ --0.5 with 4000K $\le$ 
\teff$_{Alonso}$ $\le$ 6000K (omitting the spectroscopically peculiar 
dwarfs HD134439 and HD25329).  For these 15 stars we find a negligible 
offset of $\delta$\teff\ = +32K $\pm$ 56K, in the sense UV $minus$ IRFM.  
Unfortunately, the sample consists entirely of dwarfs.  Allende~Prieto \& 
Lambert also compare their UV-flux derived values of \teff\ with those of 
Gratton \etal\ (2000\nocite{GSCB00}).  In this case, there are four 
giants in common (we omit the heavily reddened HD166161).  For these four 
stars, we find a much larger offset of $\delta$\teff\ = +94K $\pm$ 63K, in 
the sense Gratton \etal\ $minus$ Allende~Prieto \& Lambert. 

We also investigated the effects of adopting the color-\teff\ calibration 
of Sekiguchi \& Fukugita (2000\nocite{SF00}).  For stars with temperatures 
that correspond to the warmer M5 stars in our sample, the Sekiguchi \&
Fukugita color-\teff\ calibration produces temperatures $\sim$50K hotter 
than the other calibrations.  This temperature shift improves the 
situation for our hottest AGB stars but also affects our warm RGB stars.
On the other hand, for stars with temperatures that correspond to the 
coolest M5 stars in our sample, we find that their calibration produces a 
\teff-scale that is $\sim$100K {\it cooler} than that of the Alonso 
\etal\ (1996) calibration, a temperature difference that is in agreement 
with the overall findings of Sekiguchi \& Fukugita.  Thus, the overall 
effect of the Sekiguchi \& Fukugita \teff-scale is to change the slope of 
the \teff\ vs \logg\ relationship to one which is in the {\it opposite} 
sense of what is required to correct the cool AGB versus RGB + ``tip'' 
star iron abundances.   In summary, these comparisons clearly offer no 
firm evidence that the Alonso \etal\ \teff-scale requires any upward 
revision.

In an alternative approach, we abandon the Alonso \etal\ IRFM-based \teff\ 
scale and instead derive \teff\ and \logg\ from the comparison of observed 
and synthetic colors of models for low-mass, metal-poor giants. New models 
have recently been calculated by Houdashelt \etal\ (2000\nocite{HBS00}), 
in which values of $(B-V)$, $(V-K)$, (and other colors) are given as a 
function of \teff\ and \logg\ for metal abundances ranging from solar to 
[Fe/H] = --3.  To determine whether adoption of these models would in some 
way modify our conclusions, we considered a sample of M5 giants drawn from 
our Table~\ref{m5.tab1}, distributed so that RGB, AGB and ``tip'' stars 
are all represented.  We then calculated \teff\ and \logg\ for each star, 
entering the Houdashelt \etal\ tables with the observed values of \BmV0, 
and assuming as before a true distance modulus of 14.40. The BC's adopted 
in this case were those of Houdashelt \etal.  Unfortunately, this 
procedure proved difficult to apply in practice for two reasons. First, 
the expected metallicity of M5 is in the range [Fe/H] = --1.2 to --1.35, 
and the Houdashelt \etal\ tables contain entries only for [Fe/H] = --1.0 
and --2.0. Thus one must interpolate within the framework of a rather 
coarse grid. Second, at a fixed $(B-V)$, the relationship between Teff 
and [Fe/H] is non-linear, so that linear interpolation at the metallicity 
of M5 is not adequate.  However, these difficulties can be overcome by 
employing \VmK0\ as the independent variable, since \teff\ is practically 
independent of [Fe/H] at a fixed value of \VmK0, and depends very little 
on \logg. To obtain \VmK0\ for the stars in our Keck sample, we plotted
\BmV0\ vs \VmK0\ for the 25 stars observed by Frogel \etal\ 
(1983\nocite{FPC83}), and used this plot to transform \BmV0\ to \VmK0, 
retaining the more recently acquired $V$ magnitudes and $(B-V)$ colors of 
Sandquist (1996\nocite{SBSH96}; 2000, private communication). The 
color-color plot proved to be extremely tight: we estimate that the 
transformation could introduce an error of no more than 0.01 mag in 
\VmK0. This procedure permitted us to estimate values of \teff\ with 
little uncertainty due to errors in interpolation.

\begin{figure}  %FIGURE 5
\scalebox{0.65}{\includegraphics[angle=270]{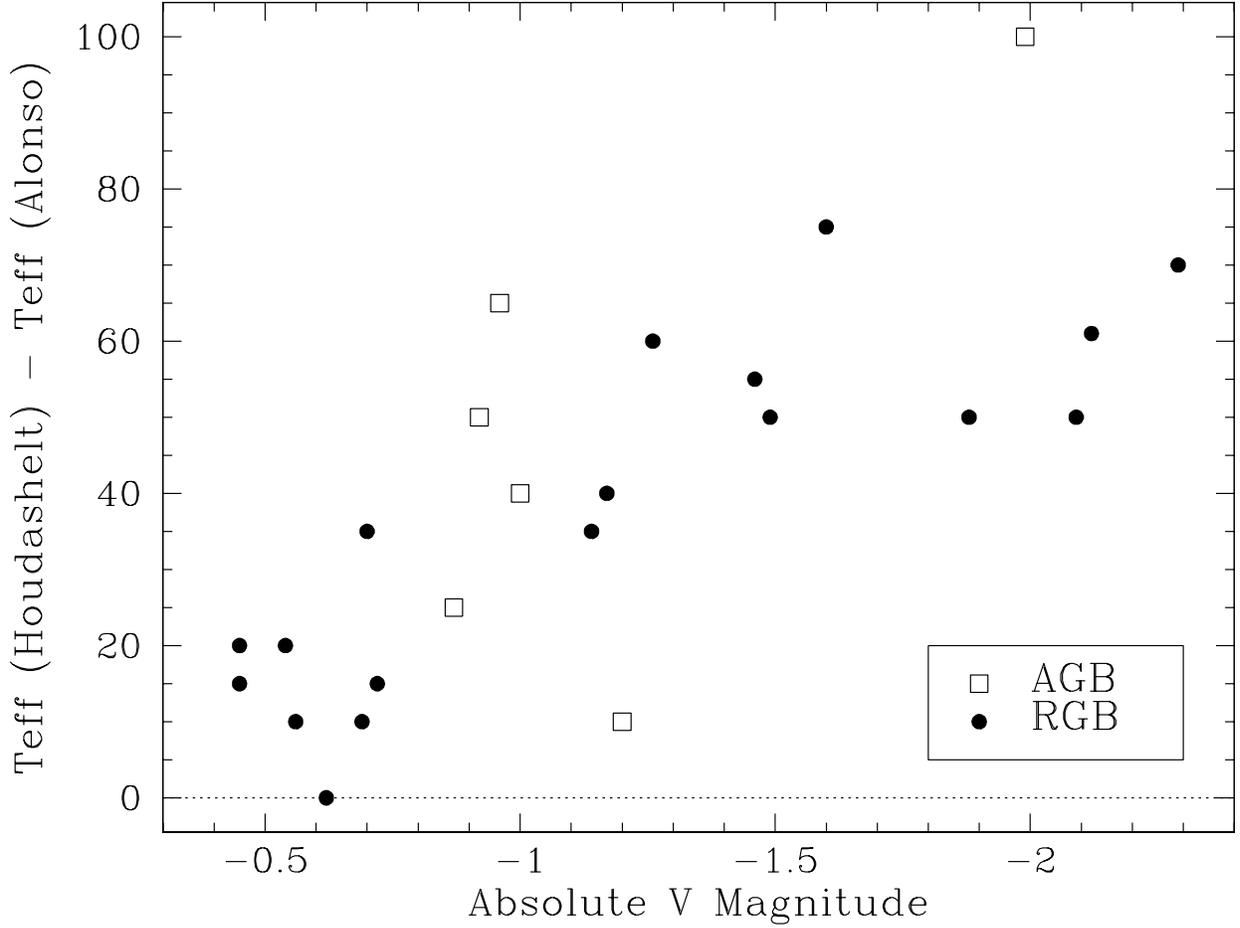}}
\caption{\teff\ derived using the Houdashelt \etal\ (2000) calibration 
$minus$ the \teff\ derived from the Alonso \etal\ (1999) calibration
as a function of the absolute V magnitude (as derived from the 
calibration of Alonso \etal\ 1999) for our M5 program stars observed
at Keck.  The symbols in the figure correspond to AGB or RGB plus
``tip'' stars.
\label{m5.fig5}}
\end{figure}

The difference between \teff\ derived from the Houdashelt \etal\ models
and \teff\ derived from the Alonso \etal\ scale is shown as a function 
of \MV0\ in Figure~\ref{m5.fig5}. The difference shows a steady 
increase with luminosity from $\sim$zero at \MV0\ = --0.5 to $\sim$+60K 
at \MV0\ = --2.5.  Results and comparisons with entries in 
Table~\ref{m5.tab2} are shown in Table~\ref{m5.tab4}.  The 
Houdashelt \etal\ values of \teff\ are higher than the Alonso \etal\ 
IRFM-based values of \teff\ by average offsets of 27K$\pm$ 21K for the 
three RGB stars, +60K $\pm$ 10K for the three ``tip'' stars, and +43K 
$\pm$ 20K for the three AGB stars (also see Figure~\ref{m5.fig5}).  
Within the errors, the offsets in \teff\ are appear comparable (the 
overall average is +43K $\pm$ 20K).  However, using the higher values 
of \teff, the average offsets in the iron abundances, $\delta$[Fe/H] 
= \eps{\ion{Fe}{1}} $minus$ \eps{\ion{Fe}{2}}, become --0.06 $\pm$ 
0.02, --0.07 $\pm$ 0.06, and --0.16 $\pm$ 0.07 for the same three groups 
of stars.  Regardless of which of the preceding \teff-scales we adopt, 
the abundance of Fe based on \ion{Fe}{2} remains essentially constant 
with evolutionary state, whereas [Fe/H] based on \ion{Fe}{1} remains 
significantly smaller on the AGB as compared with the RGB and ``tip'' 
stars. Simply adopting the hotter \teff-scale of Houdashelt \etal\ for 
the sample does not solve the overall problem of over-ionization.

As to additional sources of systematic differences between \ion{Fe}{1} vs 
\ion{Fe}{2} abundances, the referee noted that the $gf$-value zero-point for 
\ion{Fe}{2} is possibly not as well known as one would like.  Two recent 
studies of \ion{Fe}{2} $gf$-values, those of the ``critical compilation'' of 
the NIST Atomic Spectra Database (Version 2.0; http://physics.nist.gov/asd; 
Martin \etal\ 1999\nocite{Meta99}) and Schnabel \etal\ 
(1999\nocite{SKH99}) provide lines in common with those shown in the 
Appendix (Table~\ref{m5.tab10}).  The difference between our values 
and the NIST values is +0.10~dex $\pm$ 0.09~dex, in the sense of M5 
$minus$ NIST.  Adopting the NIST log~$gf$-values would produce an even 
larger disagreement between our \ion{Fe}{1} and \ion{Fe}{2} abundances.  
With the Schnabel \etal\ (1999\nocite{SKH99}) linelist, our two lines 
in common have a difference in the log~$gf$-values of --0.14~dex $\pm$ 
0.09~dex, in the sense of this study $minus$ Schnabel {\it et al}.  
However, the solar abundance of iron derived using the Schnabel \etal\ 
linelist is 7.42, not the 7.52 we have adopted here and in our previous 
work.  Normalizing the Schnabel \etal\ lines to reproduce our adopted 
solar abundance would negate the offset that the lines would otherwise 
generate.  While an increase in the log~$gf$-values of \ion{Fe}{2} by 
0.1~dex from those which we have employed in our previous Lick-Texas 
work would indeed bring the ``tip''and RGB giant [Fe/H] values for 
\ion{Fe}{1} and \ion{Fe}{2} into agreement using the Alonso \etal\ 
\teff-scale, any change ``across the board'' in \ion{Fe}{2} $gf$-values 
would not simultaneously satisfy the \ion{Fe}{1} vs \ion{Fe}{2} offsets 
for the AGB stars.

We summarize the findings of this section by noting that one of three 
procedural choices can be adopted:
\begin{enumerate}
\item We adopt the traditional methods of high resolution spectroscopy,
including setting the surface gravity from the ionization equilibrium 
of iron, in which case we find that the mean [Fe/H] value decreases by 
0.15~dex as stellar evolution advances from the RGB to the AGB.
\item We abandon the traditional approach using spectroscopic constraints, 
basing the analysis instead on values of \teff\ derived from the Alonso 
\etal\ (1999\nocite{AAM99}) relation between $(B-V)$ and \teff, which is 
in turn based on the Infra-Red Flux Method (Blackwell \etal\ 
1990\nocite{BPAHS90} and references therein), and values of \logg\ derived 
from application of stellar evolution plus knowledge of the cluster 
distance modulus.  This approach stabilizes the \ion{Fe}{2} abundance as a 
function of evolutionary state but requires acceptance of the idea that 
\ion{Fe}{1} is over-ionized and out of equilibrium with \ion{Fe}{2}, 
consistent with the non-LTE precepts described by Th\'evenin \& Idiart 
(1999\nocite{TI99}).  The over-ionization of \ion{Fe}{1} turns out to be
more severe among AGB as compared with RGB stars. Interestingly, this 
\teff-scale is in close agreement with the \teff-scale derived from the 
\ion{Fe}{1} excitation vs EW plot.
\item An alternative solution requires arbitrarily increasing the values of 
\teff\ above the Alonso \etal\ scale by $\sim$60K on the RGB and $\sim$120K 
on the AGB; these changes would bring \ion{Fe}{1} and \ion{Fe}{2} 
abundances nearly into agreement.  The recent models of metal-poor stars by 
Houdashelt \etal\ (2000\nocite{HBS00}), which predict $(B-V)$, $(V-K)$, 
and other colors from \teff\ and \logg\ for different choices of [Fe/H], 
do indeed predict higher values of \teff\ than those of the Alonso 
\etal\ scale.  Why the models give a \teff\ vs color scale that is hotter 
than the scale based on the IRFM is not clear.  The Houdashelt \etal\ 
models come close to satisfying ionization constraint requirement among 
the RGB stars, but are still too cool by $\sim$70K to rectify the 
situation for AGB stars.  And, if we make the AGB stars 120K hotter than 
the Alonso \etal\ (1999) scale, the abundance of \ion{Fe}{2} will drop 
to a level about 0.1 dex lower than its value among RGB stars.

\end{enumerate}

Here we adopt procedure (2) as one extreme, and report the results of
procedure (1), the opposite extreme, in the Appendix. The reader should
bear in mind that the ``intermediate'' solution under (3) remains an
option, but requires a fairly large systematic correction to the 
IRFM-based \teff-scale and a smaller, but still significant, correction
to the \teff-scale based on the Houdashelt \etal\ models.

%Sections 4.0
%%%%%%%%%%%%%%%%%%%%%%%%%%%%%%%%%%%%%%
\section{[el/Fe] Ratios: A Rationale} 
%%%%%%%%%%%%%%%%%%%%%%%%%%%%%%%%%%%%%%

Based on the revised approach, the determination of [el/Fe] ratios 
becomes more complex than is the case in the traditional approach. 
If Fe is over-ionized, then one might expect a corresponding 
over-ionization of elements having first ionization potentials $\lesssim$ 
to that of iron.  In the yellow-red spectral regions of globular cluster 
giants, almost all detectable transitions arise from ``metallic elements'' 
that exist predominantly in singly ionized states.  But aside from Fe, 
which has both neutral and ionized species lines available, only a few 
elements (\eg, Sc, Ba, La, Eu, and sometimes Ti) have observable 
transitions arising from their first ionized states in our stars.
Fortunately, the [el/Fe] ratios of these elements are confidently 
estimated from their [el/H] ratios and [Fe/H] ratios from \ion{Fe}{2}.
For the majority of elements with only neutral-species lines present,
estimates must be made of the degree to which the neutral populations
are depleted by over-ionization.  Oxygen is a special case: it remains 
overwhelmingly neutral and in the ground state, shielded from 
over-ionization both by its very high first ionization potential 
(13.6~eV) and the opacity corresponding to the Lyman jump.  There is 
little doubt that the [O/Fe] ratio should be based on [Fe/H] derived 
from \ion{Fe}{2}.\footnote{
This statement does not take into account the possibility of a small
effect induced by ionizations from the low-lying singlet S and D states of 
\ion{O}{1}.}

The degree of over-ionization of any particular species depends on the 
ionization potential, the term scheme and the location and strength
of the absorption transitions of that atom in relation to the flux 
distribution of the excess $UV$ photons.  The excess $UV$ photons 
envisaged by TI99 must have a complicated $UV$ energy distribution 
reflecting the highly jagged opacity distribution longward of the Lyman 
limit.  Calculating the degree of excess ionization is further 
complicated by the fact that ionizations can take place from excited 
levels as well as the ground state.  To determine accurately the degree 
of over-ionization of those species which appear in our spectra only in 
the neutral state would require the calculation of collisional and 
radiative rates for thousands of levels, as was done in the case of 
\ion{Fe}{1} and \ion{Fe}{2} by TI99.  Such calculations for similar 
elements are beyond the scope of this paper, although it is obvious that 
detailed studies need to be carried out.

In the absence of such theoretical calculations, we looked for guidance
in the empirical domain, in particular among stars with values of 
\logg\ similar to those of M5 giants, but having higher metallicities 
so that EWs of ionized lines of such species as Si, Ti, and V are large 
enough to be measured.  A sample of LMC and SMC cepheids (Luck \etal\ 
1998)\nocite{LMBG98} provides [el/H] ratios for neutral and ionized 
states of these three elements both for \logg\ based on stellar 
evolution and for \logg\ derived from the \ion{Fe}{1} versus 
\ion{Fe}{2} ionization balance.  Whenever Fe appears to be over-ionized 
as a result of adopting an ``evolutionary'' \logg, these authors 
generally find that Si, Ti, and V are excessively ionized by 
essentially the same amount as is Fe.\footnote{
Note that we are concerned here with $changes$ in the ionization as
a result of abandoning the ionization equilibrium of Fe as a means of
setting \logg.  Thus for example in the case of the SMC cepheid HV~837, 
Luck \etal\ find that [\ion{Fe}{2}/\ion{Fe}{1}] increases from --0.01 to 
+0.54 as \logg\ changes from --0.28 (spectroscopic) to +0.82 
(evolutionary).  The corresponding increases in [\ion{Ti}{2}/\ion{Ti}{1}] 
are 0.00 to +0.55 and for [\ion{Si}{2}/\ion{Si}{1}] are +0.31 to +0.84.  
The changes are essentially the same as for Fe.  However, we note that 
when Fe is in equilibrium, Si is not.}   
A similar over-ionization effect is found by Kovtykh \& Andrievsky 
(1999)\nocite{KA99} in $\delta$~Cep.

If this situation applies also in the M5 giants considered here, then 
the abundance ratios of [Si/Fe], [Ti/Fe], and [V/Fe] can be estimated from 
the assumption that the degree of over-ionization of these species 
is the same as that of Fe. In that case these elements must be referenced 
to the abundance of Fe based on \ion{Fe}{1}.  In the absence of detailed 
calculations we broaden this procedure to include all elements which 
present themselves in the neutral state except for oxygen, which for 
reasons already cited, we reference to Fe based on \ion{Fe}{2}.

We summarize our estimates of the [el/Fe] ratios for the 19 RGB plus
``tip'' and 6 AGB stars in question in Table~\ref{m5.tab5}. 
Columns~3 and 4 contain the values of [Fe/H] estimated independently for 
\eps{\ion{Fe}{2}} and \eps{\ion{Fe}{1}} and column~5 contains [O/Fe], 
assuming that \ion{Fe}{2} yields the correct abundance of Fe. 
For the remaining elements up through the Fe-peak group (except for Sc), 
we list [el/Fe] on the simple assumption that it is ``correct'' to ratio 
\eps{\ion{el}{1}} to \eps{\ion{Fe}{1}}.  For the heavy elements (and Sc), 
we ratio \eps{\ion{el}{2}} to \eps{\ion{Fe}{2}}.  Mean values of [el/Fe] 
are found at the bottom of Table~\ref{m5.tab5}, individually 
calculated for RGB and AGB stars.   These should be compared with mean 
values from Table~\ref{m5.tab11} in the Appendix, which are based on 
the ``traditional'' method of analysis.

%Section 5.0
%%%%%%%%%%%%%%%%%%%%%%%%%%%%%%%%%%%%%%%%%%%%%%%%%%%%%%%%%%%%%%%%%%%%%%%%%%%%%
\section{Adopted [el/Fe] Ratios: the 25 Stars Observed with the Keck~I HIRES}
%%%%%%%%%%%%%%%%%%%%%%%%%%%%%%%%%%%%%%%%%%%%%%%%%%%%%%%%%%%%%%%%%%%%%%%%%%%%%

Following the arguments of the last two sections we assume that [el/Fe] 
ratios are properly deduced by referring neutral species abundances to 
\ion{Fe}{1} and ionized species to \ion{Fe}{2}, the only exception being 
[O/Fe] derived from [\ion{O}{1}], which is referred to \ion{Fe}{2}. 
Following this precept, we add to the 19 RGB and AGB giants of 
Table~\ref{m5.tab5} the six stars near the red giant tip and display 
the resultant [el/Fe] ratios also in Table~\ref{m5.tab5}.

Table~\ref{m5.tab5} lists the means of [Fe/H] derived from \ion{Fe}{2}
and \ion{Fe}{1}, and the means for the [el/Fe] ratios based on the above 
discussion, where we have divided the material into four groups: 13 RGB, 
6 AGB, 19 RGB plus ``tip'' stars and finally, all 25 stars observed with 
the Keck~I HIRES.  Except for the differences in [Fe/H] derived from 
\ion{Fe}{1} versus \ion{Fe}{2}, there are few surprises.  O, Na, and Al 
abundances have a substantial spread of the kind exhibited by most 
globular clusters (see the reviews cited in \S1), and the 
$\alpha$-elements Si, Ca, and Ti have their usual abundance enhancements 
of $\sim$+0.2 to +0.35 dex.  Sc, V, and Ni have [el/Fe] ratios not far  
from 0.0, [Mn/Fe]~$\simeq$ --0.25 as expected (see \eg, McWilliam 
1997\nocite{Mc97}), and the ratio [Ba/Eu]~$\simeq$ --0.27 is similar to 
that found in field giants (Shetrone 1996, McWilliam 1997) and field 
subdwarfs (Fulbright 2000\nocite{F2000}, 2001\nocite{F2001}) having the 
metallicity of M5.

Somewhat disconcerting is the slight run of [Ba/Eu] toward larger values 
in the most advanced evolutionary state -- the AGB.  That this is 
probably not a manifestation of slow neutron-capture nucleosynthesis occuring
within the stars themselves follows from the fact that [La/Eu] exhibits 
the opposite behavior.  The slight runs seen here in Ba are likely due 
to the choice of microturbulent velocity: the iron line constraint is 
satisfied but the same microturbulent velocity may not be appropriate 
for the atmospheric layers where the Ba lines are formed in the lower 
density AGB stars.

%Section 6.0
%%%%%%%%%%%%%%%%%%%%%%%%%%%%%%%%%%%%%%%%%%%%%%%%%%%%%%
\section{Re-analysis of the Earlier Lick Observations}
%%%%%%%%%%%%%%%%%%%%%%%%%%%%%%%%%%%%%%%%%%%%%%%%%%%%%%

The abundances reported by S92-M5 refer to 13 M5 giants observed with 
the Lick Hamilton Echelle, two of which (II-85 and IV-47) overlap with 
the 25 stars observed at Keck.  Most of these 13 Lick stars lie near 
the RGB tip and therefore provide a valuable supplement to the Keck 
sample.  The earlier analysis employed the ``traditional'' approach; 
we consider here a re-analysis of the same data based on our revised 
approach.  Allowances must be made, however, for the lower spectral 
resolution and more limited free spectral range of the earlier Lick 
observations.

\begin{figure}  %FIGURE 6
\epsscale{1.0}
\plotone{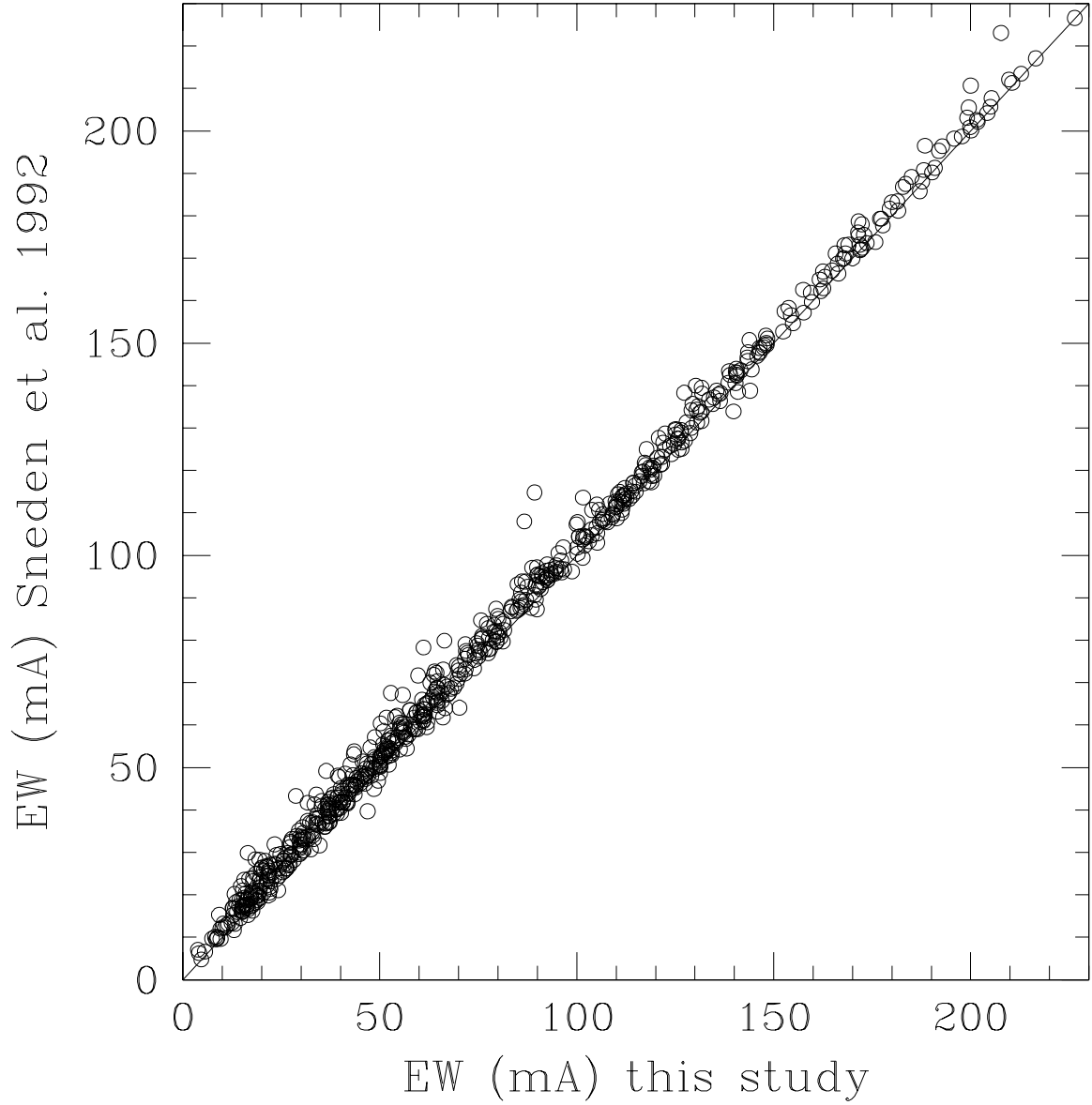}
\caption{Equivalent widths for M5 giants taken from the original Lick 
study (Sneden \etal\ 1992) are plotted against EWs for the same lines 
re-measured in the Lick spectra for this study.
\label{m5.fig6}}
\end{figure}

\begin{figure}  %FIGURE 7
\epsscale{1.0}
\plotone{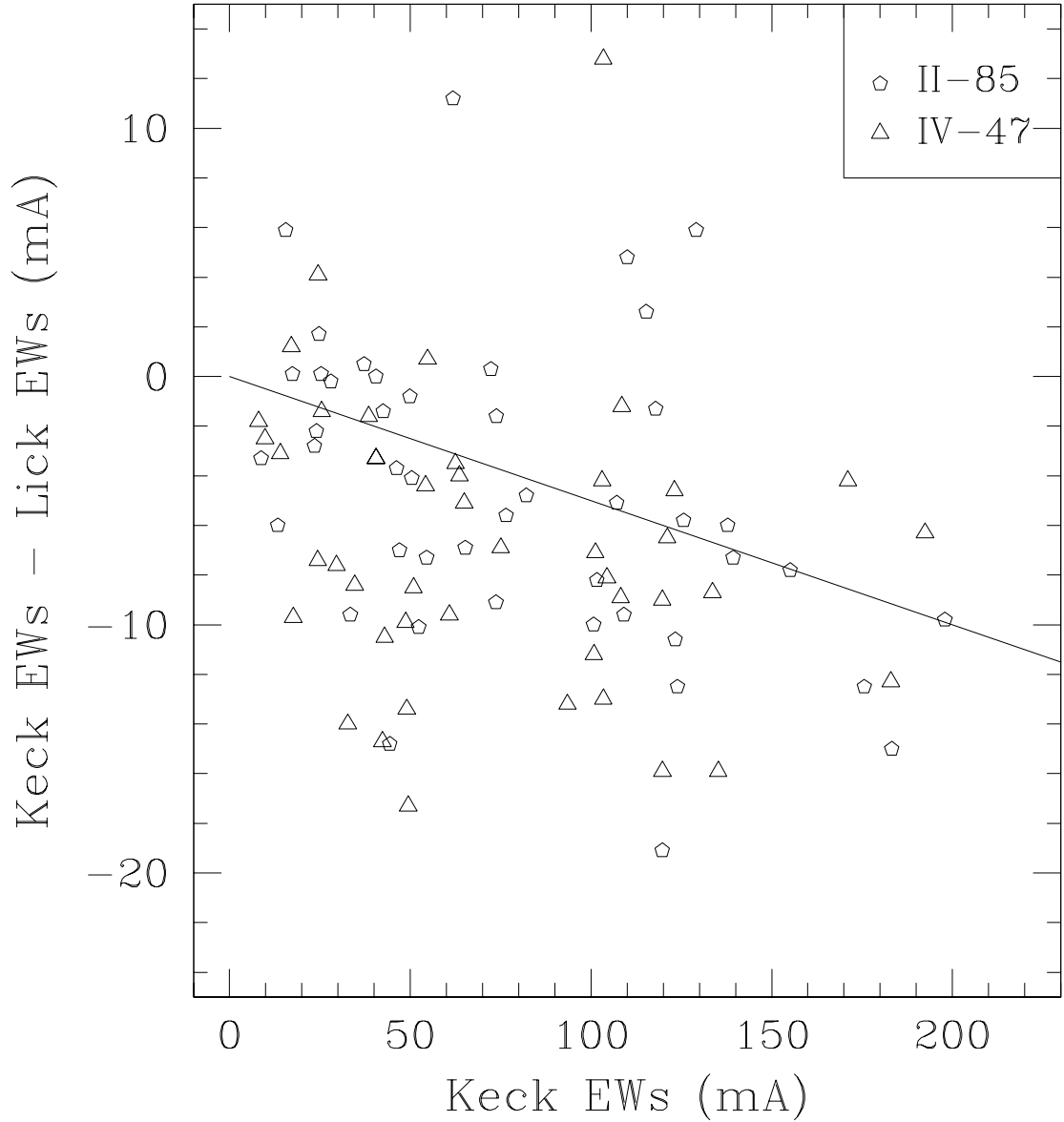}
\caption{Re-measured Lick EWs for data taken prior to installation of 
the new Hamilton spectrograph corrector {\it minus} the Keck EWs as a 
function of Keck EWs for the two stars II-85 and IV-47.  The straight 
line shows the 5\% correction that is required to put the two data 
sets on to the same system.
\label{m5.fig7}}
\end{figure}

We first re-measured the Lick EW's to be sure that continuum levels and
line fitting procedures were consistent with the norms established in
dealing with the Keck I observations.  A plot of original versus 
re-measured Lick EWs is shown in Figure~\ref{m5.fig6}, from which it 
is clear that there is no significant difference between them.  We then 
used the two stars that had been observed both at Keck~I and Lick to 
compare the EW scales of their spectrographs.  In 
Figure~\ref{m5.fig7} we plot the difference between the Keck~I and 
(re-measured) Lick EWs as a function of the Keck~I EWs.  The straight 
line fit illustrated in this figure indicates that the Lick EWs must be 
reduced systematically by 5\% to get on the system defined by the 
Keck~I HIRES spectrograph, that is 
EW$_{\rm Keck}$ = 0.95$\times$EW$_{\rm Lick}$~$\pm$ 0.09~m\AA\ 
($\sigma$~= 6.1~m\AA) for the 42 lines of II-85 and the 51 lines of IV-47
in common between the data sets.\footnote{
This 5 percent correction applies only to Lick Hamilton spectrograph 
observations made prior to 1995, at which time the optics were upgraded
and the 800$\times$800 TI chip was replaced with a 2048$\times$2048 chip. 
EWs from the upgraded Hamilton are known to be on the system of the Keck~I 
HIRES spectrograph (Shetrone 1996\nocite{Sh96}, Johnson 
1999\nocite{Jo99}.)}
 
In Table~\ref{m5.tab6} we give the ``new'' values of \teff, \logg, 
$v_t$, [Fe/H] based on the abundances of \ion{Fe}{1} and \ion{Fe}{2}, 
and the [el/Fe] ratios, all following the modified procedures outlined 
in \S3.2, and employing the revised EWs discussed above, reduced by 5\%. 
We omit star II-9 from further consideration: the S/N of the observed 
Lick spectrum we now consider unacceptably low.  The combined Lick/Keck 
sample therefore contains 35 giants.  The offsets between 
\eps{\ion{Fe}{1}} and \eps{\ion{Fe}{2}} for the entire sample, as 
functions of \teff\ and \logg, are illustrated in 
Figures~\ref{m5.fig2} and \ref{m5.fig3}, and the corresponding 
\eps{\ion{Fe}{2}} values are shown in Figure~\ref{m5.fig4}.

We compare the entries in Table~\ref{m5.tab6} with the ``all star'' 
means of Table~\ref{m5.tab5}, since the Lick sample contains a mixed 
group of tip, RGB, and AGB stars.  There are no differences in [Fe/H] 
exceeding 0.03~dex, and no differences in [el/Fe] ratios exceeding a 
1-$\sigma$ error except in the case of [Sc/Fe] (derived from \ion{Sc}{2} 
lines), where the difference approaches the 2-$\sigma$ level.  The Lick 
spectra also provided access to \ion{Sc}{1} as well as \ion{Sc}{2} 
lines.  However, the [Sc/Fe] ratios from the two stages of ionization 
are in poor agreement.  Literature values for the oscillator strengths 
for the \ion{Sc}{1} lines vary by $\sim$0.4~dex (see S92-M5 for 
discussion) and thus the lines are dropped from this study as well.
The only other element in which two stages of ionization are exhibited 
is Ti, but unfortunately in only two stars, II-85 and IV-47.  In these 
stars, the two [Ti/Fe] stages are in rough agreement even though the 
result is based on only one \ion{Ti}{2} line: 
[\ion{Ti}{1}/\ion{Fe}{1}] and [\ion{Ti}{2}/\ion{Fe}{2}] are 0.30 and 
0.14 for IV-47, and 0.18 and 0.24 for II-85.  Except for the case of 
Sc, the agreement between the Lick- and Keck~I-based abundances appears 
therefore to be excellent.

%Section 6.1
%^^^^^^^^^^^^^^^^^^^^^^^^^^^^^^^^^^^^^^^^^^^^^^^^^^
\subsection{Re-analysis of Other Lick Observations}
%^^^^^^^^^^^^^^^^^^^^^^^^^^^^^^^^^^^^^^^^^^^^^^^^^^

In order to expand our analysis to include more M5 abundances of Mg, Al
and Eu (observed by us in only 2, 23, and 25 of the 36 stars, 
respectively), we have sought out EWs in the literature.  Fortunately, 
Shetrone's (1996) study, based on post-1995 Lick Hamilton spectra,  
included five M5 stars in common with this study (excluding II-9 which we 
omitted earlier due to S/N considerations).  All of the stars in common 
are also part of our Lick sample.  

In Table~\ref{m5.tab7}, we present the results of applying our 
new models (the right hand columns of Table~\ref{m5.tab2}) to the EWs 
of Shetrone (1996) for Mg, Al, and Eu for the 5 stars in common with our
study.  For Mg, we consider only the atomic lines, neglecting the results 
from the MgH features.  The averages and standard deviations of the 
results using the Keck data taken of the two stars for which we are able 
to derive Mg and Al abundances are also shown.  Since both sets of M5 
results are within 1-$\sigma$, in subsequent figures we will treat the 
elemental abundances derived using data we acquired with equal weight to 
those derived from our re-analysis of the Shetrone (1996) EWs.

Finally in Table~\ref{m5.tab8} we present the mean values of the 
[el/Fe]-ratios for AGB, RGB, and RGB plus ``tip'' stars, averaged over the 
Keck and Lick observations, plus ``grand'' mean values averaged over all 35 
stars (the means exclude II-9 as discussed in \S6) taken together.

\begin{figure}  %FIGURE 8
\epsscale{1.0}
\plotone{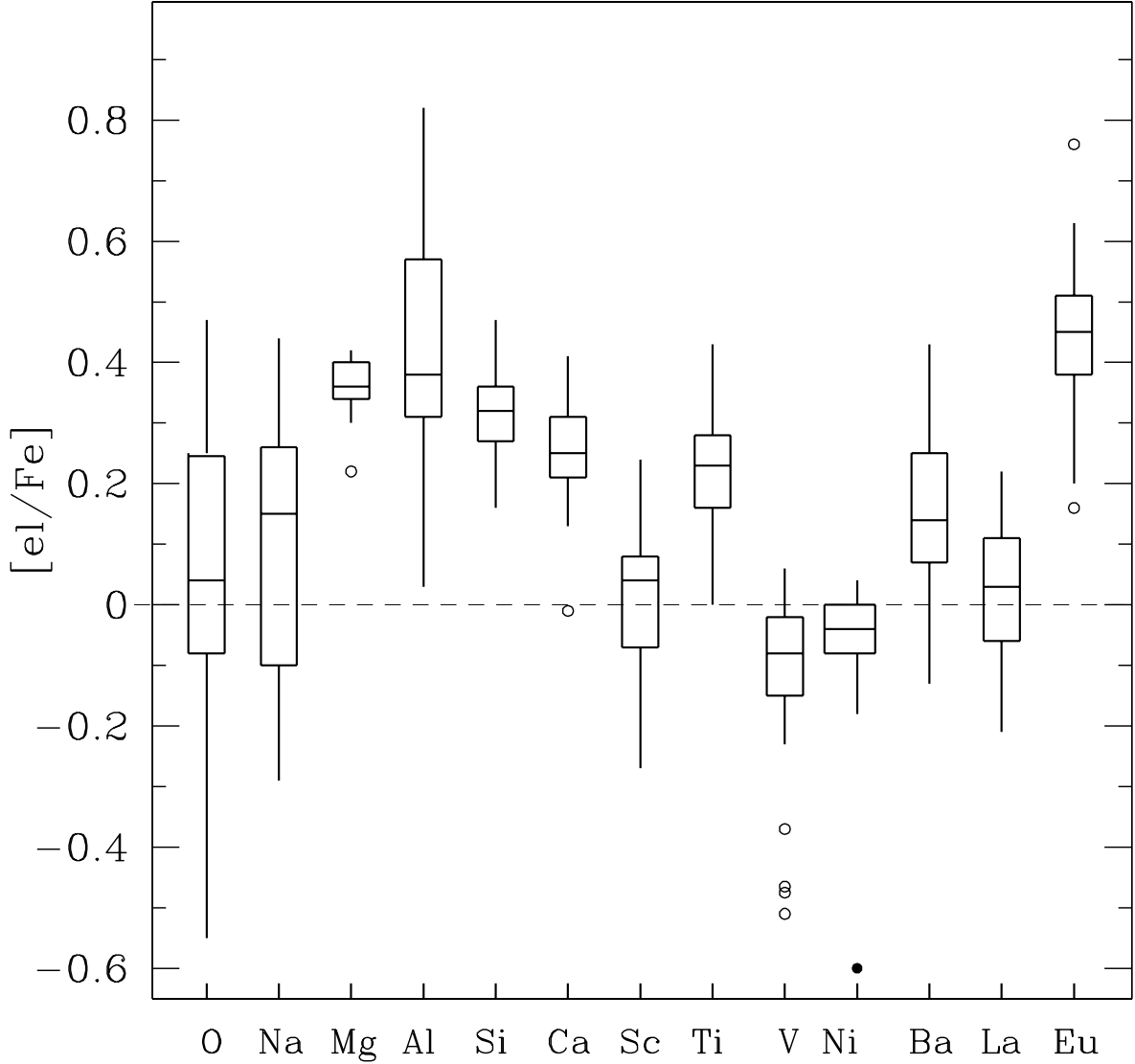}
\caption{Boxplot of the M5 giant star element abundances.   For all 
of the individual abundance boxes, the ``box'' contains the middle 
50\% of the data (ie.~the interquartile range) and the  horizontal 
line inside the box indicates the median value of a particular 
element.  The vertical tails extending from the boxes indicate the 
total range of abundances determined for each element, excluding 
outliers.  Mild outliers (those between 1.5 and 3 times the 
interquartile range) are denoted by open circles.  Severe outliers 
(those greater than 3 times the interquartile range) are denoted 
by filled circles.
\label{m5.fig8}}
\end{figure}

In Figure~\ref{m5.fig8} we present a ``boxplot'' to summarize the 
mean and scatter of each element we analysed in M5.  The Keck and Lick 
results are both represented and the Mg, Al, and Eu abundances include the 
results obtained by putting the Shetrone (1996) EWs on to our system.  This 
boxplot illustrates the median, data spread, skew, and distribution of the 
range of values we derived for each of the elements from our program stars,
as well as possible outliers.  As can be seen in the figure, the abundance 
range for elements sensitive to proton-capture nucleosynthesis is large 
whereas the star-to-star abundance variations for all of the heavier 
elements is quite small, and consistent with the normal scatter resulting 
from observational error.

%Section 7.0
%%%%%%%%%%%%%%%%%%%%%%%%%%%%%%%%%%%%%%%%%%%%%%%%%%%%%%%%%%%%%%%%%%%%%%%%%%%%%
\section{Relationships among [O/Fe], [Na/Fe], [Al/Fe], and CN Band Strengths} 
%%%%%%%%%%%%%%%%%%%%%%%%%%%%%%%%%%%%%%%%%%%%%%%%%%%%%%%%%%%%%%%%%%%%%%%%%%%%%

%Section 7.1
%^^^^^^^^^^^^^^^^^^^^^^^^^^^^^^^^^^^^^^^^^^^^^^^^^^^^^^^^^^^^^^^^^^
\subsection{The Distributions of [O/Fe] and [Na/Fe] with Respect to 
Evolutionary State}
%^^^^^^^^^^^^^^^^^^^^^^^^^^^^^^^^^^^^^^^^^^^^^^^^^^^^^^^^^^^^^^^^^^

\begin{figure}  %FIGURE 9
\epsscale{1.0}
\plotone{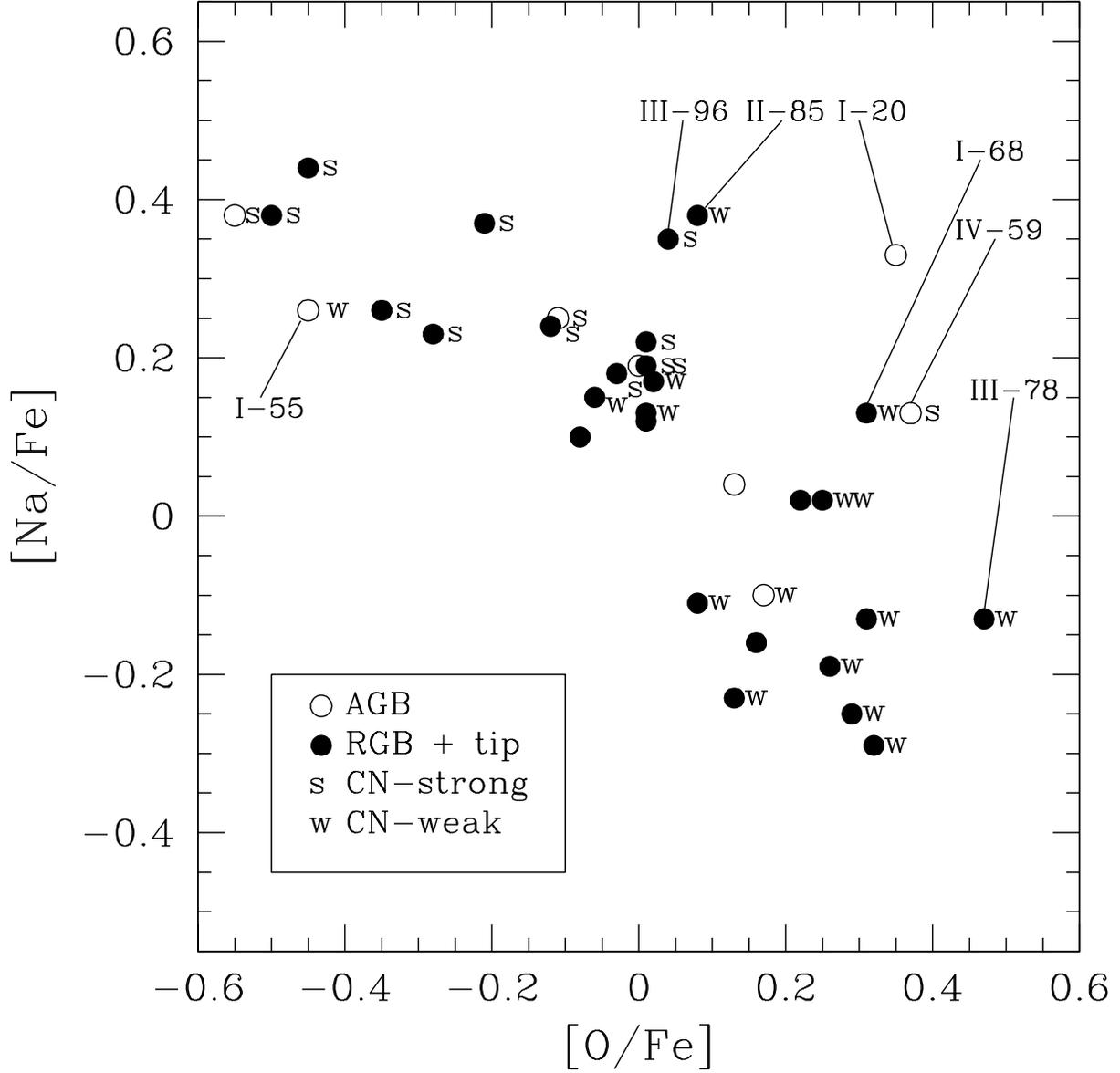}
\caption{M5 sodium abundances (determined by averaging the 
abundances derived by syntheses of the $\lambda\lambda$5682, 
5688~\AA\ and EW abundance derived from $\lambda\lambda$6154, 
6161~\AA\ features) plotted versus oxygen abundances 
(determined by spectrum syntheses of the $\lambda\lambda$6300, 
6364~\AA\ features).  We bin all of the stars by evolutionary 
state (AGB or RGB and ``tip'') and 30 of the stars by CN 
bandstrength (CN-strong or CN-weak).  Some stars stand out, 
are marked with individual star names and are discussed in 
\S7.1 and 7.2.
\label{m5.fig9}}
\end{figure}

\begin{figure}  %FIGURE 10
\epsscale{1.0}
\plotone{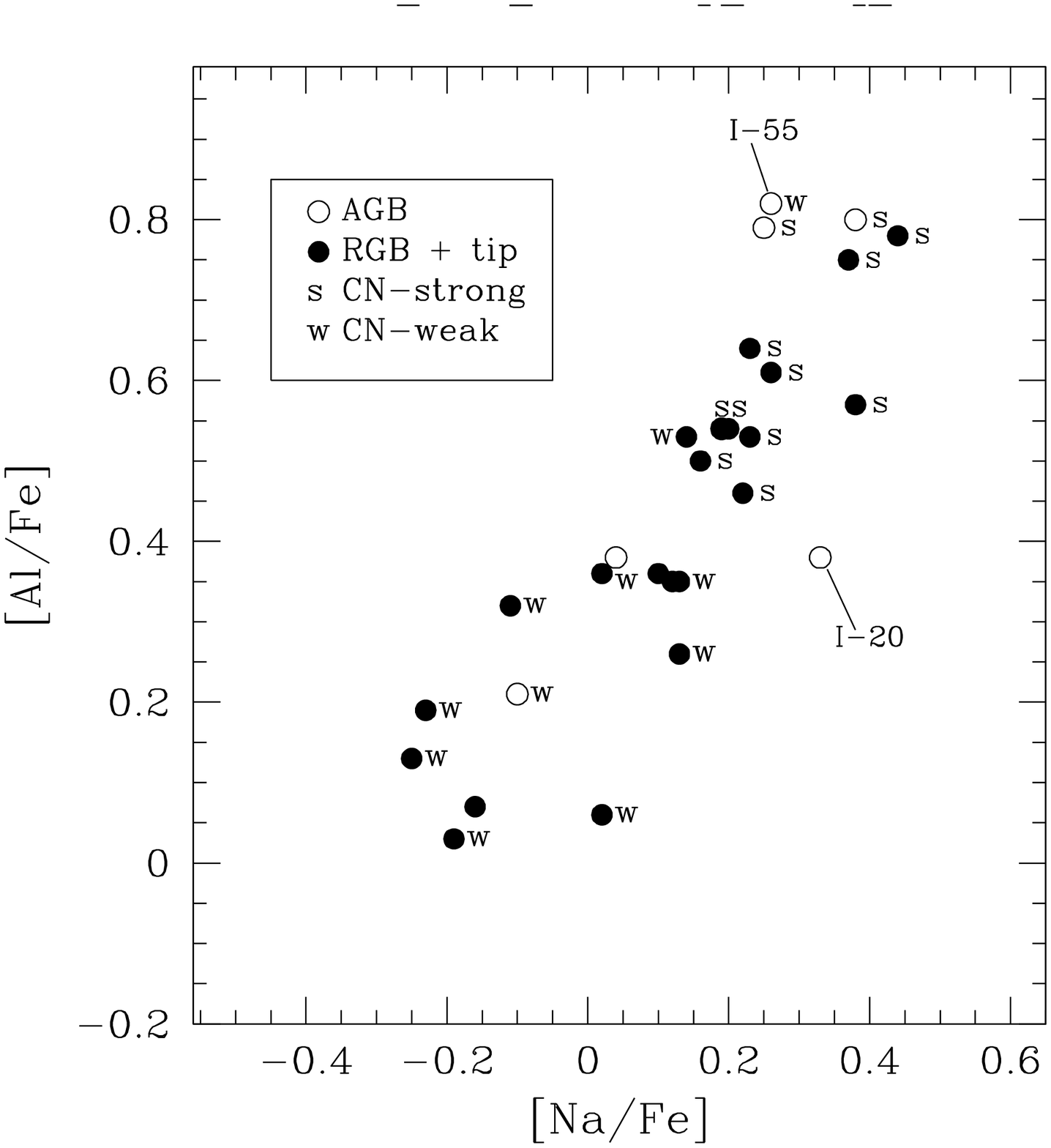}
\caption{M5 aluminum abundances plotted versus sodium.  Our 
M5 program stars show the ``classic'' anti-correlations and 
correlations seen in the brighter stars, as well as in other 
clusters observed by the Lick-Texas group.  Stars are marked 
as those of Figure~\ref{m5.fig9}.
\label{m5.fig10}}
\end{figure}

In Figure~\ref{m5.fig9} we plot [Na/Fe] versus [O/Fe] and in 
Figure~\ref{m5.fig10}, [Al/Fe] versus [Na/Fe].  The results are 
consistent with the expected anti-correlation of Na with O and the 
correlation of Al with Na (see reviews cited in \S1).  The shape of 
these relationships generally follows that seen earlier in M15 
(Sneden \etal\ 1997)\nocite{SKSSLP97}, M10 (Kraft \etal\ 
1995)\nocite{KSLSB95}, M3 (Kraft \etal\ 1992)\nocite{KSLP92}, 
M92 (Sneden \etal\ 1991\nocite{SKPL91}, Shetrone 1996\nocite{Sh96}) 
and M13 (Shetrone 1996, Kraft \etal\ 1997\nocite{KSSSLP97}, Cavallo \&
Nagar 2000).  The ranges of O and Na in M5 are comparable with the large
range seen in M13, although the range of Al is distinctly smaller,
more in keeping with the other clusters cited.  On the other hand the 
range in O and Al is larger than is found in M4, a cluster with 
metallicity similar to M5, although the range in Na is about the same 
(I99-M4).

One star that stands well off the relationships shown in these two figures
is I-20, the coolest of the AGB stars in our sample.  It has an unusually 
low \ion{Fe}{2} abundance, a fairly low \ion{Fe}{1} abundance and by far 
the largest microturbulent velocity.  The low Fe abundances, however, 
cannot alone account for the high [O/Fe] and [Na/Fe] values, while 
simultaneously yielding an [Al/Fe] ratio that is too low for its [O/Fe]. 
Possibly the star is an unresolved binary, consisting of a pair of RGB 
stars, but this would require that the two components differ substantially 
in $V$, since the combined light is only 0.2~mag above the RGB.  In that 
case, the line profiles might indeed be widened, thus accounting for the 
large $v_t$ value, but would also be unsymmetrical, contrary to their 
actual appearance.   However, I-20 does exhibit the largest H-$\alpha$ 
emission among our sample for which H-$\alpha$ was recorded on the chip.  
We note that if I-20 is a pair of RGB stars disguised as an AGB, the 
\logg\ for these stars ($\sim$1.65) would lead to a {\it further 
reduction} of 0.1~dex in the [V/Fe]-ratio (see Table~\ref{m5.tab3}).
At the moment we conclude that I-20 has elevated O and Na abundances 
compared with other M5 stars, exhibiting in exaggerated form the excess 
O and Na abundances found previously in IV-59 (Kraft \etal\ 1992, Briley 
\& Smith 1993\nocite{BS93}, Smith \etal\ 1997).  The additional 
anomalous stars noted in Figure~\ref{m5.fig9} are discussed further 
in \S7.2.

\begin{figure} %FIGURE 11
\epsscale{1.0}
\plotone{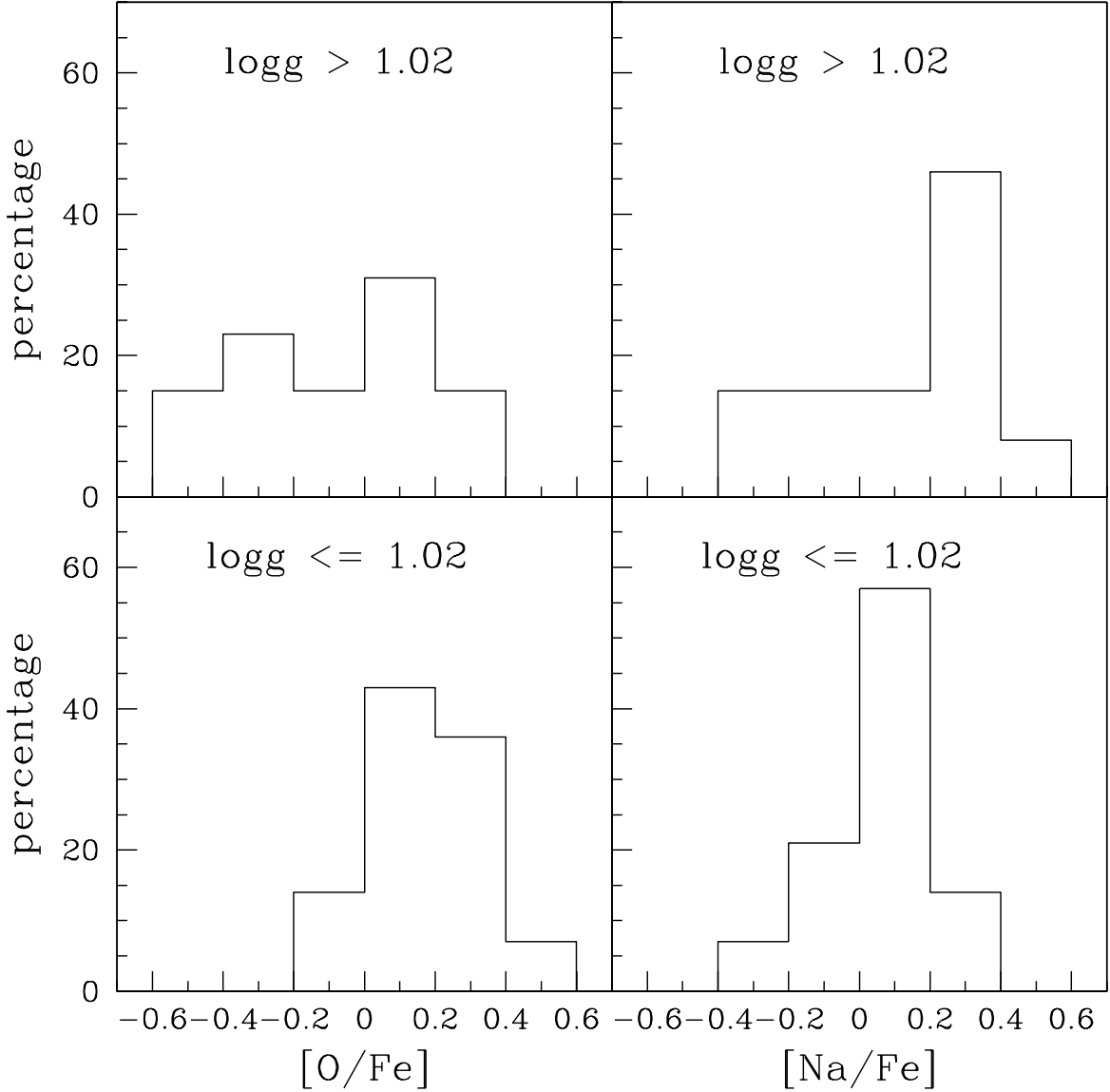}
\caption{Four histograms of [O/Fe] (on the left) and [Na/Fe] 
(on the right) abundance ratio distributions for our sample 
of M5 RGB and ``tip'' stars.  The top panels illustrate the 
percentage of the 13 stars with \logg~$>$ 1.02 and the bottom 
panels illustrate the percentage of the 14 stars with 
\logg~$\leq$ 1.02.  One can easily see how the distributions 
in these two groups change.
\label{m5.fig11}}
\end{figure}

When a large sample of M13 giants is divided at \logg~= 1.02, there is a 
clear shift of Mg, Na, and O abundances in support of the evolutionary
scenario (Kraft \etal\ 1997, Hanson \etal\ 1998). If the present sample of
M5 RGB and ``tip'' stars is divided by evolutionary state, do we find a
similar shift that supports the evolutionary picture? In response to this
inquiry, we explored the distributions of [O/Fe] and [Na/Fe] when the
RGB plus ``tip'' giants are divided into two virtually equal evolutionary
groups: 14 giants having \logg~$\leq$ 1.02 and 13 having \logg~$>$ 1.02. 
The results are shown in histogram form in Figure~\ref{m5.fig11}. We 
performed a Mann-Whitney U-Test (equivalent to a Wilcoxon Rank-Sum Test)
to test the hypothesis that the two samples have the same mean of
oxygen abundance ratio distribution against the hypothesis that they 
differ.  There is a less than 1\% probability that the two samples have 
the same mean of distribution.  Thus we confirm statistically what can be 
discerned by eye: the two samples have significantly different means of 
distribution.  In contrast with M13, the distributions above and below  
evolutionary \logg\ of 1.02 are differently skewed, but in a sense opposite 
to that expected in the evolutionary scenario. This result, although not 
incompatible with the primordial scenario, does not rule out the notion 
that the required evolutionary change could have taken place among giants 
fainter than those probed by our sample.

%Section 7.2
%^^^^^^^^^^^^^^^^^^^^^^^^^^^^^^^^^^^^^^^^^^^^^^^^^^^^^^^^^^^^^^^^^^^
\subsection{Relationship of CN Band Strengths to [O/Fe] and [Na/Fe]}
%^^^^^^^^^^^^^^^^^^^^^^^^^^^^^^^^^^^^^^^^^^^^^^^^^^^^^^^^^^^^^^^^^^^

Measurements of CN band strengths in a substantial number of M5 RGB and
AGB stars have been carried out by Smith \& Norris (1983, 1993) and by 
Briley \& Smith (1993).  Like many other globular clusters, M5 giants 
present a bimodal distribution of CN band strengths. Smith \etal\ (1997) 
found that strong CN bands among M5 giants are generally driven by a 
high abundance of N, and that the index S(3839), which is a measure of
the flux in the \wave{3883} CN molecular band relative to that in a nearby 
comparison region, is anti-correlated with [O/Fe] and correlated with 
[Na/Fe].  For a scenario in which the star-to-star abundance variations 
seen here are a result of proton-capture nucleosynthesis, such a behavior of CN 
band strengths is expected, since C will likely have been processed to N 
when O is transmuted to N and Ne to Na (Langer \etal\ 1993, Cavallo \etal\ 
1998).

\begin{figure} %FIGURE 12
\epsscale{1.0}
\plotone{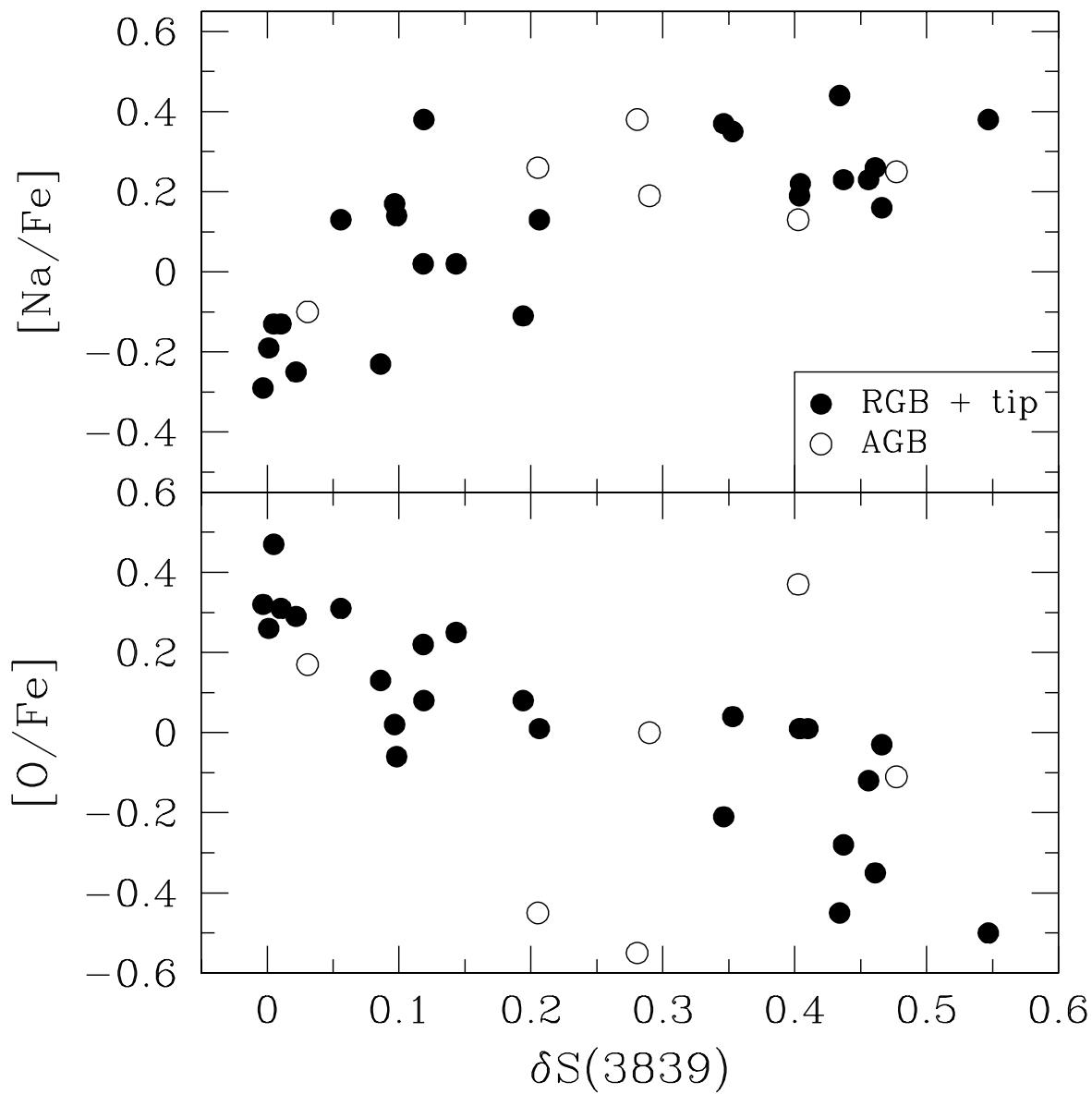}
\caption{[Na/Fe] and [O/Fe] plotted as a function of the CN 
bandstrength index $\delta$S(3839), which has been derived
from the data from Smith \& Norris (1983, 1993), Briley \& 
Smith (1993) and Smith \etal\ (1997), and ``detrended'' for 
temperature effects.   Stars are depicted by AGB or RGB and 
``tip'' evolutionary state.
\label{m5.fig12}}
\end{figure}

The giants studied here add further weight to this picture.  In 
Figure~\ref{m5.fig12}, we plot [Na/Fe] and [O/Fe] as a function of the 
CN bandstrength index $\delta$S(3839) for 30 giants.  The S(3839) values
were taken from Smith \& Norris (1993), supplemented by Smith \& Norris
(1983), Briley \& Smith (1993) and Smith \etal\ (1997).  Where available, 
the Smith \& Norris (1993) values were adopted.  In the case of multiple 
measurements from the other sources, an average value was adopted, 
subsequent to employing the transformations described in Smith \etal\ 
(1997).  The ``raw'' S(3839) values are shown in Table~\ref{m5.tab1}.
Because some of the CN bandstrength measured by S(3839) is 
sensitive to temperature (given the same C and N abundances, cooler stars 
have intrinsically larger S(3839) indices than their hotter counterparts), 
we ``detrended'' the raw S(3839) index. We fitted a ``baseline'' as a 
function of \teff\ to the S(3839) results, and formed a differential CN 
strength index,
$$
\delta S(3839) = S(3839) - (0.991 - 1.95 \times 10^{-4} \times \teff).
$$
We further binned the stars into ``CN-strong'' and ``CN-weak'' groups using 
Smith \& Norris (1993) as a guide, with the exception that we designated 
the Smith \& Norris ``intermediate'' strength stars, and other stars with 
similar $\delta$S(3839) measures, as CN-strong.  

We find the $\delta$S(3839) index to be correlated with Na and 
anti-correlated with O, as shown in Figure~\ref{m5.fig12}.  Returning 
to Figure~\ref{m5.fig9}, where we note the stars as being CN-strong (s) 
or CN-weak (w), we see that the CN-strong stars lie in the low O, high Na
part of the diagram, with CN-weak stars dominating the high O, low Na 
portion.  As seen in previous studies of other globular clusters, the Na 
and O abundance patterns also correlate with the CN strength as inferred 
by the $\delta$S(3839) index.  A few stars plotted in 
Figure~\ref{m5.fig9} remain anomalous; we have already noted I-20 and 
IV-59.  The two stars I-68 and III-78 appear to be a bit oxygen-rich for 
their sodium abundances.  II-85 has high [O/Fe] but only an intermediate
value of $\delta$S(3839).  I-55 has diminished oxygen and enhanced 
sodium, opposite to what would be expected from its CN bandstrength 
index.  For this star, the abundances of the two pairs of Na lines have 
a standard deviation of 0.06~dex and the two oxygen line syntheses are 
in excellent agreement.  However, there is some ambiguity about its CN 
bandstrength classification: Smith \& Norris (1993\nocite{SN93}) give a 
small S(3839) value, corresponding to that of a CN-weak star but list 
I-55 as a CN-strong star.  We suggest that a closer examination of the 
CN measurements of this particular AGB star is required to resolve the 
ambiguity.  The star III-96 appears to have an excessive Na abundance, 
but the values of [Na/Fe] derived from the $\lambda\lambda$5682, 
5688~\AA\ pair and the $\lambda\lambda$6154, 6161~\AA\ pair are in poor 
agreement. 

Similar correlations between CN-strength and location in the Al-Na 
correlation are found in Figure~\ref{m5.fig10}, where we find, in 
general, that the CN-strong stars lie in the high Al, high Na part of the 
diagram.  But again there are a few anomalous stars like I-20 and I-55. 
We conclude that, within the errors of our abundance determinations, most 
M5 giants follow the expected pattern of proton-capture nucleosynthesis with 
only a few exceptions, but these exceptions are compatible with the 
existence of primordial abundance variations among at least a few members 
of M5.

Figure~\ref{m5.fig12} also demonstrates that within both the CN-weak 
and CN-strong groups there is an intrinisc spread in $\delta$S(3839), 
[O/Fe], and [Na/Fe].  Even the CN-weak stars themselves show an intrinsic 
dispersion in $\delta$S(3839), [O/Fe], and [Na/Fe].   The same is true for 
the CN-strong stars.  If these observations are not a result of scatter
due to observational error, then they are consistent with a ``primordial''
scatter of Na and O, produced by proton-capture nucleosynthesis in an 
earlier generation of stars, which has subsequently been modified by deep 
mixing in the sample stars themselves.

%Section 8.0
%%%%%%%%%%%%%%%%%%%%%%%%%%%%%%%%%%%%%%%%%%%%%%%%%%%%%%
\section{A Comparison of [el/Fe] Ratios in M4 and M5}
%%%%%%%%%%%%%%%%%%%%%%%%%%%%%%%%%%%%%%%%%%%%%%%%%%%%%%

Although many clusters have giants with variations in C, N, O, Na, Al, and
Mg that are attributable to the proton-capture process, they also usually 
yield [el/Fe] ratios of the Fe-peak elements, $\alpha$-capture elements 
such as Si, Ca, and Ti and heavy neutron-capture elements such as Ba and Eu 
that are stable and ``normal'' with respect to typical halo field stars. 
In this respect, M5 is no exception.  In Table~\ref{m5.tab9} we list 
[el/Fe] ratios for M5 together with the mean of a large sample of halo 
field subdwarfs at [Fe/H]~$\simeq$ --1.2 (Fulbright 2000\nocite{F2000}, 
2001\nocite{F2001}), for most of the elements, supplemented by field giants 
(Gratton \& Sneden 1994\nocite{GS94}) in the case of La.  In the upper part 
of this table, we consider only those elements not subject to 
proton-capture nucleosynthesis.  Among these elements, the agreement between M5 
and the halo field subdwarfs is good, the difference never exceeding 
1-$\sigma$ (for M5).  In the lower part of this table  we add rows for 
[Na/Fe] and [Al/Fe].  In contrast to cluster stars, field stars do not show 
evidence for enhanced Na and Al as a result of proton captures on Ne and 
Mg, respectively (Hanson \etal\ 1998, Kraft 1999\nocite{Kr99}, Gratton 
\etal\ 2000), so it is not surprising that M5 giants on the average show 
higher [Na/Fe] and [Al/Fe] values than are found in halo field subdwarfs.

We wish to compare [el/Fe]-ratios in M5 with M4, a cluster having nearly 
the same metallicity as M5.  However, the M5 [el/Fe] ratios listed in 
Table~\ref{m5.tab9} come from the revised method of this paper.  The 
abundances in M4 had been derived assuming that \logg\ could be set from 
the ionization equilibrium of \ion{Fe}{1} and \ion{Fe}{2}.  
Unfortunately, for the M4 stars we cannot set \logg\ from the Alonso 
\etal\ (1999) relation between $(B-V)$ and \teff\ plus application of 
stellar evolution.  M4 is heavily and differentially reddened, and the 
ratio of total to selective absorption  $R_V$~= $A_V/E(B-V)$ is not the 
normal value of 3.2, but is estimated to lie in the range 3.1 to 4.0 
(Dixon \& Longmore 1993; I99-M4). Thus $A_V$ is not accurately known.

To gain an idea of the effect on [el/Fe] ratios in M4 if we were to adopt 
the same approach consistent with the non-LTE precepts of TI99, we turn 
the problem around: we adopt the geometrical distance (1.7~kpc) given by 
Peterson \etal\ (1995) and increase the traditionally derived values of 
\logg\ by an amount that offsets [Fe/H] from \ion{Fe}{1} and \ion{Fe}{2} 
by the same amount as we have derived for M5, \ie, 0.09~dex. From Table~2 
of I99-M4, we find that the values of \logg\ of giants in M4 need to be 
increased by 0.12~dex.  Again from Table~\ref{m5.tab3} we see that the 
effect of this increase on [el/Fe]-ratios is actually quite small: for O, 
Na, Mg, Al, Si, Ca, Sc, Ti, La, and Eu, the corrections do not exceed 
0.02~dex.  The reduction in [Ba/Fe] is a bit larger: 0.04~dex.  These 
values, therefore slightly revised from those given in Table~5 of I99-M4, 
are listed in the last column of Table~\ref{m5.tab9}.  We conclude 
that any comparison of [el/Fe] ratios between these two clusters is 
nearly independent of the analysis technique.  But two consequences of 
the revised approach are noteworthy.  First, for M4, $<$[Fe/H]$>$~= 
--1.17 and --1.08 if derived from \ion{Fe}{1} and \ion{Fe}{2}, 
respectively.  Second, for a geometrical distance of 1.7~kpc and the 
revised values of \logg, we obtain a ratio of total to selective 
absorption $R_V$~= 3.9 (\S3.4 of I99-M4) if $<E(B - V)>$~= 0.37 (Dixon 
\& Longmore 1993).

To put the I99-M4 results on the M5 system of analysis used in this
study, we have applied the $\delta$[el/Fe] corresponding to the 0.12~dex 
increase in \logg.  Most of the mean values of [el/Fe] listed in 
Table~\ref{m5.tab9} for M4 and M5 are derived from 24 and 35 stars 
respectively, and therefore the standard deviations are generally quite 
small. In addition, systematic errors are likely not to be a problem since 
the methods of analysis are basically the same.  Among most Fe-peak and 
$\alpha$-element abundances, the agreement between M4 and M5 is good, 
generally within 1-$\sigma$.  The main exception seems to be [Si/Fe]: the 
Si abundance in M4 exceeds that of M5 by about 3-$\sigma$.  Similar 
3-$\sigma$ overabundances in M4 compared with M5 (and the field) are found 
in Ba and La.  Al is also higher in M4 than in M5 (and the field), but 
the significance of this result is less clear since Al is a product of 
proton-capture nucleosynthesis and known to be highly variable among cluster 
stars.  

Based on our large stellar sample and the updated I99-M4 results, we
extend the work of I99-M4 and confirm what Brown \& Wallerstein 
(1992\nocite{BW92}) found from a small sample of M4 stars: Si, Al, Ba
and La abundances are unusually large in M4.  We confirm also the 
earlier study of S92-M5 which showed that M5 has ``normal'' abundances. 
In the comparison of M5 with M4, we did not find the difference in Mg 
reported by I99-M4 but, in our observations, the \ion{Mg}{1} lines were 
recorded in the spectra of only two M5 giants.  The mean abundance for 
our two stars is 0.39 $\pm$ 0.03 ($\sigma$ = 0.07) which is comparable 
to the mean in I99-M4, 0.42 $\pm$ 0.02 ($\sigma$ = 0.08).  In the I99-M4
study, the difference reported with respect to M5 was determined
using the results of Shetrone (1996).  Employing Shetrone's published EWs 
and the revised models we have derived in this analysis, we obtain an 
average [Mg/Fe]-ratio of +0.32 $\pm$ 0.03 ($\sigma$ = 0.07) for the 
Shetrone data, which, combined with the two stars from our sample, gives 
an average [Mg/Fe]-ratio of 0.34 $\pm$ 0.03 ($\sigma$ = 0.07).  We find
that the difference in Mg between M4 and M5 is a bit more than 1-$\sigma$.

\begin{figure} %FIGURE 13
\epsscale{1.0}
\plotone{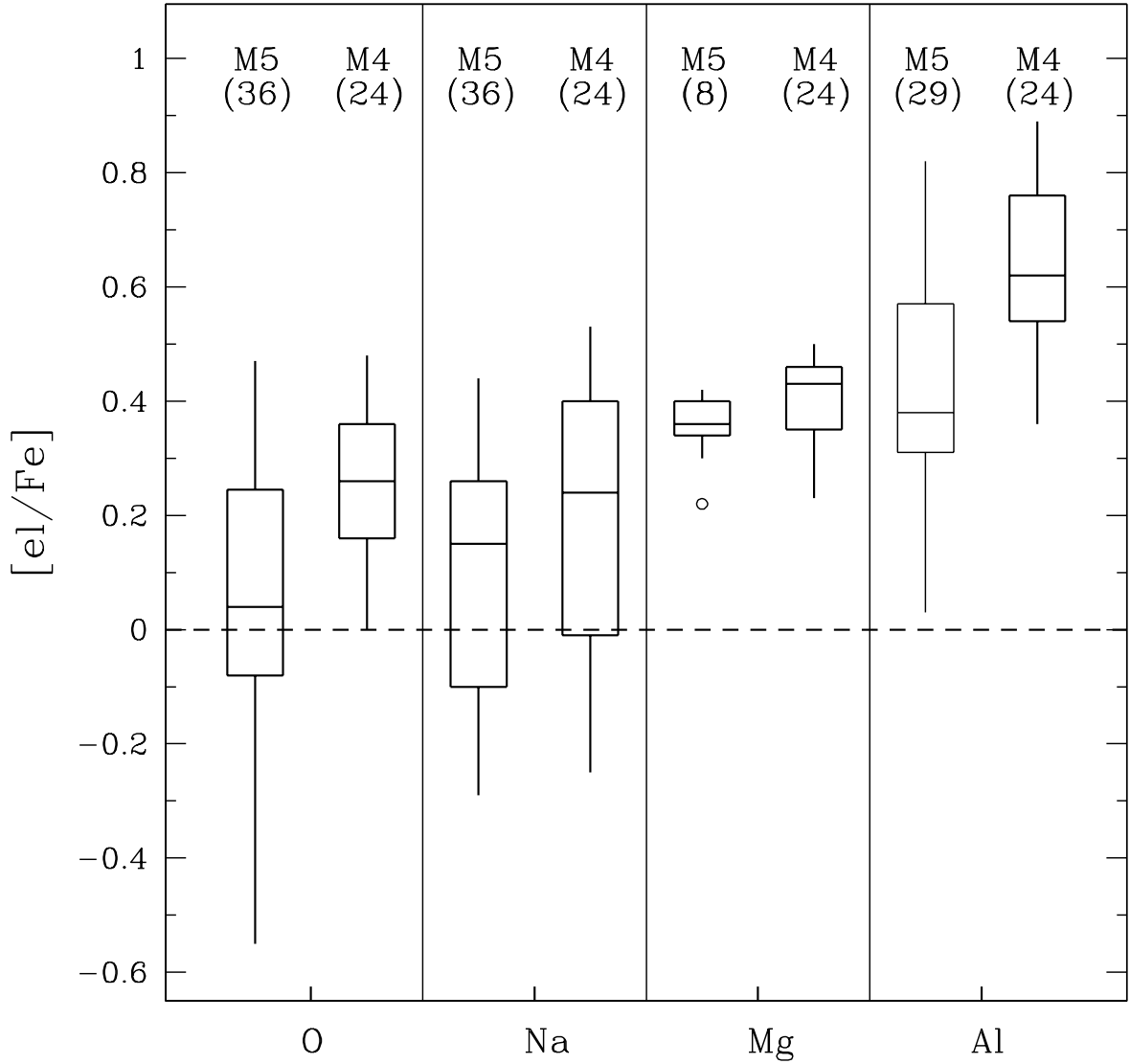}
\caption{Boxplot of the M4 and M5 giant star abundances for 
elements which may be sensitive to proton-capture 
nucleosynthesis.  The statistical abundance distributions 
represented by each box's vertical boundaries, etc., are as 
described in Figure~\ref{m5.fig8}.
The number of stars included in each boxplot is noted in parentheses.
\label{m5.fig13}}
\end{figure}

\begin{figure} %FIGURE 14
\epsscale{1.0}
\plotone{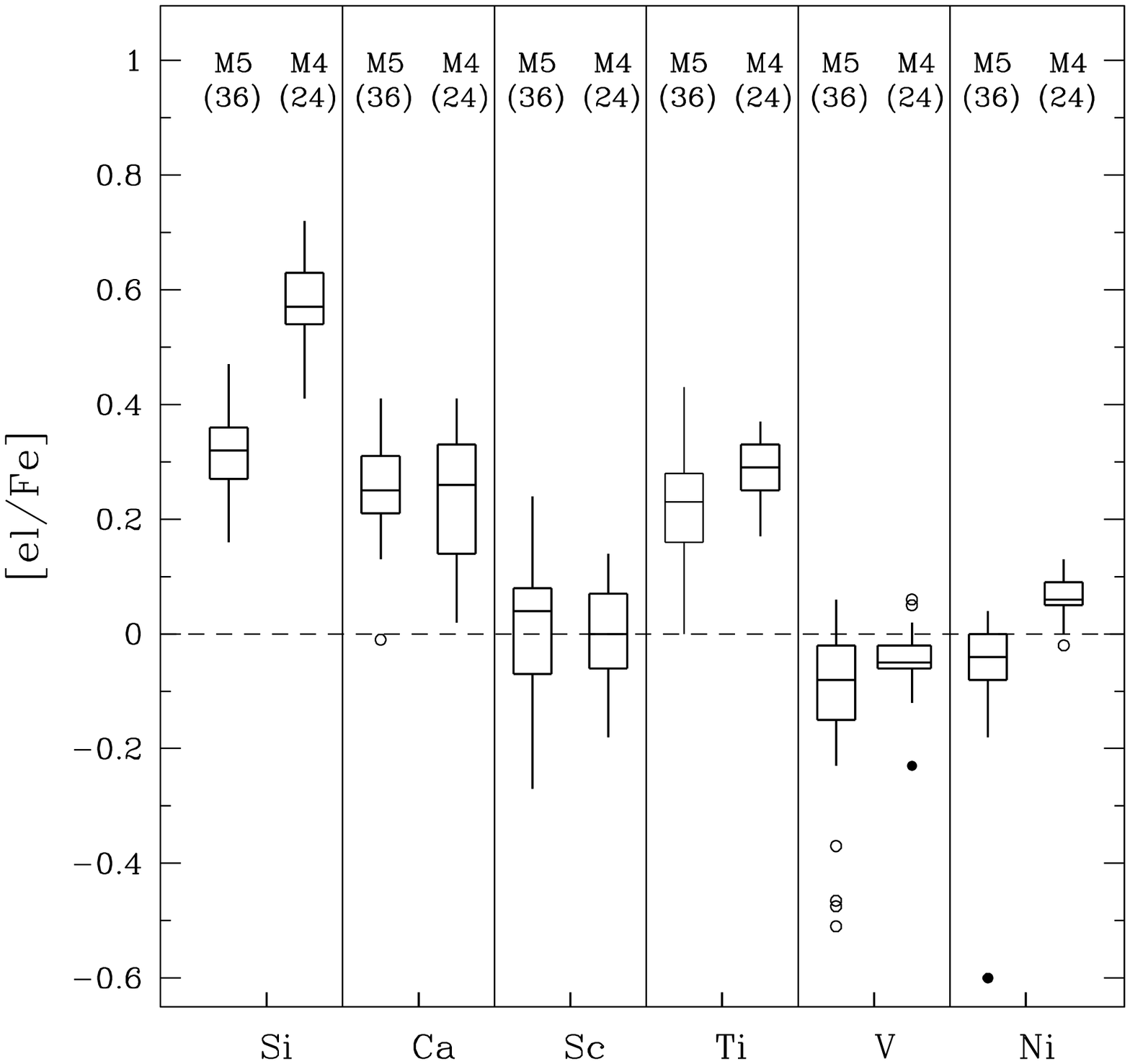}
\caption{Boxplot of the M4 and M5 giant star abundances for 
heavier $\alpha$- and Fe-peak elements.  The statistical 
abundance distributions represented by each box's vertical 
boundaries, etc., are as described in Figure~\ref{m5.fig8}.
The number of stars included in each boxplot is noted in parentheses.
\label{m5.fig14}}
\end{figure}

\begin{figure} %FIGURE 15
\epsscale{1.0}
\plotone{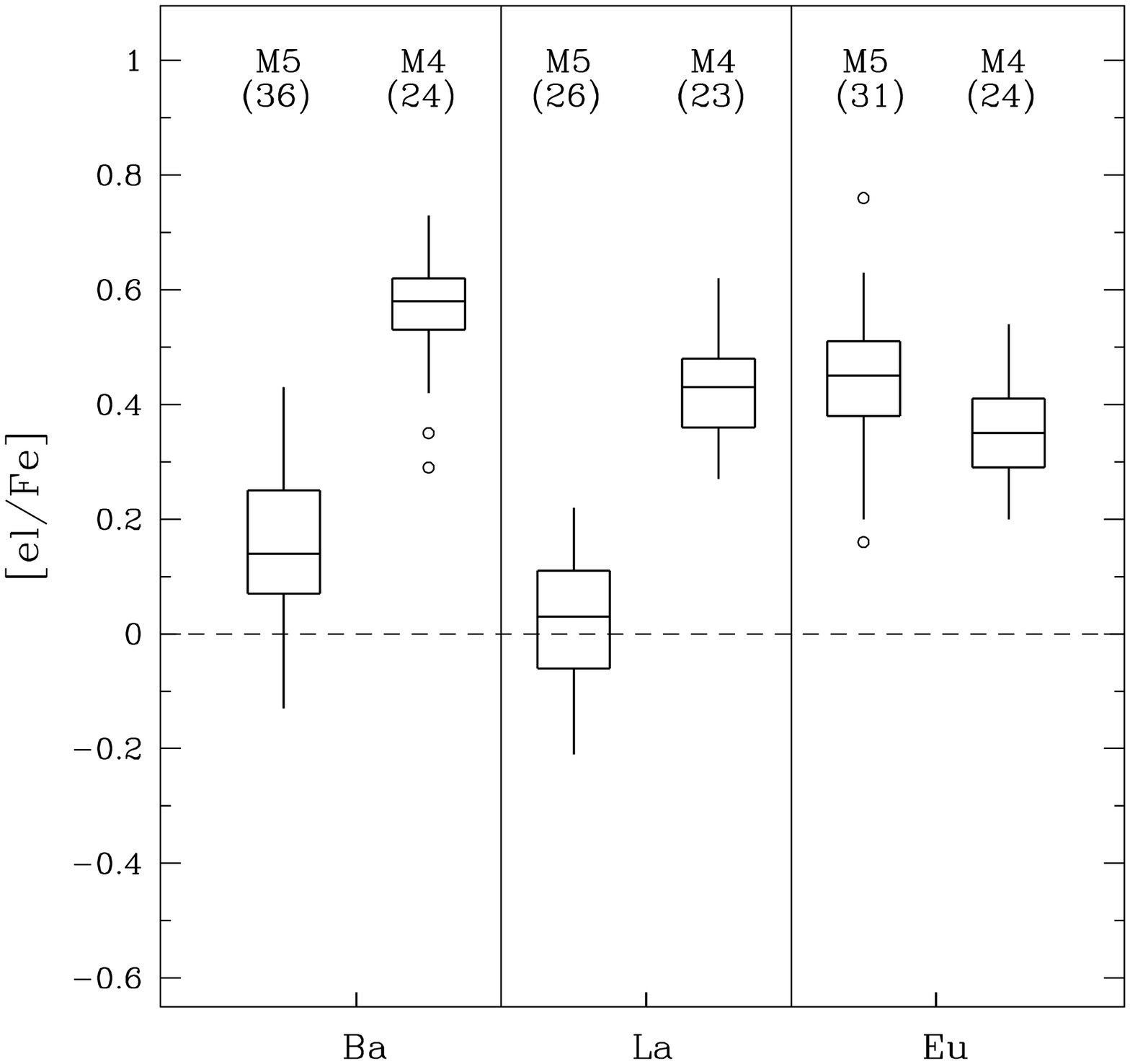}
\caption{Boxplot of the M4 and M5 giant star abundances for 
$s$- and $r$-process elements.  The statistical abundance 
distributions represented by each box's vertical boundaries, 
etc., are as described in Figure~\ref{m5.fig8}.  The number 
of stars included in each boxplot is noted in parentheses.
\label{m5.fig15}}
\end{figure}

In the following boxplots, we show the range of abundances found in both 
M4 and M5: Figure~\ref{m5.fig13} illustrates the elements which may 
be sensitive to proton-capture nucleosynthesis, Figure~\ref{m5.fig14} 
shows the heavier $\alpha$- and Fe-peak elements, and 
Figure~\ref{m5.fig15} illustrates the $s$- and $r$-process element 
abundances.  In these plots, it is apparent that the Al ``floor'' in M4 
is elevated, while the range of Al abundances is roughly equal to that 
of M5; Si is elevated; and both Ba and La are elevated.  The other 
elements show sufficient overlap to be consistent with showing no 
significant differences in the abundance ratios.  Could the enhanced Ba 
and La abundances in M4 found by Brown \& Wallerstein (1992) and I99-M4 
be simply due to the weighting of the AGB stars in their samples?   
First, we point out that not a single star in M5 (AGB, RGB or ``tip'') 
has been found to possess a La abundance that is as large as even the 
lowest value found in M4.  Next, we note that in the I99-M4 study, the 
mean Ba and La abundances for the AGB stars (as determined using their 
Figure~12) are the same as those in the rest of the sample.  Since the 
Ba and La enhancements in M4 are not a result of slow neutron-capture 
synthesis occuring in the stars themselves, they must be signatures of 
primordial enrichments of the material out of which the M4 stars formed.

%Section 9.0
%%%%%%%%%%%%%%%%%%%%%%%%%%%%%%%%%%%%%%%%%
\section{Comparisons with Other Clusters}
%%%%%%%%%%%%%%%%%%%%%%%%%%%%%%%%%%%%%%%%%

\begin{figure} %FIGURE 16
\epsscale{1.0}
\plotone{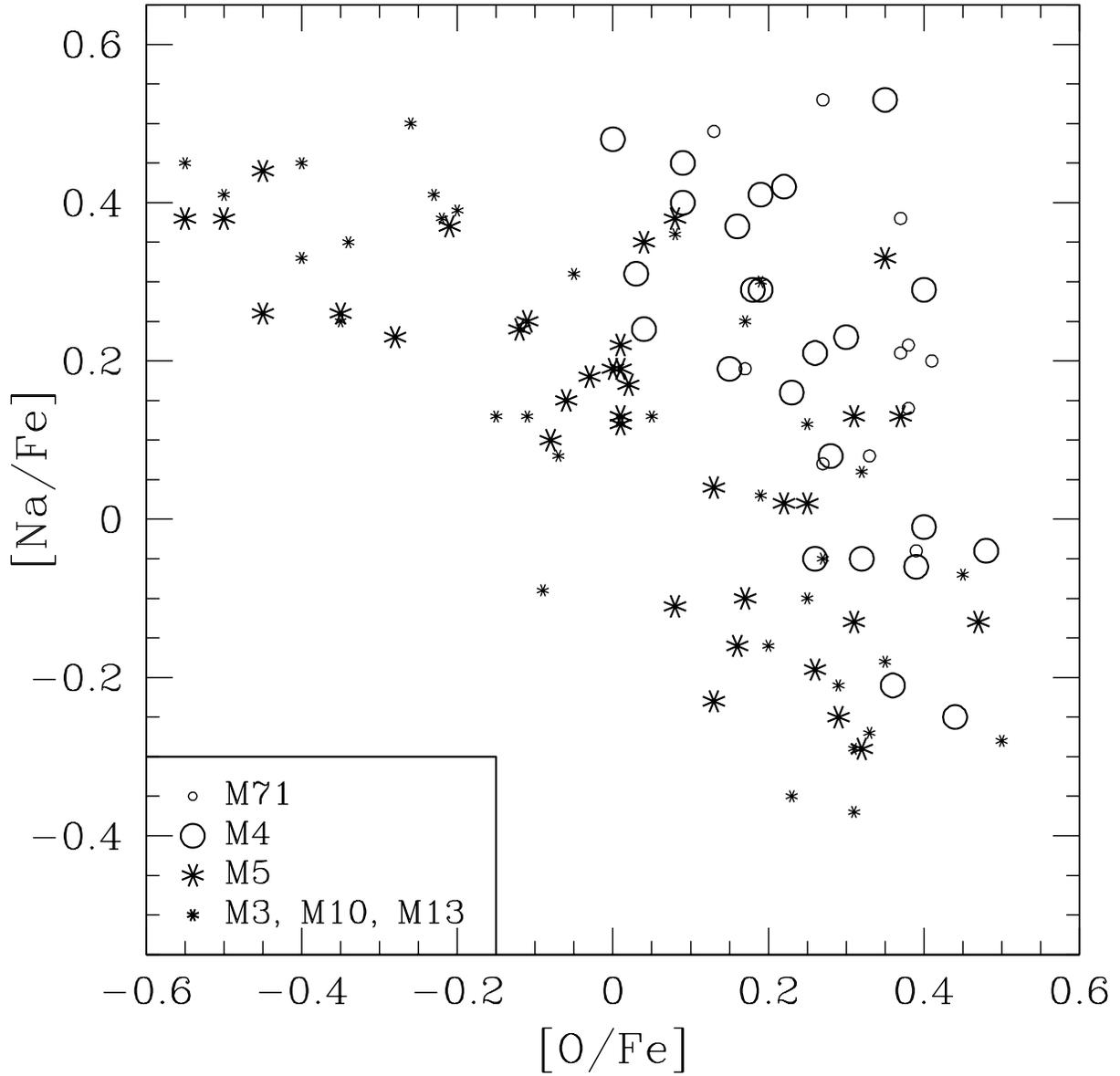}
\caption{[Na/Fe] versus [O/Fe] for M5 and M4 and globular clusters 
previously studied by the Lick-Texas group that bracket M5 and M4
in metallicity. The abundance ratio anti-correlation is divided 
into two groups, and the symbols chosen accordingly, one for the 
M4-like clusters and one for the M5-like clusters.
\label{m5.fig16}}
\end{figure}

In Figure~\ref{m5.fig16}, we plot [Na/Fe] versus [O/Fe] values for 
M4 and M5 and globular clusters that bracket M4 and M5 in metallicity.  
The plot illustrates a difference in abundance patterns that can be 
divided into two groups.  The O versus Na anti-correlation found in M5 
resembles that found in the slightly more metal-poor clusters M3, M10 
and M13 ($<$[Fe/H]$>$ --1.5 to  --1.6).  The pattern is quite different 
from that found in M4.  Instead, M4's behavior seems to be much more 
like that of M71, a disk cluster of much higher metallicity 
($<$[Fe/H]$>$ $\sim$ --0.8; Sneden \etal\ 1994\nocite{SKLPS94}).  We 
note that these clusters can also be binned by horizontal branch 
morphology.  According to the catalog compiled by Harris 
(1996\nocite{Har96}; June 22, 1999 revision), M3, M5, M10, and M13 all 
have $(B-R)(B+V+R) >$ 0 (where $B$, $V$, and $R$ represent the number 
of stars on the blue end of the HB, in the Hertzsprung gap, and on the 
red end, respectively) whereas M4 and M71 have $(B-R)(B+V+R) <$ 0.  In 
addition, Shetrone \& Keane (2000\nocite{SK00}) note that the Na-O and 
Al-O anticorrelations between the slightly more metal-poor clusters 
NGC~288 ($<$[Fe/H]$>$ = --1.39) and NGC~362 ($<$[Fe/H]$>$ = --1.33) 
resemble those of M4 (I99-M4) and M5 (S92-M5), respectively.  

The important conclusion established by this investigation is that 
there is no definitive ``single'' value of [el/Fe] at a given [Fe/H]
for at least some $\alpha$-capture, odd-Z, and $s$-process elements, 
in this case, Si, Al, Ba and La.  It is therefore not clear that one 
can claim some ``exact'' value of (say) [Si/Fe] that characterizes 
globular cluster or halo field stars at a given [Fe/H].  Rather, 
there is a spread that is certainly real and not a result simply of 
observational or analytical error.  Our result is therefore consistent 
with an increasingly large body of evidence (\eg, Sneden 2000) that in 
the halo [el/Fe] ratios are not universal at a given metallicity. 
For example, the outer halo clusters Rup~106 and Pal~12 have very low 
(close to solar) $\alpha$-capture element abundances (Brown \etal\ 
1997\nocite{BWZ97}) as do a few subdwarfs having unusually large absolute 
angular momenta (Carney 1999\nocite{Car99}, Fulbright 2001\nocite{F2001}, 
and references therein). Very high Si abundances ($<$[Si/Fe]$>$ 
$\sim$0.6) are observed in the very metal-poor ($<$[Fe/H]$>$ 
$\sim$--2.4) globular cluster M15 (Sneden \etal\ 1997).  The 
overabundance of Ba in M15 compared with M92, two clusters of very
similar metallicity, is well established (Sneden \etal\ 2000\nocite{SPK00}). 
It has also been known for some time that Ba and other $s$-process species 
are greatly enhanced in metal-rich giants of $\omega$~Cen compared with 
field halo and globular cluster stars of comparable metallicity 
(Vanture \etal\ 1994\nocite{VWB94}; Norris \& Da Costa 1995).   In fact, 
there exists in $\omega$~Cen a stellar sample which possesses nearly 
identical elemental overabundances of [Si/Fe], [Al/Fe], and [Ba/Fe] as seen 
in M4.  However, in $\omega$~Cen, the sample consists of stars possessing a 
range of metallicities (--0.5 $<$ [Fe/H] $<$ --2.0).  Even more striking is 
the Eu deficiency found among $\omega$~Cen giants of intermediate metallicity 
(Smith \etal\ 2000\nocite{SSCGBLS00}).

%Section 10.0
%%%%%%%%%%%%%%%%%%%%%%%%%%%%%%%%%
\section{Summary and Conclusions} 
%%%%%%%%%%%%%%%%%%%%%%%%%%%%%%%%%

We have analysed the chemical abundances of 36 M5 giant stars by two 
different techniques. We employed ``traditional'' spectroscopic analysis 
procedures, setting \teff\ by satisfying the constraint of iron
excitation potential equilibrium, setting $v_t$ by satisfying the 
constraint of iron equivalent width equilibrium, setting \logg\ by 
satisfying the constraint of iron ionization equilibrium, and satisfying 
the additional constraint that the derived [Fe/H] does not vary 
systematically over the range of \teff\ and \logg\ represented by the 
cluster program stars.  Satisfying these constraints led to models whose 
spectroscopic \logg\ values were $\sim$0.5~dex lower than expected, which 
in turn led to $<$[Fe/H]$>$ ratios 0.15~dex lower among our AGB sample 
than among our RGB sample.  These outcomes are consistent with known 
problems that result from applying LTE assumptions to non-LTE atmospheres.  

We investigated a number of alternative approaches to the analysis, seeking 
to resolve the outcomes without resorting to invoking non-LTE effects.  
However, regardless of which \teff-scale or \logg-scale we adopted, the 
[Fe/H] based on \ion{Fe}{1} remained significantly smaller on the AGB as 
compared with the RGB and ``tip'' stars.  Accordingly, we adopted an 
analysis consistent with the non-LTE precepts as discussed by Th\'evenin \& 
Idiart (1999), employing ``new'' models with evolutionary values of \logg\ 
on the same system as those of previous M5 work. These results yielded 
$<$[Fe/H]$>$  = --1.21 ($\sigma$~= 0.06), that were neither dependent on 
\teff\ nor \logg\ and are in good agreement with previously derived values 
for the metallicity of M5 in the literature (\eg, $<$[Fe/H]$>$~= --1.4, 
Zinn \& West 1984\nocite{ZW84}; $<$[Fe/H]$>$~= --1.17, Sneden \etal\ 
1992\nocite{SKPL92}).  Applying the same procedures to the M4 results of 
Ivans \etal\ (1999\nocite{Ivetal99}), we re-determine the metallicity of 
M4 to be $<$[Fe/H]$>$  = --1.08 ($\sigma$ = 0.02). The remaining abundances 
in M4 are offset by an amount equivalent to an increase in \logg\ of 
0.12~dex (see Table~2 of Ivans \etal\ 1999).  Applying this increase in 
\logg\ to the M4 stars, we derive the ratio of total to selective 
absorption $R_V$~= 3.9 (see \S3.4 of Ivans \etal\ 1999).

With the revised method of analysis, we find good agreement between M5 and
M4 (and the field) in most of the Fe-peak and $\alpha$-element abundances.  
The exception is silicon.  The [Si/Fe] abundance in M4 exceeds that of M5 
by $\sim$3-$\sigma$.  Ba and La are similarly overabundant in M4 with 
respect to M5 (and the field), as is aluminum.  However, since Al is 
sensitive to proton-capture nucleosynthesis, the range of aluminum 
abundances in both clusters mask the overall difference in the ``floor'' 
abundances.  Based on these large stellar samples, we confirm and extend 
the previous findings for both of these clusters: Si, Al, Ba, and La are 
enhanced in M4, whereas M5 has ``normal'' abundances.

In M5, we find the classic anti-correlation of O and Na abundances, and
correlated Al and Na abundances.  And, the behavior of these abundances
is further correlated with the CN strength index, $\delta$S(3839): stars
with larger CN indices also have larger Al abundances, larger Na abundances 
and lower O abundances than stars with lower CN indices.  This behavior is 
consistent with that seen in previous studies of other globular clusters
and follows the expected pattern of proton-capture nucleosynthesis (\ie,
low oxygen abundances are usually accompanied by low carbon and enhanced 
nitrogen abundances.  Thus, stronger CN bands, reflecting higher N 
abundances, belong to stars that are more highly CNO-processed).

We binned the M5 RGB and ``tip'' giants into two evolutionary groups by 
\logg, and find that the O and Na abundances are different for the two 
groups:  the stars with lower \logg\ have higher O and lower Na abundances 
on average than the stars with higher values of \logg.  Thus, in M5, the 
dependence of the abundance variations on \logg\ is in the {\em opposite} 
sense to that found in M13 by Kraft \etal\ (1997), where the relationship 
provided strong evidence in support of the evolutionary scenario.  The 
present analysis of M5 giants does not necessarily rule out the 
evolutionary scenario, but it neither provides support for it nor is it 
incompatible with the primordial scenario.  In fact, both may be at work.  
Our observations of the spread in [O/Fe], [Na/Fe], and [Al/Fe] ratios in 
both the CN-strong and CN-weak groups are consistent with the idea that 
an earlier generation of stars may have enriched some of the material out 
of which the current sample formed then, once on the RGB, the stars were 
subject to deep mixing, further altering the abundances.  Thus, deep 
mixing on the RGB would explain the spreads within the CN-strong and
CN-weak groups, and primordial enrichment the difference between the two
groups.

In comparison with clusters that bracket M4 and M5 in metallicity, we
find that the abundance patterns can be divided into two groups: the
O vs Na anti-correlation found in M5 resembles the pattern seen in 
slightly more metal-poor globular clusters M3, M10, and M13 
($<$[Fe/H]$>$ = --1.5 to  --1.6) whereas the anti-correlation found in M4 
resembles that of the more metal-rich disk cluster M71
($<$[Fe/H]$>$ $\sim$ --0.7).  These similarities extend to the HB
morphology of the clusters: according to the catalog compiled by Harris 
(1996\nocite{Har96}), M3, M5, M10, and M13 all have $(B-R)(B+V+R) >$ 0 
(where $B$, $V$, and $R$ represent the number of stars on the blue end of 
the HB, in the Hertzsprung gap, and on the red end, respectively) whereas 
M4 and M71 have $(B-R)(B+V+R) <$ 0.  

We conclude that there is no ``single'' value of [el/Fe] at a given [Fe/H] 
for at least some $\alpha$-capture, odd-Z, and $s$-process elements, in 
this case, Si, Al, Ba, and La.  The spread is real and not a result due to 
observational or analytical error.  Our result is therefore consistent 
with an increasingly large body of evidence (\eg, Sneden 2000) that in the 
halo [el/Fe] ratios are not universal at a given metallicity.  The 
dichotomy between M4 and M5 established here adds more evidence favoring 
the existence of considerable abundance diversity in the Galactic halo.

%%%%%%%%%%%%%%%%%%%%%%%%%%%
%\section{Acknowledgements}
%%%%%%%%%%%%%%%%%%%%%%%%%%%

We are happy to acknowledge that this research was partially funded by 
NSF grants AST-9618351 to R.P.K.~and G.H.S.~and AST-9618364, AST-9987162 
to C.S.~and has made use of NASA's Astrophysics Data System 
Bibliographic Services.  
We are indebted to Eric Sandquist for supplying the exquisite photometry 
of M5 as well as for responding to queries regarding individual stars.  
Andy McWilliam has our gratitude for sending along his most recent 
calculations of V hfs and for sharing his code and expertise.  
We thank Kirk Gilmore for sharing with us his 2$\mu$m image of the cluster.  
We appreciate Earl Luck's help in tracking down some of the laboratory log 
$gf$-values and Jennifer Johnson's \& Jon Fulbright's assistance in 
transporting SPECTRE onto a different platform.  
John Norris, Gary Da Costa, and George Wallerstein have our appreciation 
for both taking the time to read a draft of the paper and subsequently 
offering thoughtful comments and useful suggestions that improved it.  
Ruth Peterson also has our thanks for helpful discussions regarding 
Kurucz atmospheres. 
The anonymous referee's detailed comments and thoughtful suggestions helped 
improve the paper and are much appreciated.  
I.I.I.~gratefully acknowledges the financial support of a Continuing 
Fellowship from the University of Texas at Austin and the Audrey Jorss 
Commemorative Fellowship from the Australian Federation of University Women 
Queensland Branch during the time that this work was performed and 
thanks sincerely the members of the Research School of Astronomy and 
Astrophysics of the Australian National University at Mount Stromlo 
Observatory for their hospitality during the preparation of this paper.

%%%%%%%%%%%%%%%%%%%%%%%%%%%%%%%%%%%%%%

%%%%%%%%%%%%%%%%%%%
\section*{Appendix}
%%%%%%%%%%%%%%%%%%%

Table~\ref{m5.tab10} shows the atomic parameters of the lines used in 
this study, adopting the linelist employed in Ivans \etal\ (1999; 
``I99-M4'').  Most of the lines used in the I99-M4 study are the same as 
those used in earlier papers of the Lick-Texas group.  In I99-M4, however, 
the metallicity of M4 metallicity forced a culling of many blended lines 
which left a list of only fairly strong lines (see I99-M4 for details).  
Ivans \etal\ then added to the list clean Fe lines of low to medium 
strength as well as additional other elemental lines for which laboratory 
values of the atomic parameters were available from other M4 studies.  
Also added to the I99-M4 linelist were three La lines for which 
astrophysical $gf$-values had been derived by the inverse solar method by 
Brown \& Wallerstein (1992\nocite{BW92}).  However, one of these La 
lines ($\lambda$6390~\AA) now has an up-to-date laboratory $gf$-value 
determination (Lawler \etal\ 2001\nocite{LBS01}) which we employed here 
as well as in the updated I99-M4 results.  By maintaining the same 
linelists in both studies, we are able to compare the results of M4 and 
M5 directly. 

Table~\ref{m5.tab11} illustrates the abundances derived from a 
traditional spectroscopic analysis as outlined in \S3.1.  These should be 
compared with mean values from Table~\ref{m5.tab5}.  For specific 
species, we examine the ``split'' between RGB and AGB $<$[el/Fe]$>$ 
ratios, based on the ``new'' vs ``traditional'' analysis. 
\begin{enumerate}
\item Oxygen: $<$[O/Fe]$>$ shows the same split of 0.06~dex between the 
RGB and AGB in both the ``new'' and ``traditional'' analyses but the 
scatter is large because of the Na versus O anti-correlation (see \S5 and 
S92-M5).
\item Sodium and aluminum: there is a large scatter but, as with O, shows
no significant difference in mean values between RGB and AGB.  The 
scatter probably reflects true star-to-star abundance differences.
\item Silicon: $<$[Si/Fe]$>$ is increased by 0.01~dex over the 
``traditional'' value.  However, the split between the RGB and AGB 
stays the same.
\item Calcium: the split in $<$[Ca/Fe]$>$ between the RGB and AGB 
remains the same; there is only a slight change in $<$[Ca/Fe]$>$ overall.
\item Titanium: the difference in $<$[Ti/Fe]$>$ between the AGB and RGB 
is somewhat reduced in the ``new'' analysis.
\item Vanadium: in the ``traditional'' analysis $<$[V/Fe]$>$ = --0.12,
taken over all stars, but $<$[V/Fe]$>$ is more negative by 0.17~dex on the 
AGB compared with its value on the RGB.  The ``new'' analysis produces a 
smaller difference of 0.09~dex, and a slightly more positive overall mean 
of --0.10.
\item Manganese:  The ``new'' analysis reduces the split between the 
branches by 0.03~dex while the overall mean was reduced by 0.02~dex.
\item Nickel: The ``new'' and ``traditional'' values of $<$[Ni/Fe]$>$ are 
both --0.05, and the two branches are in very close agreement in both cases.
\item Heavy elements: The difference in $<$[el/Fe]$>$ between the two 
branches in the ``new'' analysis is reduced for Ba (by 0.05~dex), La
(by 0.04~dex) and Eu (by 0.04~dex), as compared to the ``traditional''
analysis.

\end{enumerate}

%%%%%%%%%%%%%%%%%
% LIST OF TABLES
%%%%%%%%%%%%%%%%%

\begin{dummytable}
\label{m5.tab1} % Table 1
\end{dummytable}

\begin{dummytable}
\label{m5.tab2} % Table 2
\end{dummytable}

\begin{dummytable}
\label{m5.tab3} % Table 3 
\end{dummytable}

\begin{dummytable}
\label{m5.tab4} % Table 4
\end{dummytable}

\begin{dummytable}
\label{m5.tab5} % Table 5
\end{dummytable}

\begin{dummytable}
\label{m5.tab6} % Table 6
\end{dummytable}

\begin{dummytable}
\label{m5.tab7} % Table 7
\end{dummytable}

\begin{dummytable}
\label{m5.tab8} % Table 8
\end{dummytable}

\begin{dummytable}
\label{m5.tab9} % Table 9
\end{dummytable}

\begin{dummytable}
\label{m5.tab10} % Table 10
\end{dummytable}

\begin{dummytable}
\label{m5.tab11} % Table 11
\end{dummytable}

\clearpage

%%%%%%%%%%%
% CAPTIONS
%%%%%%%%%%%

\clearpage

\clearpage

\clearpage

\clearpage

\clearpage

\clearpage

\clearpage

\clearpage

\clearpage

\clearpage

\clearpage

\clearpage

\clearpage

\clearpage

\begin{figure} %TABLE 1
\epsscale{1.0}
\plotone{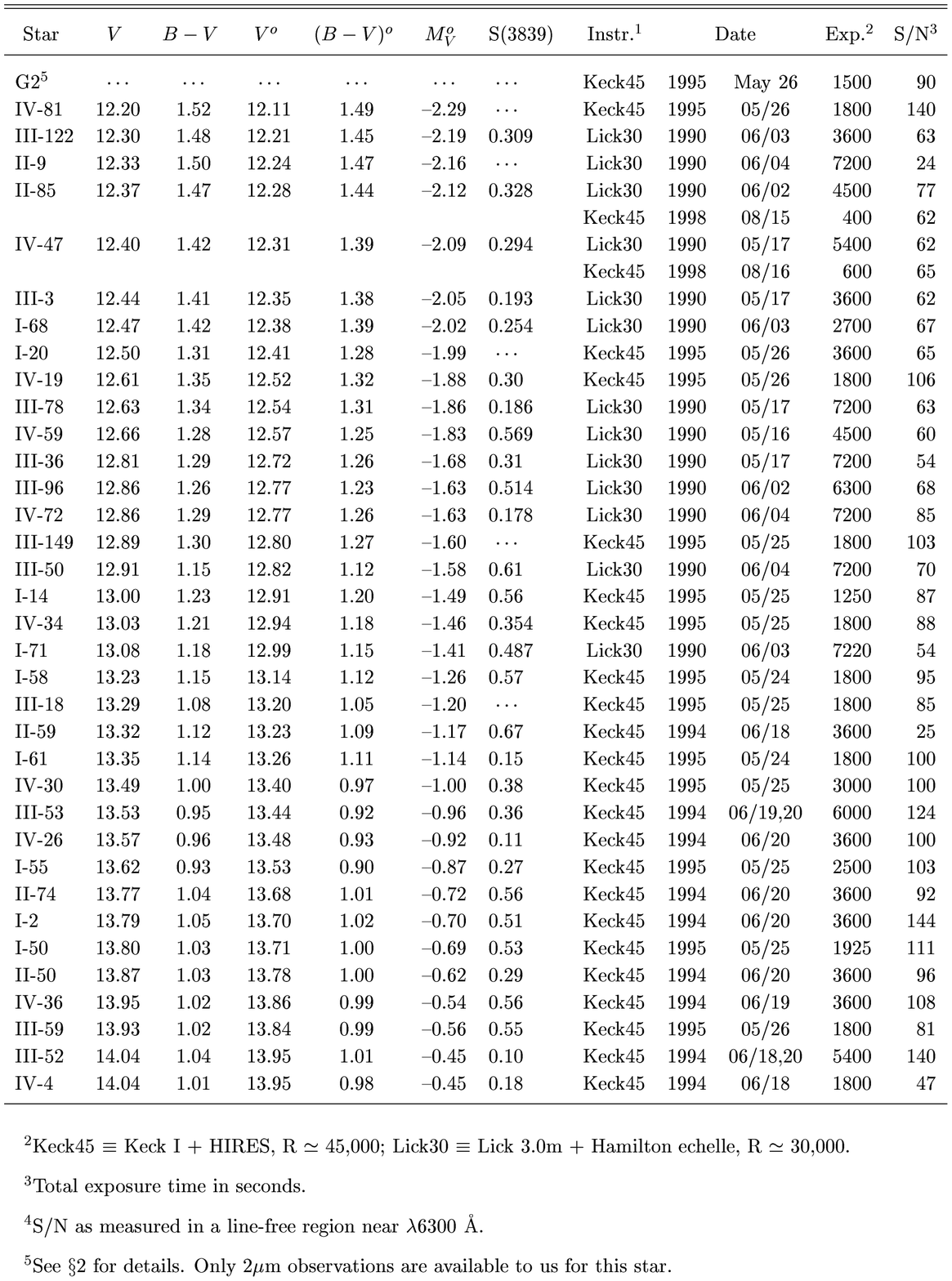}
\end{figure}

\clearpage

\begin{figure} %TABLE 2
\epsscale{1.0}
\plotone{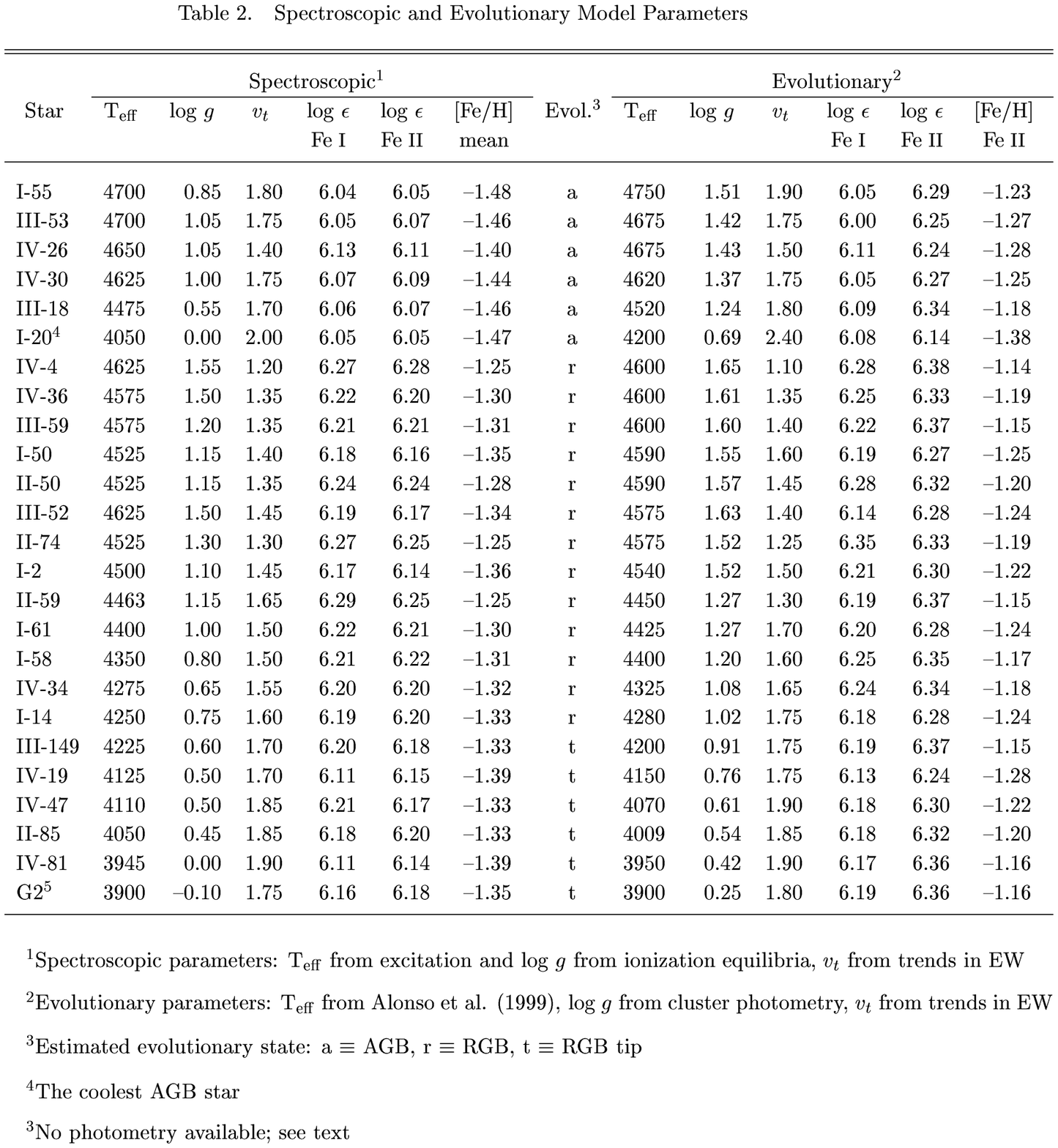}
\end{figure}

\clearpage

\begin{figure} %TABLE 3
\epsscale{1.0}
\plotone{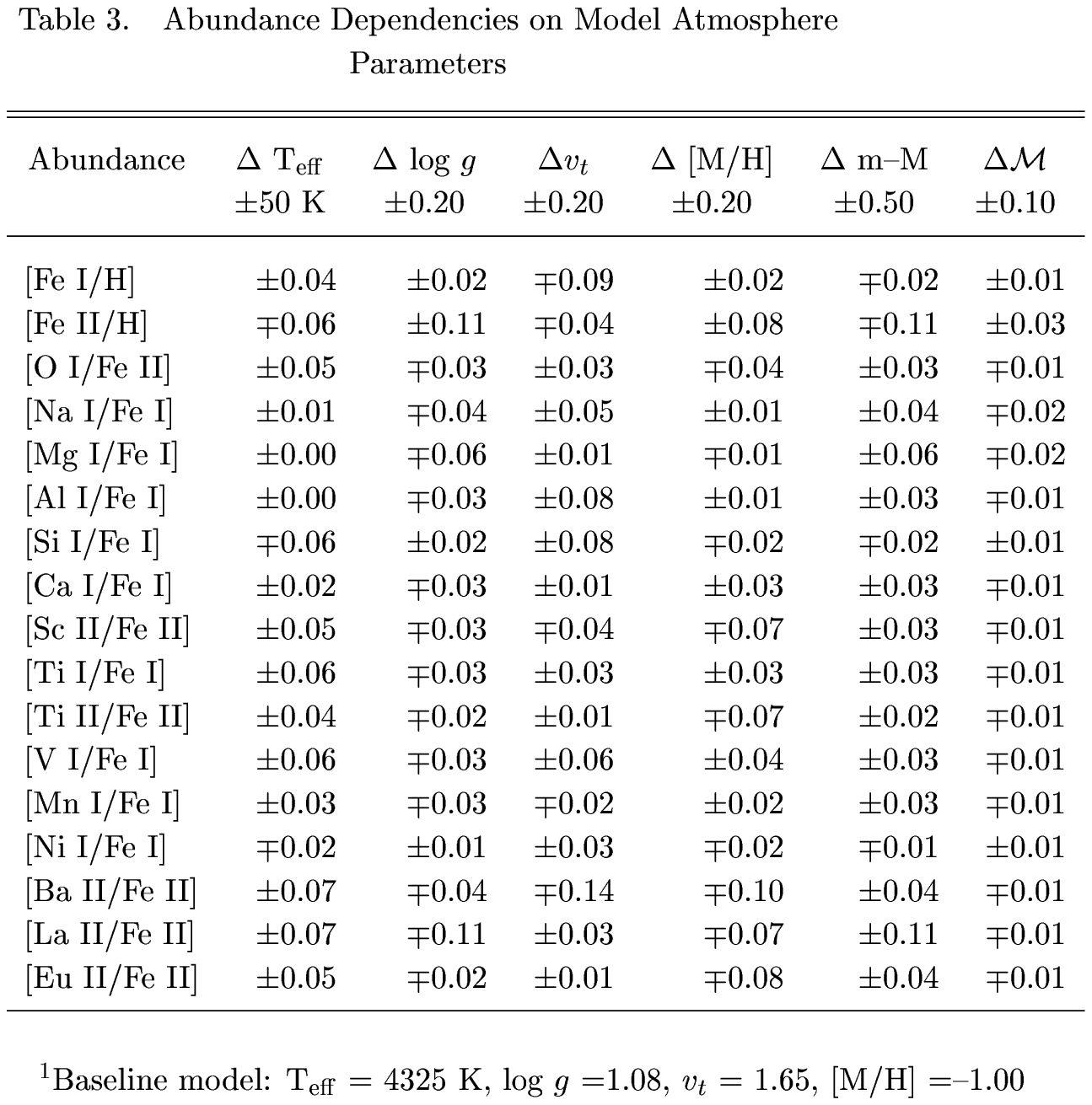}
\end{figure}

\clearpage

\begin{figure} %TABLE 4
\epsscale{1.0}
\plotone{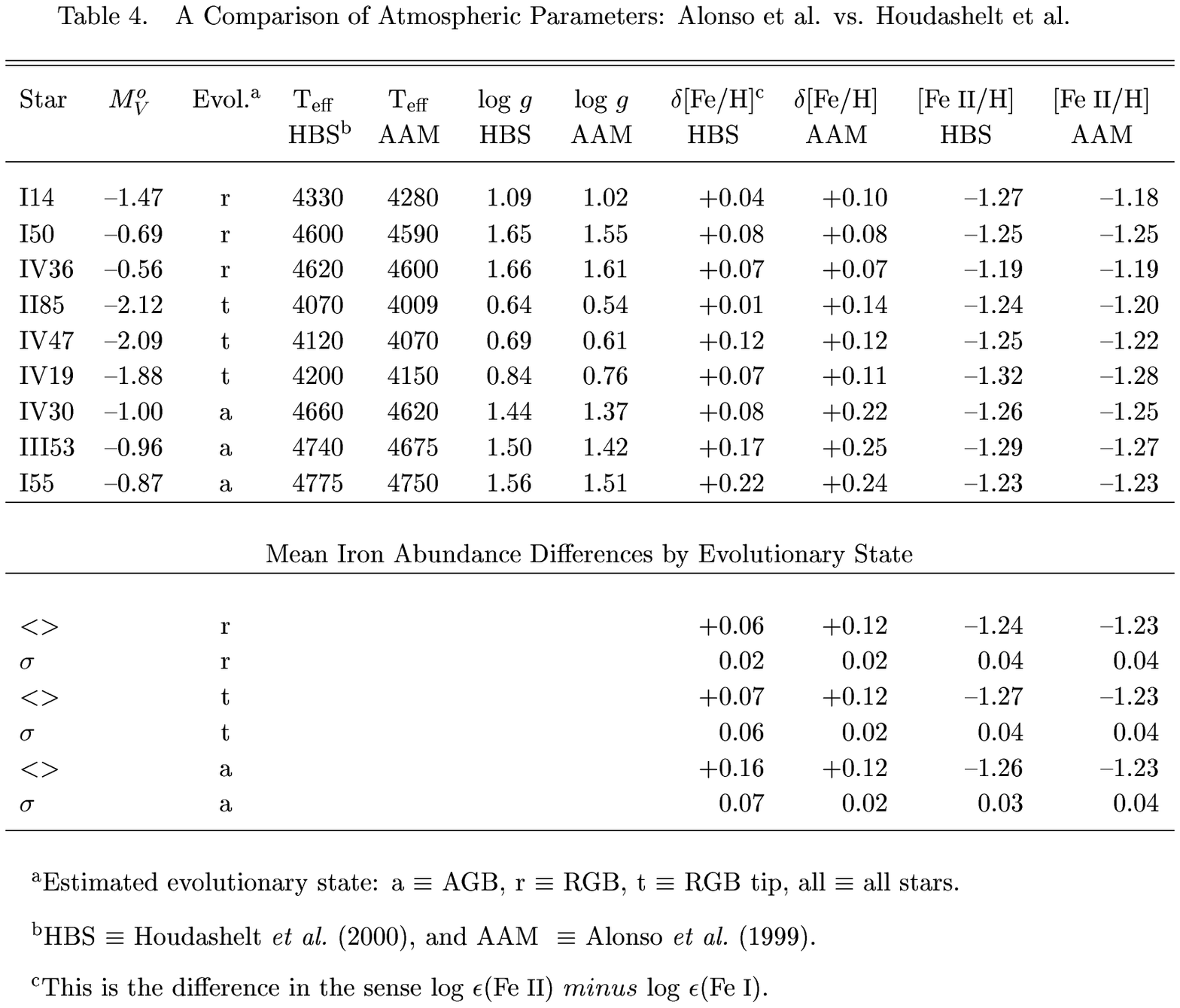}
\end{figure}

\clearpage

\begin{figure} %TABLE 5
\epsscale{1.0}
\plotone{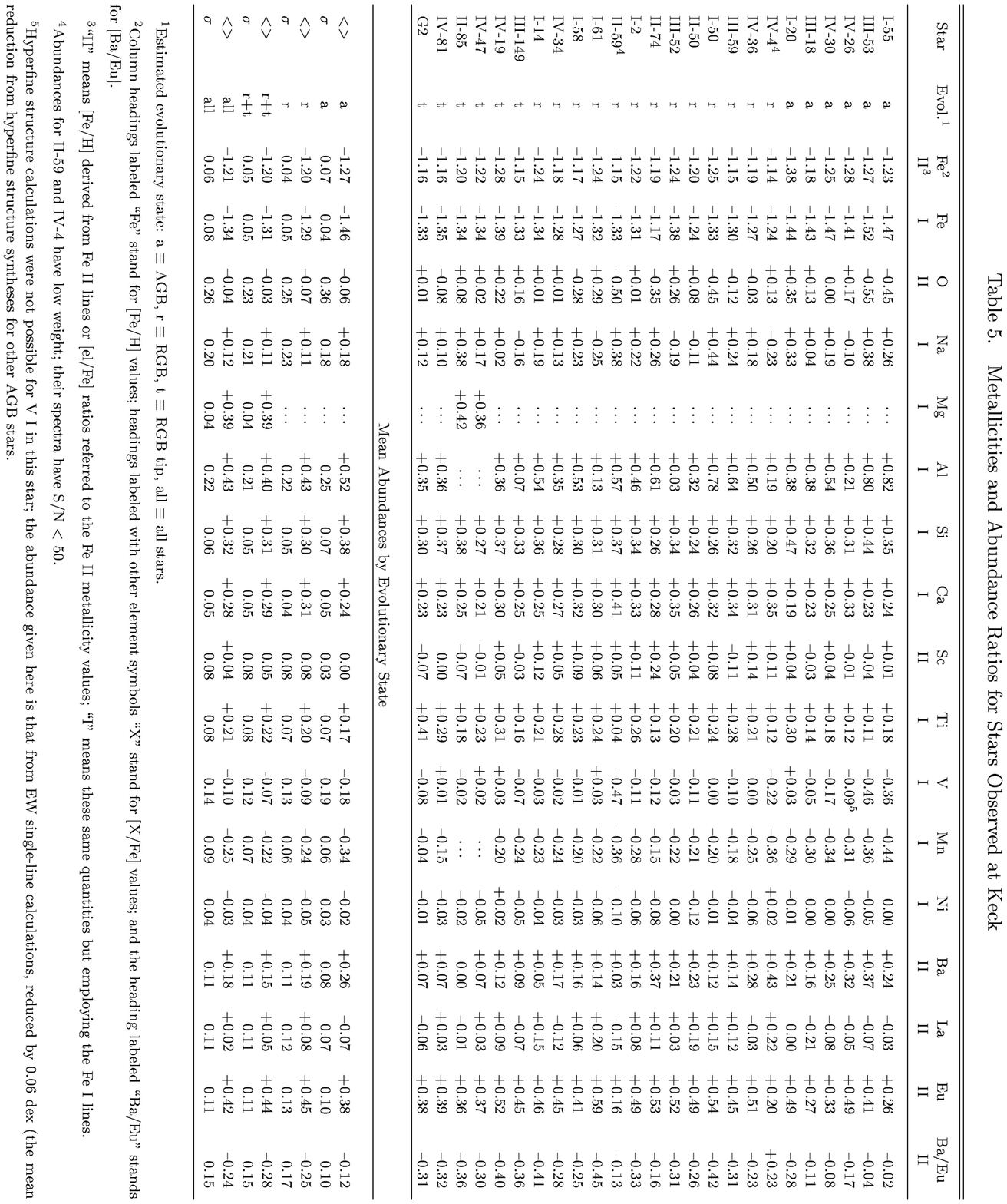}
\end{figure}

\clearpage

\begin{figure} %TABLE 6
\epsscale{1.0}
\plotone{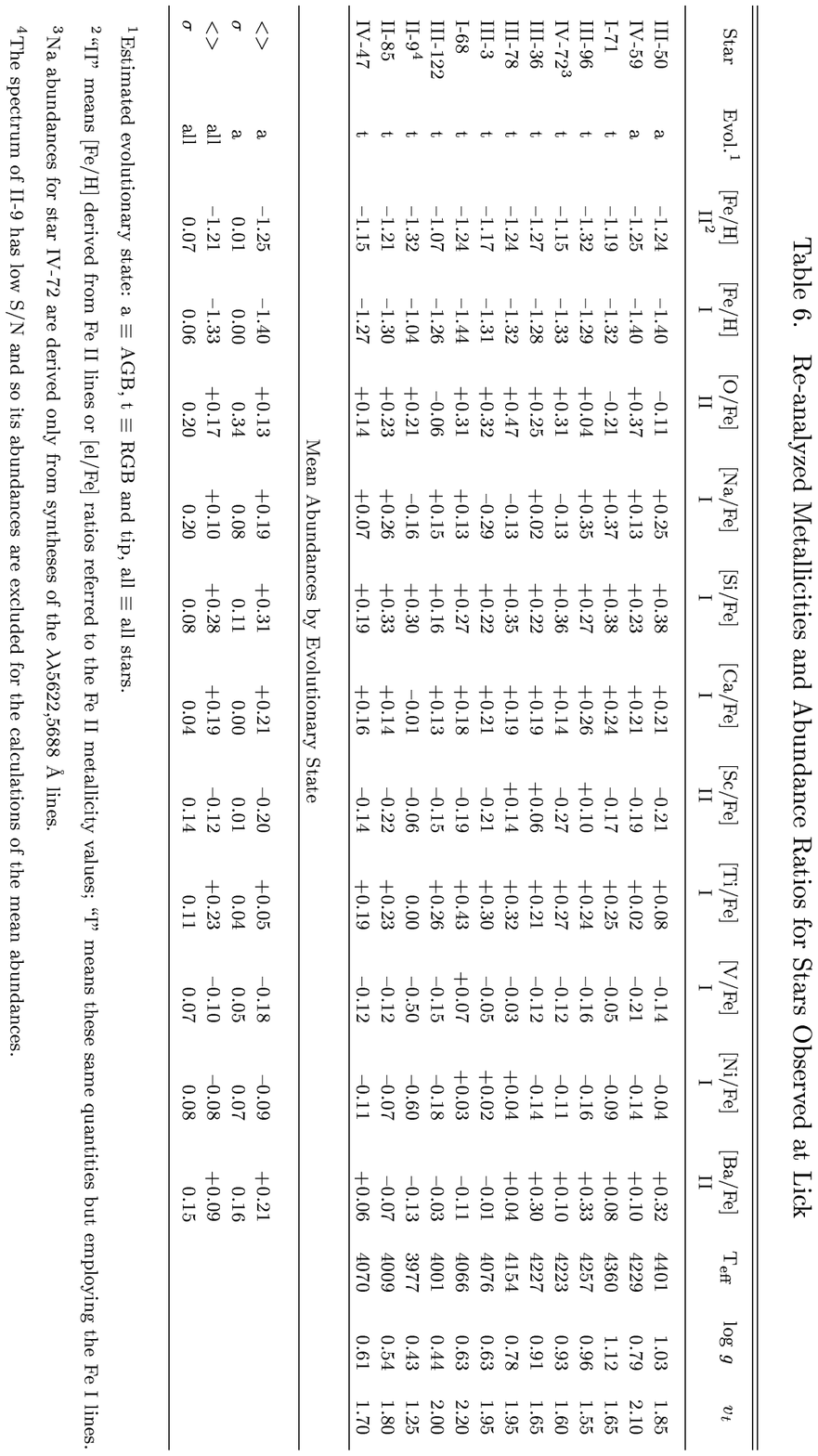}
\end{figure}

\clearpage

\begin{figure} %TABLE 7
\epsscale{1.0}
\plotone{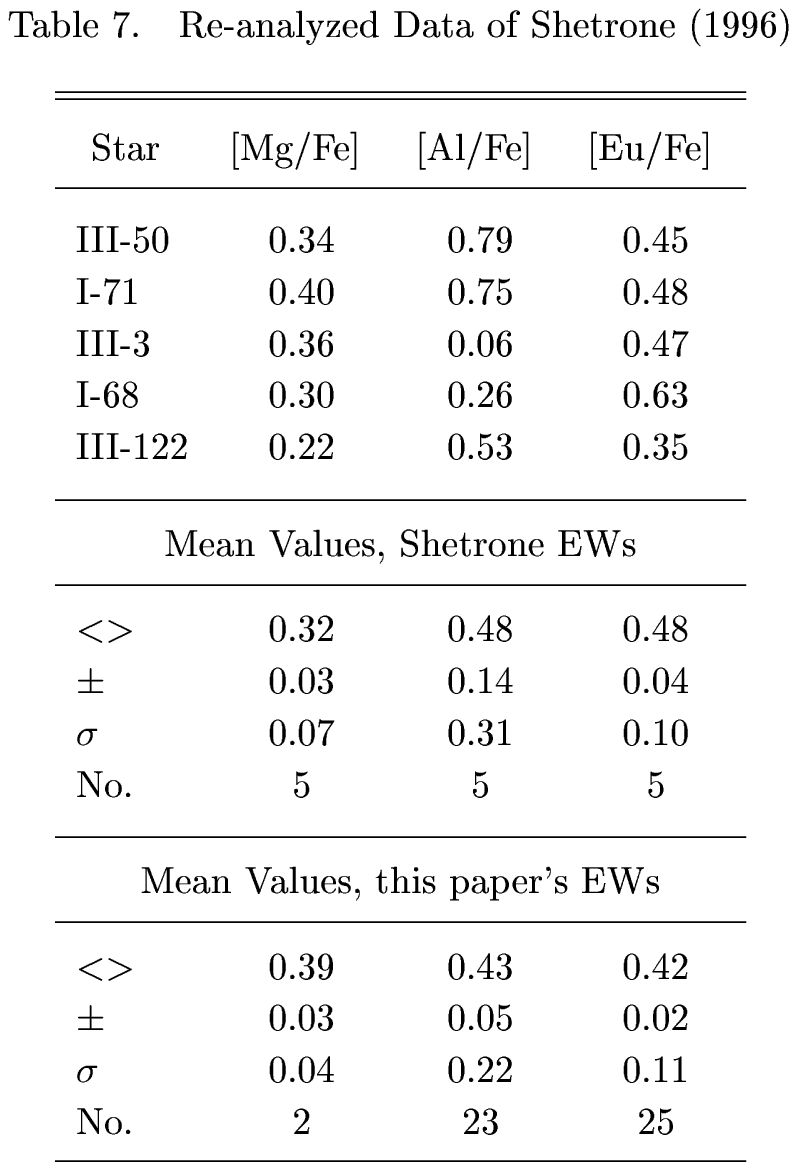}
\end{figure}

\clearpage

\begin{figure} %TABLE 8
\epsscale{1.0}
\plotone{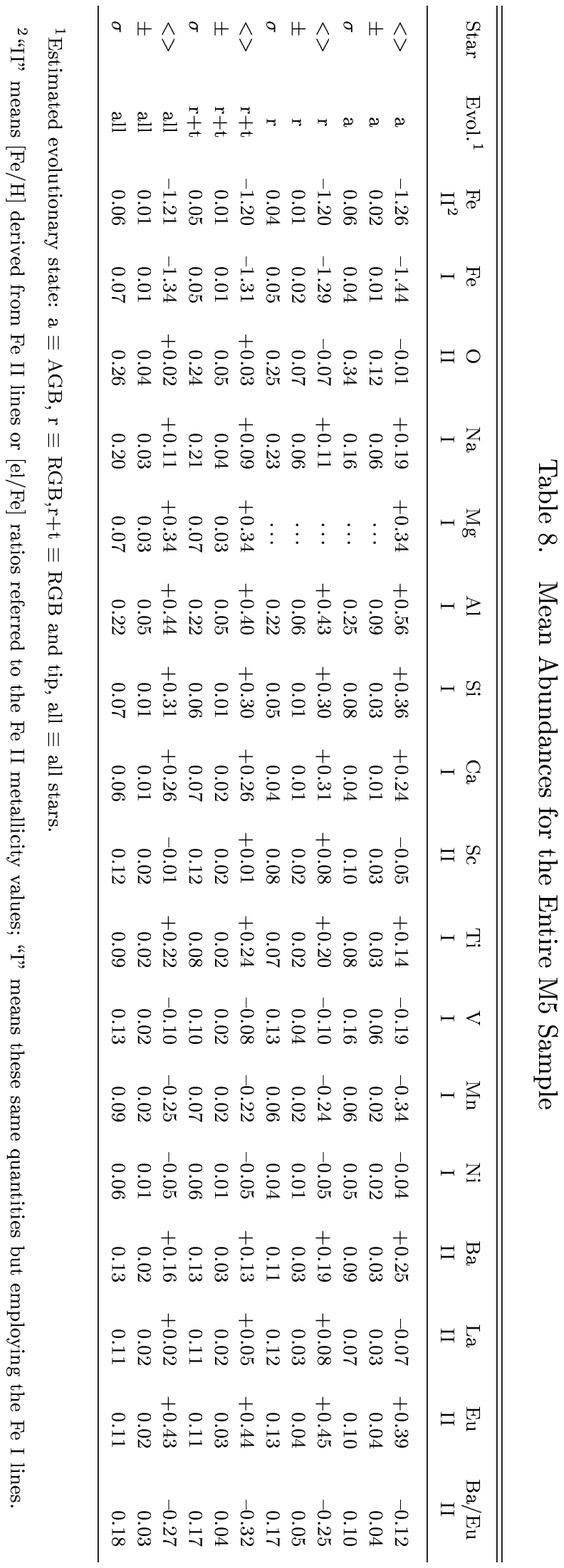}
\end{figure}

\clearpage

\begin{figure} %TABLE 9
\epsscale{1.0}
\plotone{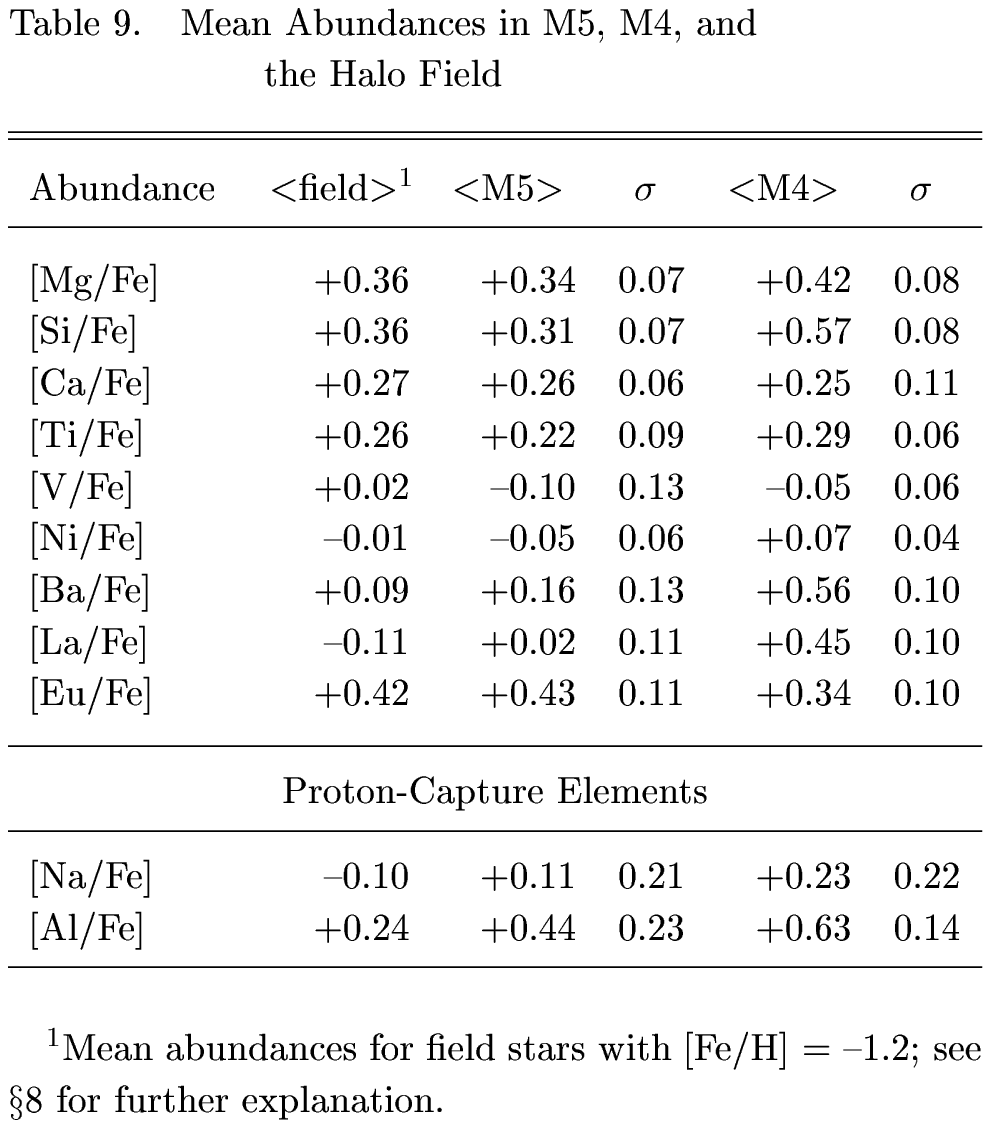}
\end{figure}

\clearpage

\begin{figure} %TABLE 10
\epsscale{1.0}
\plotone{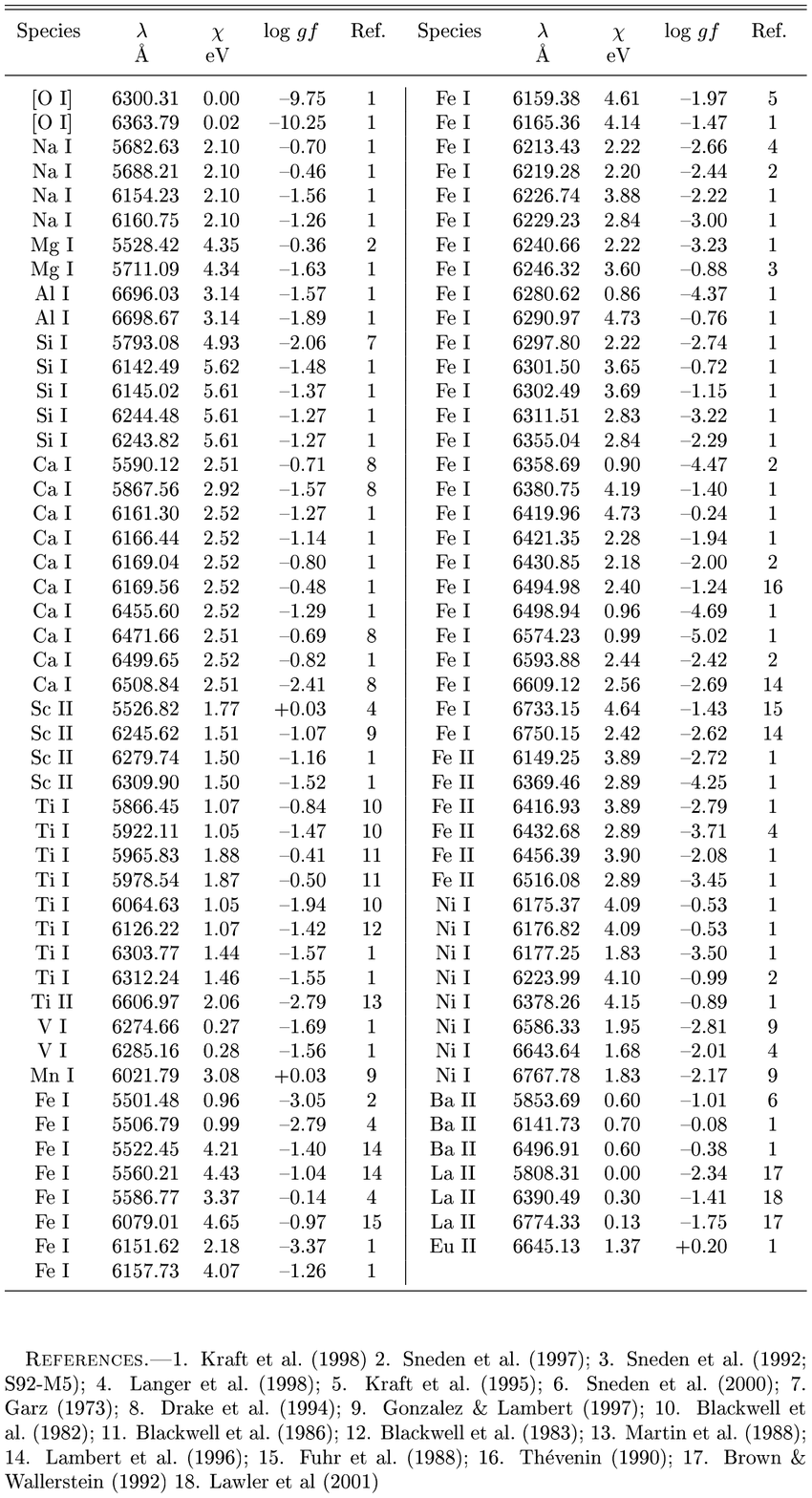}
\end{figure}

\clearpage

\begin{figure} %TABLE 11
\epsscale{1.0}
\plotone{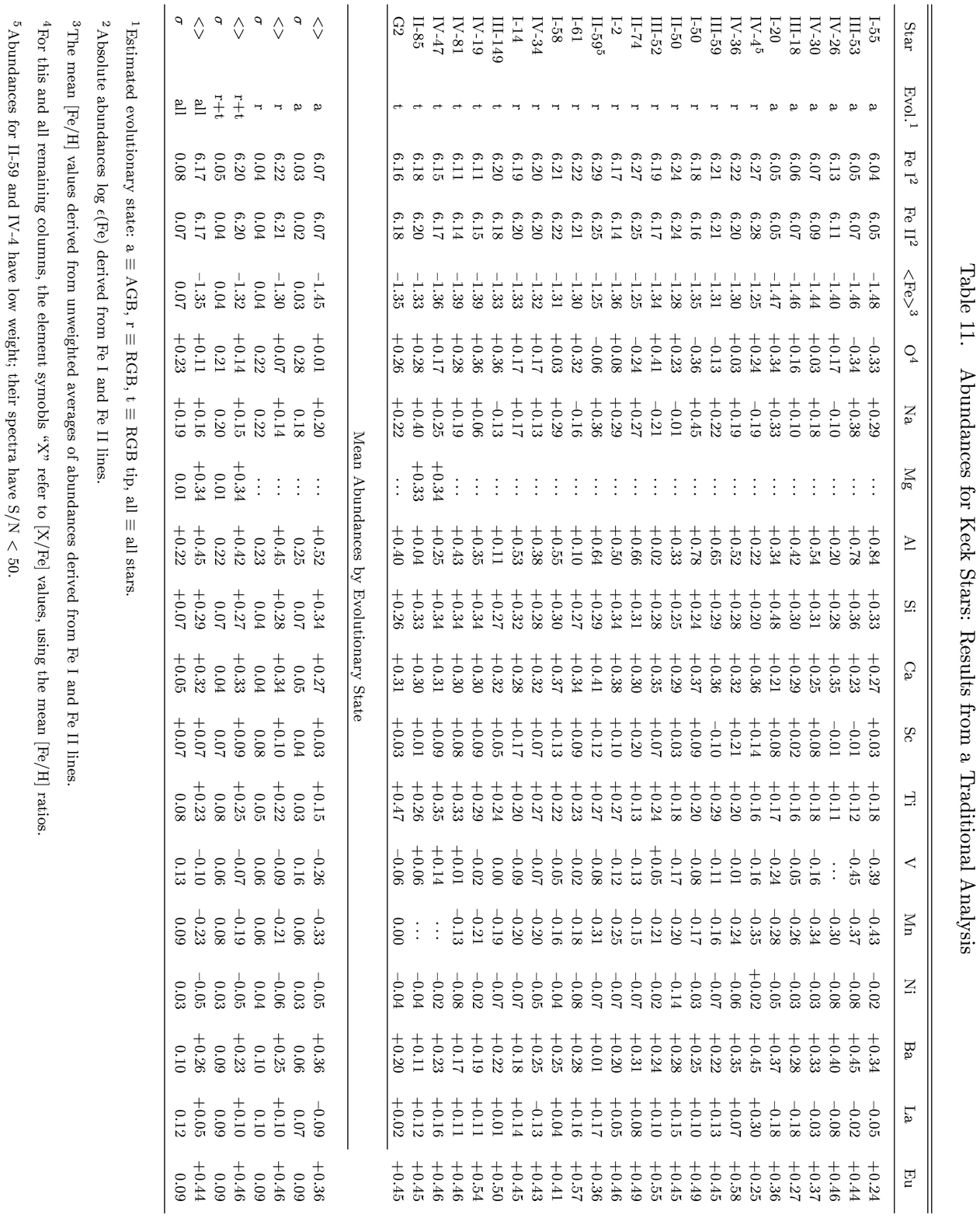}
\end{figure}


\begin{thebibliography}
%%%%%%%%%%%%%%%%%%%%%%%%%%%%%%%%%%%%%%

\bibitem[Allende Prieto \etal\ (1999)]{AGLG99}
Allende~Prieto, C., Garc\'ia~L\'opez, R., Lambert, D.~L.~\& Gustafsson,
B.~1999, \apj, 527, 879

\bibitem[Allende Prieto \& Lambert (2000)]{AL00}
Allende~Prieto, C. \& Lambert, D.~L.~L.~2000, \aj, 119, 2445

\bibitem[Alonso \etal\ (1996)]{AAM96}
Alonso, A., Arribas, S., Mart\'inez-Roger, C.~1999, \aaps, 117, 227

\bibitem[Alonso \etal\ (1999)]{AAM99}
Alonso, A., Arribas, S., Mart\'inez-Roger, C.~1999, \aaps, 140, 261

\bibitem[Bates \etal\ (1993)]{BKM93}
Bates, B., Kemp, S.~N.~\& Montgomery, A.~S.~1993, \aaps, 97, 937

\bibitem[Bell \& Gustafsson (1978)]{BG78}
Bell, R.~A.~\& Gustafsson, B.~1978, \aaps, 34, 229

\bibitem[Blackwell \etal\ (1990)]{BPAHS90}
Blackwell, D.~E., Petford, A.~D., Arribas, S., Haddock, D.~J.~\& Selby,
M.~J.~1990, \aap, 232, 396

\bibitem[Briley \& Smith (1993)]{BS93}
Briley, M.~M.~\& Smith, G.S.~1993, \pasp, 105, 1260

\bibitem[Briley \etal\ (1994)]{BBHS94}    
Briley, M.~M., Bell, R.~A., Hesser, J.~E.~\& Smith, G.~H.~1994, 
Can.~J.~Phys., 72, 772

\bibitem[Briley \etal\ (1995)]{Betal95}
Briley, M., Smith, V.~V., Suntzeff, N.~B., Lambert, D.~L., Bell, R.~A.~\&
Hesser, J.~E.~1995, Nature, 383, 604

\bibitem[Brown \& Wallerstein (1992)]{BW92}
Brown, J.~A.~\& Wallerstein 1992, \aj, 104, 1818

\bibitem[Brown \etal\ (1997)]{BWZ97}
Brown, J.~A., Wallerstein, G.~\& Zucker, D.~1997, \aj, 114, 180

\bibitem[Buonanno \etal\ (1981)]{BCF81}
Buonanno, R., Corsi, C.~E., Fusi Pecci, F.~1981, \mnras, 196, 435

\bibitem[Carney (1999)]{Car99}
Carney, B.~W.~1999, in The Third Stromlo Symposium: The Galactic Halo,
ed.~B.~K.~Gibson, T.~S.~Axelrod, \& M.~E.~Putnam, ASP Conf.~Ser.~165, 230

\bibitem[Carretta \& Gratton (1997)]{CG97}
Carretta, E.~\& Gratton, R.~G.~1997, \aaps, 121, 95

\bibitem[Cavallo \& Nagar (2000)]{CN00}
Cavallo, R.M.~\& Nagar, N.~M.~2000, \aj, 120, 1364

\bibitem[Cudworth  (1979)]{Cu79}
Cudworth, K.~M.~1979, \aj, 84, 1866

\bibitem[Cudworth \& Hanson (1993)]{CH93}
Cudworth, K.~M.~\& Hanson, R.~B.~1993, \aj, 105, 168

\bibitem[Da~Costa (1997)]{DaC97}
Da~Costa, G.~S.~1997, in Fundamental Stellar Properties: The Interactions 
Between Observations and Theory, ed.~T.~R.~Bedding, A.~J.~Booth \& 
J.~Davis, IAU Symp.~189, 193

\bibitem[Denissenkov \etal\ (1998)]{DDNW98}
Denissenkov, P.~A., Da~Costa, G.~S., Norris, J.~E.~\& Weiss, A.~1998,
\aap, 333, 926

\bibitem[Djorgovski (1993)]{Dj93}
Djorgovski, S.~1993, in Structure and Dynamics of Globular Clusters,
ed.~S.~G.~Djorgovski \& G.~Meylan, ASP Conf.~Ser.~50, 373

\bibitem[Dumont \etal\ (1975)]{DHJP75}
Dumont, S., Heidman, N., Jeffries, J.~T.~\& Pecker, J.-C.~1975, \aap, 40, 127

\bibitem[ESA (1997)]{ESA97}
ESA, 1997, The Hipparcos and Tycho Catalogues (ESA SP-1200; Noordjwik: ESA)

\bibitem[Fitzpatrick \& Sneden (1987)]{FS87}      
Fitzpatrick, M.~J.~\& Sneden, C.~1987, \baas, 19, 1129

\bibitem[Frogel \etal\ (1983)]{FPC83}
Frogel, J.~A., Persson, S.~E., \& Cohen, J.~G.~1983, \apjs, 53, 713

\bibitem[Fuhr \etal\ (1988)]{Fuh88}
Fuhr, J.~R., Martin, G.~A.~\& Weise, W.~L.~1988, J.~Phys.~Chem.~Ref.~Data, 
17, Suppl.~No.~4, 1

\bibitem[Fuhrmann \etal\ (1997)]{FPFRG97}
Fuhrmann, K., Pfeiffer, M., Frank, C., Reetz, J.~\& Gehren, T.~1997, \aap,
323, 909

\bibitem[Fulbright (2000)]{F2000}
Fulbright, J.~2000, \aj, 120, 1841

\bibitem[Fulbright (2001)]{F2001}
Fulbright, J.~2001, \aj, Submitted

\bibitem[Fulbright \& Kraft (1999)]{FK99}
Fulbright, J.~P.~\& Kraft, R.~P.~1999, \aj, 118, 527

\bibitem[Gratton \& Sneden (1991)]{GS91}
Gratton, R.~G.~\& Sneden, C.~1991, \aap, 241, 501

\bibitem[Gratton \& Sneden (1994)]{GS94}
Gratton, R.~G.~\& Sneden, C.~1994, \aap, 287, 927

\bibitem[Gratton \etal\ (1997)]{Git97}
Gratton, R.~G., Fusi Pecci, F., Carretta, E., Clementini, G., Corsi, C.~E., 
\& Lattanzi, M.~1997, \apj, 491, 749.

\bibitem[Gratton \etal\ (2000)]{GSCB00}
Gratton, R.~G., Sneden, C., Carretta, E.~\& Bragaglia, A.~2000,
\aap, 354, 169

\bibitem[Gratton \etal\ (2001)]{Git01}
Gratton, R.~G., Bonifacio, P., Bragaglia, A., Carretta, E., Castellani, 
V., Centurion, M., Chieffi, A., Claudi, R., Clementini, G., D'Antona, F.,
Desidera, S., Francois, P., Grundahl, F., Lucatello, S., Molaro, P.,
Pasquini, L., Sneden, C., Spite, F., \& Straniero, O.~ 2001, \aap, 369, 87

\bibitem[Gustafsson \etal\ (1975)]{GBEN75}    
Gustafsson, B., Bell, R.~A., Ericksson, K.~\& Nordlund, A.~1975, \aap, 42, 407

\bibitem[Hanson \etal\ (1998)]{HSKF98}    
Hanson, R.~B., Sneden, C., Kraft, R.~P.~\& Fulbright, J.~1998, \aj, 116, 1286

\bibitem[Harris (1996)]{Har96}
Harris, W.~E. 1996, \aj, 112, 1487

\bibitem[Hearnshaw (1976)]{Hea76}
Hearnshaw, J.~B.~1976, \aap, 51, 71

\bibitem[Houdashelt \etal\ (2000)]{HBS00}
Houdashelt, M.~L., Bell, R.~A. \& Sweigart, A.~V.~2000, \aj, 119, 1448

\bibitem[Ivans \etal\ (1999)]{Ivetal99}
Ivans, I.~I., Sneden, C., Kraft, R.~P., Suntzeff, N.~B., Smith, V.~V., 
Langer, G.~E.~\& Fulbright, J.~P.~1999, \aj, 118, 1273 (I99-M4)

\bibitem[Johnson (1999)]{Jo99}
Johnson, J.~1999, Ph.D.~thesis, Univ.~California, Santa Cruz

\bibitem[Jones \etal\ (1988)]{JCL88}
Jones, R.~V., Carney, B.~W.~\& Latham, D.~W.~1988, \apj, 332, 206

%\bibitem[Kelleher \etal\ (1999)]{Keta99}
%Kelleher, D.~E., Martin, W.~C., Wiese, W.~L., Sugar, J., Fuhr, J.~R., Olsen, 
%K., Musgrove, A., Mohr, P.~J., Reader, J.~\& Dalton, G.~R.~1999, in
%Ultraviolet Atmospheric and Space Remote Sensing: Methods and 
%Instrumentation II, ed.~G.~R.~Carruthers \& K.~F.~Dymond, SPIE 3818, 170
%National Institute of Standards and Technology, 

\bibitem[Kovtykh \& Andrievsky (1999)]{KA99}
Kovtykh, V.~V.~\& Andrievsky, S.~M.~1999, \aap, 351, 597

\bibitem[Kraft (1994)]{Kr94}      
Kraft, R~ P.~1994, \pasp, 106, 553

\bibitem[Kraft (1999)]{Kr99}
Kraft, R.~P.~1999, \apss, 265, 153

\bibitem[Kraft (2001)]{Kr01}
Kraft, R.~P.~2001, in Highlights of Astronomy, ed.~H.~Rickman (A.S.P., Provo,
Utah), Vol.~12, in press

\bibitem[Kraft \etal\ (1992)]{KSLP92}    
Kraft, R.~P., Sneden, C., Langer, G.~E.~\& Prosser, C.~F.~1992, \aj, 104, 645

\bibitem[Kraft \etal\ (1995)]{KSLSB95}   
Kraft, R.~P., Sneden, C., Langer, G.~E., Shetrone, M.~D., 
\& Bolte, M.~1995, \aj, 109, 2586

\bibitem[Kraft \etal\ (1997)]{KSSSLP97}  
Kraft, R.~P., Sneden, C., Smith, G.~H., Shetrone, M.~D., Langer, G.~E., 
\& Pilachowski, C.~A.~1997, \aj, 113, 279

\bibitem[Lambert \etal\ (1996)]{LHLD96}
Lambert, D.~L., Heath, J.~E., Lemke, M.~\& Drake, J.~1996, \apjs, 103, 183

\bibitem[Lawler \etal\ (2001)]{LBS01}
Lawler, J.~E., Bonvallet, G. \& Sneden, C.~2001, \apj, Submitted

\bibitem[Layden \etal\ (1996)]{LHHKH96}
Layden, A.~C., Hanson, R.~B., Hawley, S.~L., Klemola, A.~R.~\& Hanley, C.~J.
1996, \aj, 112, 2110

\bibitem[Luck \& Lambert (1985)]{LL85}
Luck, R.~E.~\& Lambert D.~L.~1985, \apj, 298, 782

\bibitem[Luck \etal\ (1998)]{LMBG98}
Luck, R.~E., Moffett, T.~J., Barnes, T.~G., III \& Gieren, W.~P.~
1998, \aj, 115, 605

\bibitem[Martin \etal\ (1999)]{Meta99}
Martin, W.~C., Fuhr, J.~R., Kelleher, D.~E., Musgrove, A., Sugar,
J., Wiese, W.~L., Mohr, P.~J.~\& Olsen, K.~(1999) NIST
Atomic Spectra Database (version 2.0). Available: 
http://physics.nist.gov/asd [1999 March 22]. National Institute
of Standards and Technology, Gaithersburg, MD.

%\bibitem[Martin \etal\ (2001)]{Met01}
%Martin, W.~C., Fuhr, J., Kelleher, D., Mohr, P., Musgrove, A., Olsen, 
%K., Podobedova, L., Reader, J., Saloman, E., Sansonetti, C., 
%Sansonetti, J.~\& Wiese, W.~2001, in Atomic and Molecular Data for 
%Astrophysics: New Developments, Case Studies and Future Needs, IAU
%Joint Discussion 1, in press

\bibitem[McWilliam (1997)]{Mc97}    
McWilliam, A., 1997, \araa, 35, 503

\bibitem[McWilliam (1998)]{Mc98}    
McWilliam, A., 1998, \aj, 115, 1640

\bibitem[McWilliam \& Rich (1994)]{MR94}
McWilliam, A.~\& Rich, R.~M.~1994, \apjs, 91, 749

\bibitem[McWilliam \etal\ (1995)]{MPSS95}    
McWilliam, A., Preston, G.~W., Sneden, C.~\& Searle, L.~1995, \aj, 109, 2757

\bibitem[Norris \& Da Costa (1995)]{ND95}
Norris, J.~E.~\& Da Costa, G.~S.~1995, \apj, 441, 81

\bibitem[Pilachowski \etal\ (1996a)]{PSK96}     
Pilachowski, C.~A., Sneden, C.~\& Kraft, R.~P., 1996, \aj, 111, 1689 (1996a)

\bibitem[Pilachowski \etal\ (1996b)]{PSKL96}     
Pilachowski, C.~A., Sneden, C.~Kraft, R.~P., Langer, G.~E.~1996, \aj, 
112, 545 (1996b)

\bibitem[Rees (1993)]{Re93}
Rees, R.~F., Jr.~1993, \aj, 106, 1524

\bibitem[Reid (1997)]{Re97}
Reid, I.~N.~1997, \aj, 114, 161

\bibitem[Ryan \etal\ (1996)]{RNB96}
Ryan, S.~G., Norris, J.~E.~\& Beers, T.~C.~1996, \apj, 471, 254

\bibitem[Sandage \& Cacciari (1990)]{SC90}
Sandage, A.~\& Cacciari, C.~1990, \apj, 350, 645

\bibitem[Sandquist \etal\ (1996)]{SBSH96}
Sandquist, E.~L., Bolte, M., Stetson, P.~B., Hesser, J.~E.~1996, \apj, 470, 910

\bibitem[Schnabel \etal\ (1999)]{SKH99}
Schnabel, R., Kock, M.~\& Holweger, H.~1999, \aap, 342, 610

\bibitem[Sekiguchi \& Fukugita (2000)]{SF00}
Sekiguchi, M.~\& Fukugita, M.~2000, \aj, 120, 1072

\bibitem[Shetrone (1996)]{Sh96}     
Shetrone, M.~D.~1996, \aj, 112, 2639

\bibitem[Shetrone \& Keane (2000)]{SK00}
Shetrone, M.~D.~\& Keane, M.~J.~2000, \aj, 119, 840

\bibitem[Smith \& Norris (1983)]{SN83}
Smith, G.~H.~\& Norris, J.~E.~1983, \apj, 264, 215

\bibitem[Smith \& Norris (1993)]{SN93}
Smith, G.~H.~\& Norris, J.~E.~1993, \aj, 105, 173

\bibitem[Smith \etal\ (1997)]{SSBCB97}   
Smith, G.~H., Shetrone, M.~D., Briley, M.~M., Churchill, C.~W.~\& 
Bell, R.~A.~1997, \pasp, 109, 236

\bibitem[Smith \etal\ (2000)]{SSCGBLS00}
Smith, V.~V., Suntzeff, N.~B., Cunha, K., Gallino, R., Busso, M., Lambert, 
D.~L.~ \& Straniero, O.~2000, \aj, 119, 1239

\bibitem[Sneden (1973)]{Sn73}      
Sneden, C.~1973, \apj, 184, 839

\bibitem[Sneden (1999)]{Sn99}
Sneden, C.~1999, \apss, 265, 145

\bibitem[Sneden (2000)]{Sn00}
Sneden, C.~2000, in 35th Liege Int.~Ap.~Coll., The Galactic Halo,
from Globular Clusters to Field Stars, ed.~A.~Noels, P.~Magain, D.~Caro,
E.~Jehin, G.~Parmentier, \& A.~Thoul (Li\`ege Belgium: Institut
d'Astrophysique et de G\'eophysique), p.~159

\bibitem[Sneden \etal\ (1991)]{SKPL91}    
Sneden, C., Kraft, R.~P., Prosser, C.~F.~\& Langer, G.~E.~1991, \aj, 102, 2001

\bibitem[Sneden \etal\ (1992)]{SKPL92}    
Sneden, C., Kraft, R.~P., Prosser, C.~F.~\& Langer, G.~E.~1992, \aj, 104,
2121 (S92-M5)

\bibitem[Sneden \etal\ (1994)]{SKLPS94}
Sneden, C., Kraft, R.~P., Langer, G.~E., Prosser, C.~F.~\& Shetrone, M.~D.
1994, \aj, 107, 1773

\bibitem[Sneden \etal\ (1997)]{SKSSLP97}  
Sneden, C., Kraft, R.~P., Shetrone, M.~D., Smith, G.~H., Langer, G.~E., 
\& Prosser, C.~F.~1997, \aj, 114, 1964

\bibitem[Sneden \etal\ (2000)]{SPK00}
Sneden, C., Pilachowski, C.~A.~\& Kraft, R.~P.~2000, \aj, 120, 1351

\bibitem[Suntzeff (1993)]{Su93}      
Suntzeff, N.~B.~1993, in The Globular Cluster-Galaxy Connection, 
ed.~G.~H.~Smith \& J.~B.~Brodie, ASP Conf.~Ser.~48, 167

\bibitem[Th\'evenin (1990)]{Th90}
Th\'evenin, F.~1990, \aap, 82, 179

\bibitem[Th\'evenin \& Idiart (1999)]{TI99}
Th\'evenin, F.~\& Idiart, T.~P.~1999, \apj, 521, 753 (TI99)

%\bibitem[Th\'evenin \etal\ (2001)]{Teta01}
%Th\'evenin, F., Charbonnel, C., de Freitas Pacheco, J.~A., Idiart, T.~P.,
%Jasniewicz, G., de Laverny, P., \& Plez, B.~2001, \aap, Submitted, 
%astro-ph/0105166

\bibitem[Vanture \etal\ (1994)]{VWB94}
Vanture, A.~D., Wallerstein, G.~\& Brown, J.~A.~1994, \pasp, 106, 835

\bibitem[Vogt (1987)]{Vo87}      
Vogt, S.~S.~1987, \pasp, 99, 1214

\bibitem[Vogt \etal\ (1994)]{Voetal94}
Vogt, S.~S., Allen, S.~L., Bigelow, B.~C., Bresee, L., Brown, B., 
Cantrall, T., Conrad, A., Couture, M., Delaney, C., Epps, H.~W., Hilyard, D., 
Hilyard, D.~F., Horn, E., Jern, N., Kanto, D., Keane, M.~J., Kibrick, R.~I., 
Lewis, J.~W., Osborne, J., Pardeilhan, G.~H., Pfister, T., Ricketts, T., 
Robinson, L.~B., Stover, R.~J., Tucker, D., Ward, J.~\&  Wei, M.~Z.
1994, SPIE, 2198, 362

\bibitem[Wallerstein (1962)]{Wal62}
Wallerstein, G.~1962, \apjs, 6, 407

\bibitem[Wallerstein \etal\ (1997)]{Wetal97}   
Wallerstein, G, Iben, I.~Jr., Parker, P., Boesgaard, A.~M., Hale, G.~M., 
Champagne, A.~E., Barnes, C.~A., K{\"a}ppeler, F., Smith, V.~V.,
Hoffman, R.~D., Timmes, F.~X., Sneden, C., Boyd, R.~N., Meyer, B.~S., 
\& Lambert, D.~L.~1997, Rev.~Mod.~Phys., 69, 995

\bibitem[Zinn \& West (1984)]{ZW84}      
Zinn, R.~\& West, M.~J.~1984, \apjs, 55, 45

\end{thebibliography}
\end{document}